%% file: feasible_contact_tracing.tex
\documentclass[11pt]{article}
\usepackage[toc,page,titletoc]{appendix}
\usepackage[bottom]{footmisc}
\usepackage{natbib}
\usepackage[margin=1in]{geometry}
\usepackage{graphicx}
\usepackage{comment}
\usepackage{wrapfig, multirow}
\usepackage{amsmath, amsfonts, amssymb, amsthm}
\usepackage{algorithm, algpseudocode}
\usepackage{subfiles}
\usepackage{subcaption}
\usepackage{setspace}
\usepackage{chngcntr}
\usepackage{xcolor}
\usepackage{enumitem, tabularx, booktabs, ragged2e}
\usepackage{cite} % Continuous number of citations
\usepackage{authblk}
\usepackage[colorlinks=true,citecolor=green,linkcolor=brown,urlcolor=blue,breaklinks]{hyperref}

\setcitestyle{authoryear}

\allowdisplaybreaks
\counterwithout{figure}{section} %%%Gives a continuous ordering of figures.

\usepackage{setspace}
\definecolor{orange}{rgb}{1,0.5,0}
\definecolor{blue}{rgb}{0.22, 0.58, 0.82}
\definecolor{green}{rgb}{0.2, 0.65, 0.32}
\definecolor{red}{rgb}{0.91, 0.26, 0.2}
\definecolor{purple}{rgb}{0.46, 0.21, 0.68}
\definecolor{brown}{rgb}{0.5, 0, 0.13}
\makeatletter
\def\mathcolor#1#{\@mathcolor{#1}}
\def\@mathcolor#1#2#3{%
	\protect\leavevmode
	\begingroup
	\color#1{#2}#3%
	\endgroup
}
\makeatother

\newcommand{\instructions}[1]{}
\usepackage[rightbars, color]{changebar}
\cbcolor{blue}
\newcommand{\abs}[1]{\left\lvert#1\right\rvert}
\newcommand{\E}{\mathbb{E}}
\newcommand{\muopt}{\mu^{\star}}
\newcommand{\prob}{\mathbb{P}}
\newcommand{\g}[1][G]{\mathcal{#1}}

\DeclareMathOperator*{\argmax}{\text{argmax}} % no space, limits underneath in displays

\newtheorem{theorem}{Theorem}[section] % <-- numbers by section
\newtheorem{lemma}[theorem]{Lemma}
\newtheorem{corollary}[theorem]{Corollary}

\theoremstyle{definition}
\newtheorem{definition}[theorem]{Definition}
\newtheorem{exampleplain}[theorem]{Example}
\newenvironment{example}
 {\pushQED{\qed}\exampleplain}
 {\popQED\endexampleplain}

\newcommand{\continuation}{??}

\theoremstyle{remark}

\theoremstyle{plain}% default
\newtheorem{innercustomgeneric}{\customgenericname}
\providecommand{\customgenericname}{}

\newenvironment{proofof}[1][]{\begin{trivlist}
\item[\hskip \labelsep {\bfseries Proof of #1.}]}{\hfill{}$\square$\end{trivlist}}
   
\begin{document}

\author[1]{Aparajithan Venkateswaran}
\author[2]{Jishnu Das}
\author[1]{Tyler H. McCormick}
\affil[1]{University of Washington}
\affil[2]{Georgetown University}
% \author[1]{Author 1}
% \author[2, 3]{Author 2}
% \author[1, 4]{Author 3}
% \affil[1]{Affiliation 1}
% \affil[2]{Affiliation 2}
% \affil[3]{Affiliation 3}
% \affil[4]{Affiliation 4}

%\title{Leveraging heterogeneity in infectivity to improve efficiency in contact tracing}
% \title{Heterogeneous reward distributions and regret in bandit algorithms: Evidence from graph exploration via contact tracing}
\title{Feasible contact tracing}
\date{\today}
% \date{}
\maketitle

\begin{abstract}
    Contact tracing is one of the most important tools for preventing the spread of infectious diseases, but as the experience of COVID-19 showed, it is also next-to-impossible to implement when the disease is spreading rapidly. We show how to substantially improve the efficiency of contact tracing by combining standard microeconomic tools that measure heterogeneity in how infectious a sick person is with ideas from machine learning about sequential optimization. Our contributions are twofold.  First, we incorporate heterogeneity in individual infectiousness in a multi-armed bandit to establish optimal algorithms.
    At the heart of this strategy is a focus on learning. In the typical conceptualization of contact tracing, contacts of an infected person are tested to find more infections.  Under a learning-first framework, however, contacts of infected persons are tested to ascertain whether the infected person is likely to be a ``high infector'' and to find additional infections only if it is likely to be highly fruitful. Second, using three administrative contact tracing datasets from India and Pakistan during COVID-19, we demonstrate that this strategy improves efficiency. Using our algorithm, we find 80\% of infections with just 40\% of contacts while current approaches test twice as many contacts to identify the same number of infections.  We further show that a simple strategy that can be easily implemented in the field performs at nearly optimal levels, allowing for, what we call, feasible contact tracing. These results are immediately transferable to contact tracing in any epidemic.  
\end{abstract}

\clearpage

\subfile{sections/introduction.tex}

\subfile{sections/related_work.tex}

\subfile{sections/theory.tex}

\subfile{sections/simulations.tex}

\subfile{sections/results_real_data.tex}

\subfile{sections/conclusion.tex}

\clearpage
\newpage
% \nocite{*}
\bibliographystyle{abbrvnat}
\bibliography{feasible_contact_tracing}

\clearpage

\appendix

\begin{center}
\textbf{\large Supplemental Materials}
\end{center}
\renewcommand*{\thesection}{S\arabic{section}}

%%%%%%%%%% Prefix a "S" to all equations, figures, tables and reset the counter %%%%%%%%%%
\setcounter{equation}{0}
\setcounter{figure}{0}
\setcounter{table}{0}
\makeatletter
\renewcommand{\theequation}{S\arabic{equation}}
\renewcommand{\thefigure}{S\arabic{figure}}
\renewcommand{\thetable}{S\arabic{table}}
%%%%%%%%%% Prefix a "S" to all equations, figures, tables and reset the counter %%%%%%%%%%

\subfile{sections/appendix_ag.tex}

\subfile{sections/appendix_pilot.tex}

\subfile{sections/appendix_asymptotics.tex}

\subfile{sections/appendix_algorithms.tex}
\clearpage

\subfile{sections/appendix_additional_sims.tex}

\subfile{sections/appendix_real_data.tex}

\subfile{sections/appendix_useful_results.tex}

\end{document}

%% file: sections/introduction.tex
\section{Introduction}
\label{section:introduction}

Contact tracing, in combination with quarantines and other pharmaceutical and non-pharmaceutical interventions, is a key weapon in the fight against infectious diseases. It proceeds as a branching process on the network of contacts \citep{huerta2002contact, lloyd2005superspreading}. Assuming that transmission events are independent of each other, we begin with an arbitrary infected person and test each of their contacts for the infection. If the test returns positive, indicating the presence of an infection, we start the process anew with a new set of potentially infected people.

Contact tracing has been deployed widely and effectively in infectious diseases that are geographically localized or spread slowly, such as Ebola or HIV \citep[for example][]{hyman2003modeling, saurabh2017role}. However, contact tracing poses substantial logistical and financial challenges for infections that spread more quickly and easily. During the COVID-19 pandemic, for example, contact tracing proved challenging, and in countries like the United Kingdom, many contacts could not be reached or tested in time. This is not surprising -- in typical networks, while most people have very few contacts, some may have thousands, especially if they are in public-facing jobs. Given the wide variation in the number of contacts, the question we ask is whether there are ways to reduce the number of contacts who need to be tested without a commensurate decline in the number of new infections uncovered. Interestingly, while attempts to improve the efficacy of contact tracing have focused on innovations that allow more contacts to be reached through the use of cell phones and other passive data \citep[see][]{danquah2019use, zhao2020accuracy}, there is little research on whose contacts should be traced. Feasible contact tracing remains an elusive goal in the face of a rapidly moving pandemic like COVID-19.

Here, we develop the insight that the testing of any contact provides new information about the \textit{infector}, which can then be leveraged to improve contact tracing. We show that substantial gains are possible if there is heterogeneity in the likelihood that someone passes on the infection to others, a term we label per-contact infectivity (PCI), which could arise from either biological or behavioral factors.
To see how information about PCI interacts with contact tracing, suppose there are only two types of people in the population: those who pass on the infection with probability 1 to every person they meet, and those who never pass the infection i.e., pass on the infection with probability 0. In such a world, a simple algorithm that tests exactly one contact of every infected person, to begin with, and tests further contacts if and only if the tested contact is positive can result in substantial cost savings because additional tests do not provide more information about the infection status of untested contacts. 

We expand on this intuition in this paper. The core message is that when there is heterogeneity in PCI, there are massive gains to learning about PCI quickly.  We further contend that strategies that prioritize learning in this way are feasible in practice, for instance by testing a subset of people living with the person. We demonstrate that the kind of heterogeneity in PCI that we need to realize significant gains in the efficiency of contact tracing is consistent with the data from three different South Asian locations during COVID-19. Our feasible algorithms would have allowed 80\% of infections to have been detected by testing only 40\% of the contacts of infected persons in two of the datasets (from Punjab, India, and southern India) and 60\% of contacts in the third (from Punjab, Pakistan). In contrast, currently employed strategies that test all contacts of an infected person need to test 80\% of the population to uncover the same number of infections.

Our paper is structured as follows. We first pose contact tracing as a multi-armed bandit and review related works in Section \ref{section:lit}. In Section \ref{section:methods}, we establish a theoretical framework and provide new results on the asymptotic optimality of different contact tracing algorithms. Turning from asymptotic results, in Section \ref{section:sims} we then consider finite samples and show that the performance of the algorithms is sensitive to the specific distribution of infections. For instance, depending on the distributional parameters, one algorithm can outperform another, allowing us to outline specific policy actionable guidelines for the appropriate choice of an algorithm. In Section \ref{section:real-data}, we take these insights to administrative contact tracing data from Punjab (Pakistan), Punjab (India), and southern India during the COVID-19 pandemic. We estimate the distribution of PCI for each dataset and show that there is a large variation in PCI among infected persons in our settings. Then, we show that the bandit algorithms we outline are far more efficient than a naive sampling strategy and reconcile our findings with those from the empirical simulations. Finally, we conclude our discussion and provide directions for future research in Section \ref{section:conclusion}.

%% file: sections/related_work.tex
\section{Contact tracing as bandits}
\label{section:lit}

Bandit algorithms are a staple tool for sequential decision-making under uncertainty.  The general setup is as follows. A decision-maker faces a choice between several options.  For each option, there is a reward for choosing that option that comes from a probability distribution.  The reward distributions are not known in advance, so the decision-maker faces a trade-off.  Continuing to choose the same option provides a reliable reward, though moving to a different option might provide an even higher reward.  Under this uncertainty, the goal is to construct a sequence of decisions amongst the set of options that maximizes the reward.  As we discuss in the next section, there is a deep and active literature that both develops and evaluates bandit algorithms in a wide range of contacts.  In this section, we focus on the connection between these algorithms and contact tracing.  In the context of contact tracing, the set of decisions represents people whose contacts could be tested and the reward refers to the number of new infections the decision-maker discovers.

The key insight in our paper is that each person infects their contacts at a different rate, called per-contact infectivity (PCI).  In bandit language, this heterogeneity means that the reward distribution across people varies widely, with some infected people likely to infect many of their contacts while others (or most, as we see in our empirical examples) infect few or none. 
This leads us to consider two things for contact tracing. First, we want to identify highly infectious individuals. Second, we do not want to spend a lot of effort trying to determine who is the most infectious i.e., it is sufficient to find someone who is more infectious than the average person. This is called the \textit{explore-exploit} trade-off in the context of bandits. We want to \textit{exploit} the infectious people to uncover more positive infections while also \textit{exploring} to identify those who may be more infectious.

Earlier, we considered a scenario where the PCI is binary, 0 or 1. When the distribution of PCI is continuous, this becomes analogous to what's known as a multi-armed bandit problem. The possible \textit{arms} are infected individuals. \textit{Pulling an arm} refers to testing a contact of an infected person. The \textit{payoff} is measured through the number of infections identified and the \textit{time horizon} is a function of the number of people tested. In the bandit setting there is a trade-off between \emph{exploiting} infectious people available in a current state versus \emph{exploring} for more infectious people with greater payoffs. The optimal strategy determines whether to continue testing the contacts of someone with a known PCI or move to testing the contacts of someone whose PCI is unknown. 

Although there are clear parallels, contact tracing differs from the standard bandit in two ways. First, unlike a standard bandit where the arms are fixed, for an infectious disease the arms appear and disappear rapidly -- new people get sick and people who were previously sick either recover or die within a fairly short period relative to the length of the epidemic. Thus, it serves little to know the precise PCI of an arm if it exists only for a limited period. Second, since each person has a fixed number of contacts, there are only so many times each \textit{arm} can be \textit{pulled} and the potential rewards differ based on the number of contacts. This variant of the standard bandit is called the \textit{mortal multi-armed bandit} \citep{chakrabarti2008mortal}.\footnote{It is worth noting that mortal multi-armed bandits are similar to another class of bandit problems called \textit{rotting bandits} \citep{levine2017rotting}. In rotting bandits, the mean reward distributions are not stationary. In particular, the expected mean reward of each arm decays as a function of the number of times the arm has been pulled. In that sense, mortal bandits can roughly be seen as a discretization of rotting bandits where the decaying function is defined in a step-wise manner. \citet{seznec2019rotting} show that when the number of arms is fixed, rotting bandits are no harder than the standard stochastic bandit. However, when there are infinitely many arms, the problem becomes significantly harder \citep{kim2022rotting}. In our analysis, we will assume that there are only finitely many arms. And we leave the scenario with infinite arms for future research.} 

The mortal multi-armed bandit changes the goal of the learning algorithm away from finding the \textit{best} arm that can be exploited indefinitely to an approach where it is sufficient to find a \textit{good-enough} arm by placing emphasis on arms that live longer i.e., people with more contacts. For this problem, two algorithms called Adaptive Greedy and Stochastic Sampling (which we call Pilot Sampling) have been shown to be able to theoretically identify the maximum number of infections per test in the long run by \citet{chakrabarti2008mortal}. We take these algorithms one step further by additionally incorporating the reward distribution (the PCI distribution) to show that they remain optimal. In doing so, we also demonstrate the value of information about the PCI distribution in determining the optimal approach through empirical simulations. One crucial advantage in our setting is that our goal is to define a strategy that is \textit{better} than current approaches to contact tracing, which generally involves testing all contacts. In that sense, our problem is easier than finding the \textit{best} i.e., the optimal strategy that most other works tackle.

\subsection{Related literature}

There is a long history of research on bandits. Although the current version of the problem was formalized only in \citet{robbins1952some}, such sequential optimization problems were considered as early as \citet{thompson1933likelihood}. Optimal solutions to these problems using index-based rules have been discussed by \citet{gittins1979bandit}. Another class of bandit algorithms that came to be known as the upper confidence bounds (UCB) was first put forth by \citet{lai1985asymptotically}. There are other strategies such as greedy algorithms and probability matching. \citet{sutton2018reinforcement} and \citet{bubeck2012regret} provide an extensive literature review of bandit algorithms.\footnote{There are several variants of the classical bandit problem besides the mortal bandit such as \textit{contextual bandits} where we observe covariates as well as rewards \citep{langford2007epoch}, \textit{adversarial bandits} where an adversary changes the reward structure \citep{auer1998line}, \textit{infinite-armed bandits} where there is an infinite number of arms to play \citep{agrawal1995continuum}, and \textit{non-stationary bandits} where there is a drift in the mean rewards \citep{besbes2014stochastic}.}

Although bandits are predominantly associated with machine learning and computer science, there is a growing literature that uses bandits to model human decisions. \citet{cohen2007should} suggest that humans make decisions in such exploration-exploitation problems using index rules, reminiscent of \citet{gittins1979bandit}. UCB-type algorithms have been used in decision-making by \citet{reverdy2014modeling} and \citet{wu2018generalization}. These have found applications in medical decision-making. For instance, \citet{frank2007custom} viewed treatment of depression as a bandit problem and \citet{currie2020understanding} show that more skilled doctors tend to favor a strategy with greater experimentation when searching for treatments for depression. Perhaps a more direct application of bandits is in designing experiments as noted by \citet{athey2019machine}. In fact, Thompson sampling, one of the earliest bandit algorithms put forth by \citet{thompson1933likelihood} was developed to guide data collection by identifying treatment arms that units should be assigned to. We add to this growing literature in economics by finding optimal contact tracing strategies using bandits.

Since 2020, the COVID-19 pandemic has sparked a body of work that explores the connection between contact tracing and bandits. For example, \citet{grushka2020framework} use a bandit-style framework by assigning risk scores and ranking individuals based on expressed symptoms and characteristics to identify people who need to be tested. However, they do not explicitly leverage heterogeneity in infectivity, which is where we show we can derive substantial gains. \citet{wang2020whom} construct an agent-based model to simulate infectious disease dynamics and use bandits to perform contact tracing. \citet{bastani2021efficient} use batched bandits to identify groups of people to test at nation borders and found 1.85 times as many asymptomatic travelers as random surveillance testing. \citet{meister2021optimizing} model the spread of infection in two distinct phases, an infection phase, and a contact tracing phase, and derive optimal strategies.
\citet{chugg2021reconciling} study a related problem of estimating the prevalence of the disease at a given time step using bandits.  

Additionally, some work has identified the heterogeneity in infectiousness that we exploit here.
\citet{hagenaars2004spatial} explore this in a spatial context.
\citet{bolzoni2007transmission} caution that disease control strategies should account for heterogeneity and \citet{miller2007epidemic} found that epidemics are more likely when variance in infectivity is large. More recently, \citet{arinaminpathy2020quantifying} quantified heterogeneity in infectivity in the transmission of COVID-19.

The key distinction in our work is that we exploit this heterogeneity to motivate a learning-first perspective for contact tracing. That is, in the presence of heterogeneity, the priority is to learn about the PCI of the \emph{infector}. The infector's PCI gives the decision-maker critical information about the reward distribution, resulting in more efficient decisions about who to test next. Based on that key shift in perspective, we make three contributions to the literature.

First, we consider active learning algorithms arising from heterogeneity in PCI in the context of contact tracing. Although previous contributions have proposed active learning algorithms and identified heterogeneity in infectivity for multiple infectious diseases, they have not been addressed jointly in the literature to date. Second, in merging the two and being the first to draw the connection to mortal bandits, we also provide novel theoretical results on the asymptotic properties of two bandit algorithms. Third, ours is also the first paper to use data on contact tracing to show that the use of bandit algorithms can lead to marked declines in the fraction of contacts who need to be sampled without a commensurate loss in the number of infected individuals identified. Taken together, our results provide policy-actionable and feasible methods for contact tracing in the field.

%% file: sections/theory.tex
\section{Methodology}
\label{section:methods}

In this section, we present new results on the optimality of different strategies. We formally introduce the mortal bandit problem and define loss functions to quantify performance called regret and Bayesian regret. Then, we present a lower bound on the Bayesian regret. Next, we introduce two commonly used algorithms for the mortal bandit setting: Adaptive Greedy and Pilot Sampling. Then, we provide novel bounds on the Bayesian regret.
In deriving these bounds, we assume that no new arms appear i.e., arms can only die. In practice, this means that a policymaker has a fixed group of infected persons and must decide how to allocate tests amongst the contacts of those individuals. Finally, we describe the intuition behind a variant of these algorithms where we sample arms by lifetime, which in the context of contact tracing corresponds to sampling based on the total number of contacts.  We show that this strategy does not change our asymptotic optimality results.

\subsection{Problem setup and notation}

Consider the bandit problem with $N$ arms (infected individuals). Pulling an arm $i$ (testing a contact of $i$) rewards the decision-maker with a reward $X_i$ ($X_i = 1$ if contact is positive and 0 otherwise). This reward comes from a distribution $P_{\mu_i}$ with unknown mean $\mu_i \in [0, 1]$ (the PCI). Each arm can only be pulled for a maximum of $L_i$ times where $L_i$ is known. We call $L_i$ the lifetime of the arm (the number of contacts of $i$) and say that arm $i$ is alive at time $t$ if we haven't already pulled it $L_i$ times at time $t$.  In other words, the infected person still has more contacts left to test. An arm is playable (more contacts can be tested) if and only if it is alive. This is in contrast to the standard bandit problem where there is no limit to the number of times an arm can be pulled.  In a setup where arms have limitless numbers of pulls, the goal of the bandit algorithm is to find the arm with maximum payoff (explore) and then play the maximally rewarding arm indefinitely (exploit).  In contact tracing this trade-off is less straightforward, since even the most productive arm will eventually die (i.e. all contacts will be tested). We will assume that $\{(\mu_i, L_i)\}_{i=1}^N$ are independently and identically distributed (i.i.d.) from some joint prior $\Gamma$. We will denote the marginal distribution of $\mu$ as $\Gamma_{\mu}$.

The agent sequentially pulls arms in order to maximize cumulative reward over $T$ turns (the number of tests available). Let $H_t = \{(a_{\tau}, X_{a_{\tau}, \tau})\}_{\tau=1}^{t-1}$ denote the history of the decision maker's actions $a_{\tau}$ and the corresponding rewards $X_{a_{\tau}, \tau}$ up to time $t-1$. Then, we define the decision maker's policy $\pi$ as a mapping from the history $H_t$ to the next action $a_t \in \{1, \dots, N\}$.

Define $\muopt_t = \max_{i} \mu_i \times \mathbb{I}\{\text{arm $i$ is alive at $t$}\}$. For a fixed set of mean rewards $(\mu_1, \dots, \mu_N)$ and some history of actions according to a policy $\pi$, we define three different kinds of regret for the decision maker as follows:
\begin{align*}
    R_T(\pi \mid \mu) &= \sum_{t=1}^T \muopt_t - X_{a_t, t} &\text{Realized regret} \\
    \E[ R_T (\pi \mid \mu) ] &= \sum_{t=1}^T \muopt_t - \mu_{a_t} &\text{Mean regret} \\
    BR_{T, N} (\pi) &= \E_{\Gamma} \E [R_T(\pi \mid \mu)] &\text{Bayesian regret}
\end{align*}

There are two sources of randomness. The first source is reward realization as we are assuming that the reward generation process is stochastic. This is reasonable even though the lifetime is finite because we do not know anything about the arm: we have a list of contacts but we have no information about how many contacts are infected. The second source of randomness is the distribution of $\mu_i$ itself. This comes from the fact that there is heterogeneity in PCI i.e., some people are more infectious than other people as shown by \citet{arinaminpathy2020quantifying} for COVID-19.

Given the stochastic nature of the problem, instead of analyzing realized regret, we \textit{average} out the randomness and analyze the resulting Bayesian regret. The goal of the policymaker is then to reduce the Bayesian regret i.e., their cumulative regret from pulling a sub-optimal arm over $T$ turns averaged across all possible mean rewards, lifetimes, and realizations of data.\footnote{Of course, one may imagine an adversarial scenario where the mean rewards of all arms are highly concentrated near the maximum mean reward. Then, regret, as defined here, may not be the right objective to minimize as it will remain small. Perhaps maximizing the number of infections identified is a better objective. We leave this adversarial setting as an open problem for future researchers.}

We use $\g[O], \Theta$ to denote the usual order asymptotics, and $\Tilde{\g[O]}$ to denote $\g[O]$ ignoring logarithmic factors. Formally, we say $f(x) = \g[O](g(x))$ if there is a $M > 0$ and $x_0 > 0$ such that for all $x \geq x_0$, $\abs{f(x)} \leq M g(x)$. In other words, asymptotically, $f(x)$ does not grow at a faster rate than $g(x)$. We say that $f(x) = \Theta(g(x))$ if there are constants $m, M > 0$ and $x_0 > 0$ such that for all $x \geq x_0$, $m g(x) \leq f(x) \leq M g(x)$. In other words, asymptotically, $f(x)$ and $g(x)$ grow at the same rate. Finally, we say $f(x) = \Tilde{\g[O]}(g(x))$ if there is a $k > 0$ such that $f(x) = \g[O](g(x) \log^k x)$. In other words, asymptotically, $f(x)$ does not grow at a faster rate than $g(x)$ up to some logarithmic factors.

\subsection{A lower bound on Bayesian regret}

The following definition, commonly used in the analysis of many-armed bandits, enables us to study the behavior of the mean rewards \citep[see][for examples]{wang2008algorithms, carpentier2015simple, bayati2020unreasonable}.

\begin{definition}[$\gamma$-regular distribution]
For $\gamma > 0$, a distribution $Q$ with support $[0, 1]$ is called $\gamma$-regular if $\prob_Q (\mu > 1 - \epsilon) = \Theta(\epsilon^{\gamma})$ when $\epsilon \to 0$.
\end{definition}

Commonly used distributions including the Beta distribution, which we will see throughout this paper, are $\gamma$-regular. In particular, for $\text{Beta}(\alpha, \beta)$, $\gamma = \alpha + \beta - 1$ whenever $\beta > 1$ and $\gamma = \alpha$ otherwise.

$\gamma$ controls the tail behavior of $\mu$ allowing us to bound regret in the worst-case scenario. Intuitively, when
\begin{enumerate}
    \item $\gamma < 1$, the density is concentrated towards 1,
    \item $\gamma = 1$, the density is (roughly) uniform near 1, and
    \item $\gamma > 1$, the density is concentrated away from 1. 
\end{enumerate}

When the density is concentrated towards 1, we expect a lot of arms to be highly rewarding as many people are highly infectious. So as $\gamma \ll 1$, the regret in the best case (the lower bound) will be smaller. The lower bound will be larger as $\gamma$ becomes large i.e., the density is concentrated away from 1. This is reflected in the theoretical lower bound in Theorem \ref{thm:lower-bound} which was originally shown for the standard bandit by \citet{bayati2020unreasonable} and we state without proof for the moral bandit case.

\begin{theorem}[Lower bound (Theorem 3.1 of \citet{bayati2020unreasonable})]
\label{thm:lower-bound}
Consider the mortal bandit setting. Suppose that mean rewards $\mu$ are drawn from a $\gamma$-regular distribution and that there is a constant $c$ such that $T, N \geq c$. Then, there is a constant an absolute constant $C$ such that for any policy $\pi$,
\begin{align}
    BR_{T, N}(\pi) &\geq C \min(N, T^{\gamma / (\gamma + 1)})
\end{align}
\end{theorem}

This theorem is important because it immediately places a benchmark against which we can measure the performance of our algorithms. If an algorithm approaches the lower bound, we know that its performance compares favorably to any other policy that may be considered. In fact, we will go one step better, by showing that both the algorithms we discuss for contact tracing asymptotically achieve the lower bound of Theorem \ref{thm:lower-bound} up to constant or logarithmic factors.

Our analysis throughout the rest of this paper will depend on (some subset of) the following assumptions:
\begin{enumerate}[label=(A\arabic*)]
    \item \label{assumption:prior} The joint distribution of the mean reward and lifetime is $(\mu, L) \sim \Gamma$ where the marginal $\Gamma_{\mu}$ is $\gamma$-regular.
    \item \label{assumption:minimum-arms} At any time during the game, we have at least $N_m \geq 1$ arms to play from.
    \item \label{assumption:subgaussian} The reward distribution $P_{\mu}$ is 1-subgaussian.
    \item \label{assumption:independence} The lifetime of an arm is independent of its mean reward i.e., $\mu_i \perp L_i$ for all arms $i$.
    \item \label{assumption:bern} The reward distribution is $P_{\mu} \equiv \text{Bern}(\mu)$
    \item \label{assumption:beta} The mean rewards for arms $\{\mu_i\}_{i=1}^N$ are i.i.d. $\text{Beta}(\alpha, \beta)$.
    \item \label{assumption:min-lifetime} The average lifetime $L \geq K$ for some $K > 0$.
\end{enumerate}

We now analyze the asymptotic properties of two algorithms that are used in this context.

\subsection{Adaptive Greedy sampling}

%=================
% Algorithm 3
%=================
\begin{algorithm}[!p]
\caption{Adaptive Greedy}\label{alg:adaptive-greedy}
\begin{algorithmic}[1]
\Require $T$ budget, $N$ arms
\State Set $t = 0$
\While {$t < T$}
    \State $t = t + 1$
    \If  {$t \leq N$}
        \State \label{line:estimate-mean} Pull arm $t$ and record reward
    \Else
        \State $X \sim \text{Bern}(\max_{i \text{ is alive}} \widehat{\mu}_i^t)$
        \If {$X = 1$}
            \State $i^* = \argmax_{i \text{ is alive}} \widehat{\mu}_i^t$
        \Else
            \State \label{line:ag-sampling} Sample $i^*$ uniformly from all available arms
        \EndIf
        \State Pull arm $i^*$ and update sample mean
        % \State \label{line:ag-branching} Add a new infection as an arm 
    \EndIf
\EndWhile
\end{algorithmic}
\end{algorithm}

% % %=================
% % % Algorithm 4
% % %=================
% % % \begin{algorithm}[!tbh]
% % % \caption{Subsampled Adaptive Greedy}\label{alg:ss-adaptive-greedy}
% % % \begin{algorithmic}[1]
% % % \Require $T$ budget, $N$ arms, $m$ sample size
% % % \State Sample $m$ arms at random from $N$ arms
% % % \State Run Adaptive Greedy algorithm (Algorithm \ref{alg:adaptive-greedy}) on the $m$ arms
% % % \end{algorithmic}
% % % \end{algorithm}

The first algorithm we discuss is called Adaptive Greedy sampling and it is based on greedy sampling \citep{chakrabarti2008mortal}. In greedy sampling, at time $t$, we pull an arm that has the largest sample mean at that time. Instead, Adaptive Greedy chooses to explore a different arm with probability $1 - \max_i \widehat{\mu}_i^t$. Here $\widehat{\mu}_i^t$ is the sample mean of arm $i$ at time $t$ and the $\max$ is taken over all arms that are alive. So if $\max_i \widehat{\mu}_i^t = 1$, then the algorithm decides to be greedy and pulls arm $\argmax_i \widehat{\mu}_i^t$ but if $\max_i \widehat{\mu}_i^t = 0$, the algorithm randomly pulls an arm that is available to play. This method is described in Algorithm \ref{alg:adaptive-greedy}.

\begin{example}[How does Adaptive Greedy work?]
\label{example:adaptive-greedy}
\emph{Suppose that mean rewards are $\mu \in [0, 1]$ and $\mu \sim \Gamma$. Suppose that there are only two arms with unknown rewards $\mu_1, \mu_2 \in [0, 1]$ and known lifetimes, $L_1, L_2$.}

Suppose that $\mu_1 = 0.1$ and $\mu_2 = 0.8$. Let $\widehat{\mu}_i$ be the current estimate of $\mu_i$. In Adaptive Greedy, we start with $\widehat{\mu}_i = 0$. Since $\max \widehat{\mu}_i = 0$, we will for sure explore the space of arms. Let's say, we pick arm 2 and pull it. Suppose we are rewarded. Then, we update $\widehat{\mu}_2 = 0.5$. Now, we explore with probability $ 1- \max \widehat{\mu}_i = 0.5$ and exploit arm 2 (current highest mean reward) otherwise. Suppose, we exploit arm 2 and are rewarded again. Then, $\widehat{\mu}_2 = 0.67$. Now, we exploit arm 2 with a larger probability, $0.67$. Maybe next time, we choose to explore arm 1 out of randomness and are not rewarded so $\widehat{\mu}_1 = 0$ still. As we can see, this algorithm can, in the long run, settle upon the arm with the largest mean reward. However, it is not strictly greedy as it does allow some random exploration based on how good our current best arm is.
\end{example}

We now present a new result in Theorem \ref{thm:ag-br} showing that Adaptive Greedy shares the same asymptotic behavior as Greedy sampling. While this is a new bound for the Bayesian regret, we note that \citet{traca2020reducing} state a novel bound for the mean regret. We provide a step-by-step technical discussion in Appendix \ref{section:proof-ag-BR}.

\begin{theorem}[Bayesian Regret for Adaptive Greedy]
\label{thm:ag-br}
Under the assumptions \ref{assumption:prior}-\ref{assumption:bern}, for $\epsilon \in (0, 1/3)$ and $N > \log T$,
\begin{align*}
    BR_{T, N}(AG) &= \begin{cases}
        \Tilde{\g[O]} \left( TN^{-1/\gamma} + N \min(\sqrt{T}, N^{1 / \gamma})^{1 - \gamma} \right), & \gamma < 1 \\
        \Tilde{\g[O]} \left( T N^{-1/\gamma} + N \right), & \gamma \geq 1
    \end{cases}
\end{align*}
\end{theorem}

Observe that when $N \leq T^{\gamma / (\gamma + 1)}$, $BR_{T, N}(\text{AG}) = \Tilde{\g[O]}(T^{\gamma / (\gamma + 1)})$ and otherwise, $BR_{T, N}(\text{AG}) = \Tilde{\g[O]}(N)$. To match the lower bound described in Theorem \ref{thm:lower-bound}, we subsample arms as in \citet{bayati2020unreasonable}. Therefore, when $N > T^{\gamma / (\gamma + 1)}$ we perform Adaptive Greedy  on a subset of $m = \Theta(T^{\gamma / (\gamma + 1)})$ arms to obtain $BR_{T, N}(\text{AG}) = \Tilde{\g[O]}(T^{\gamma / (\gamma + 1)})$. Thus, asymptotically, the Bayesian regret of adaptive greedy (and the subsampled version) matches the optimal lower bound up to some log factors.

\paragraph{Understanding Theorem \ref{thm:ag-br}}

Studying the behavior of Adaptive Greedy is difficult as both sources of randomness (that generate the mean reward and the reward realization) come into play immediately. So we can think of two complementary events. If we have bad luck, then no matter whose contacts we test (infectious person or not), we will never uncover new infections. In other words, the realized rewards are much much smaller than the true mean rewards (this can be quantified by invoking the sub-gaussian assumption). The first term in the asymptotic bound of Theorem \ref{thm:ag-br} defines the likelihood of having bad luck. If we have good luck, then there will be at least one infectious person who infects people at a rate close to their PCI. In other words, there is at least one arm whose realized rewards are close to the true mean reward. The second term in the asymptotic bound defines the regret in this situation.

To illustrate the role of $\gamma$, consider the case when $\gamma \ll 1$. Here, the PCI density is concentrated near 1. Since nearly everyone is highly infectious, the likelihood of bad luck is negligible. Therefore, the second term in the asymptotic bound plays a more important role.

\subsection{Pilot sampling}
\label{subsection:mortal-bandits}

%=================
% Algorithm 5
%=================
\begin{algorithm}[!p]
\caption{Pilot Sampling}\label{alg:pilot-sampling}
\begin{algorithmic}[1]
\Require $T$ budget, $N$ arms, $K$ pilot size
\State Set $t = 0$
\While {$t < T$}
    \State \label{line:pilot-sampling} Sample $i$ uniformly from all available arms
    \State Pull arm $i$ $\min\{T-t, K_i, L_i\}$ times
    \State $t = t + \min\{T, K_i, L_i\}$
    \If {Reward $> 0$ and $t < T$}
        \State Pull arm $i$ $\min\{T-t, L_i - K_i\}$ times
        \State $t = t + \min\{T-t, L_i - K_i\}$
    \Else
        \State Discard arm $i$
    \EndIf
    \State Add all new infections as new arms
\EndWhile
\end{algorithmic}
\end{algorithm}

% % %=================
% % % Algorithm 6
% % %=================
% % % \begin{algorithm}[!tbh]
% % % \caption{Subsampled Pilot Sampling}\label{alg:ss-pilot}
% % % \begin{algorithmic}[1]
% % % \Require $T$ budget, $N$ arms, $m$ sample size
% % % \State Sample $m$ arms at random from $N$ arms
% % % \State Run Pilot Sampling algorithm (Algorithm \ref{alg:pilot-sampling}) on the $m$ arms
% % % \end{algorithmic}
% % % \end{algorithm}

The second sampling method we consider is Stochastic sampling, which we call this \textit{Pilot Sampling} \citep{chakrabarti2008mortal}. The idea is to pull an arm a finite number of times that is smaller than its lifetime. If the sample mean of the arm based on this \textit{pilot} meets a predetermined threshold, then we deem this arm to be highly rewarding and pull it until it dies. 

In the context of contact tracing, pilot sampling means testing a pre-defined number of contacts, then only testing the remaining contacts if enough infections are found in the initial set (the \textit{pilot} set).  This approach would allow health officials to, for example, test a subset of the people living with an infected person and only test remaining contacts if enough family members test positive.  This approach also illustrates the gains of knowing the distribution of infectiousness since it prioritizes finding the most infectious individuals.  

This method is described in Algorithm \ref{alg:pilot-sampling}.\footnote{\citet{chakrabarti2008mortal} also describe a variant where an arm deemed to be highly rewarding can be discarded if its cumulative reward becomes too small. We do not consider it in our study.} We discuss how to optimally identify the pilot group size and threshold in Appendix \ref{section:pilot-group-size}.

\begin{example}[How does Pilot Sampling work?]
\label{example:pilot-sampling}
\emph{Suppose that mean rewards are $\mu \in [0, 1]$ and $\mu \sim \Gamma$. Suppose that there are only two arms with unknown rewards $\mu_1, \mu_2 \in [0, 1]$ and known lifetimes, $L_1, L_2$.}

Suppose, we set the parameters $K = 3$ and, as before, suppose that $\mu_1 = 0.1$ and $\mu_2 = 0.8$. Maybe we choose arm 1 first. We pull it $K = 3$ times and receive no reward. So, we move on to arm 2. Maybe we pull it $K = 3$ times and observe 2 rewards. Since we found at least one reward, we pull it until it dies. Then, we go to the next arm. If we are left with no more new arms, we can start playing arms from the discard pile.
\end{example}

We now present a Bayesian regret bound for Pilot sampling in Theorem \ref{thm:bayesian-regret-pilot} with the technical discussion and proofs presented in Appendix \ref{section:proof-pilot-BR}.

\begin{theorem}[Bayesian regret of pilot sampling]
\label{thm:bayesian-regret-pilot}
Suppose that assumption \ref{assumption:prior} and \ref{assumption:min-lifetime} hold. Then, $BR_{T, N}(\text{Pilot}) = \g[O](\min\{N, T\})$.
\end{theorem}

For $N \leq T^{\gamma/(\gamma + 1)}$, $BR_{T, N}(\text{Pilot}) = \g[O](N)$. For $N > T^{\gamma/(\gamma + 1)}$, running pilot sampling on a subset of $m = \Theta(T^{\gamma/(\gamma + 1)})$ arms gives $BR_{T, N}(\text{Pilot}) = \mathcal{O}(T^{\gamma/(\gamma + 1)})$. Thus, asymptotically, Pilot sampling achieves the same order as the lower bounds in Theorem \ref{thm:lower-bound}. In particular observe that this does not have any log factors, unlike Adaptive Greedy.

\paragraph{Understanding Theorem \ref{thm:bayesian-regret-pilot}}

Pilot sampling is easier to analyze. In the worst case, we go through all $N$ arms or we exhaust our budget but are not rewarded at all. It is important to emphasize that both the Adaptive Greedy and the Pilot Sampling algorithms achieve the optimal lower bound in big-O asymptotics. That is, asymptotically, the difference between the Bayesian regret of Pilot Sampling and the optimal lower bound does not grow with $N$. This does not imply that it necessarily attains the optimal lower bound -- a point that we return to below. The exact asymptotic behavior has some dependence on the distribution of the lifetimes. Since we make no distributional assumption on the lifetimes, we don't see it in the big-O bound.

While these are highly favorable results for the algorithms we propose, in Appendix \ref{section:asymp-regret}, we also investigate when the asymptotic behavior described in Theorems \ref{thm:ag-br} and \ref{thm:bayesian-regret-pilot} are achieved. We find that this generally occurs in the range of $N$ between $10^3$ and $10^6$ depending on the distribution of mean rewards. The exception is when $\alpha < \beta < 1$ we need $N > 10^{12}$ before Adaptive Greedy achieves the asymptotic behavior. For these parameter values, the Beta distribution is mostly uniform, but rising in both tails ($\alpha$ controls the behavior on the left tail and $\beta$ on the right tail). Fortunately, as we will discuss below, the size of the population $N$ does not preclude real-world applicability in terms of dramatic efficiency gains.

\subsection{Sampling arms by lifetime}

An issue that we have not introduced thus far is that we expect and see in empirical data, that the number of contacts varies substantially between individuals.  In particular, we often see a right-skewed distribution of contacts where a small fraction of individuals have a very large number and most have substantially fewer.  We see this pattern in our empirical examples and it has also been documented extensively in work on measuring weak ties networks \citep[for example][]{mccormick2010many,diprete2011segregation}. Through a minimal example, we motivate a variant of the algorithms where we sample arms proportional to their lifetimes instead of uniformly.

\begin{example}[Sampling arms by lifetime]
\label{example:sample-by-degree}
\emph{Suppose that mean rewards are $\mu \in \{0, 1\}$ and that $\prob(\mu = 1) = p$. Then, pulling an arm just once will tell us what the mean reward is. If we pull arm $i$ and obtain $X_i = 0$, then we know that the mean reward of that arm is $\mu_i = 0$ and if we obtain $X_i = 1$, then we know $\mu_i = 1$. Note that in this scenario, Adaptive Greedy sampling and Pilot sampling are identical. Suppose that there are only two possible lifetimes $L_A$ and $L_B$ with $L_A > L_B$ where $\prob(L = L_A) = q$ and $\prob(L = L_B) = 1 - q$. Suppose that there are only two arms with unknown rewards $\mu_1, \mu_2 \in \{0, 1\}$ and known lifetimes, $L_1, L_2 \in \{L_A, L_B\}$. Let the budget be $T < L_1 + L_2$ (if $T \geq L_1 + L_2$, we get to pull both arms until their death and all policies are optimal).}

Suppose we pull arm $i$ first. Then, we can show that the expected reward $\g[R]_i$ (with respect to $\mu_1, \mu_2$) is
\begin{align*}
    \g[R]_i &= p^2 T + p(1-p) \left( \min\{T, L_i\} + \min\{T-1, L_{1 + \abs{i-2}}\} \right)
\end{align*}

Consider four policies $\pi_1, \pi_2, \pi_3$, and $\pi_4$. In $\pi_1$, we pull arm 1 first. In $\pi_2$,  we pull arm 2 first. In $\pi_3$, arms are chosen uniformly at random i.e., with probability 0.5. And in $\pi_4$, we choose arms with probability proportional to the lifetime i.e., with probability $L_i / (L_1 + L_2)$. The expected rewards (with respect to $\mu_i$) of these policies are
\begin{align*}
    \g[R](\pi_1) &= p^2 T + p(1-p) \left( \min\{T, L_1\} + \min\{T-1, L_2\} \right) \\
    \g[R](\pi_2) &= p^2 T + p(1-p) \left( \min\{T, L_2\} + \min\{T-1, L_1\} \right) \\
    \g[R](\pi_3) &= p^2 T  + \frac{p(1-p)}{2} \left(\min\{T, L_1\} + \min\{T, L_2\}  + \min\{T-1, L_1\} + \min\{T-1, L_2\} \right) \\
    \g[R](\pi_4) &= p^2T + \frac{p(1 - p)}{L_1 + L_2} \left( L_1 \left( \min\{T, L_1\} + \min\{T-1, L_2\} \right) +  L_2 \left( \min\{T, L_2\} + \min\{T-1, L_1\} \right) \right).
\end{align*}

Without loss of generality, assume $L_1 > L_2$. It is easy to see that $\g[R](\pi_1) \geq \g[R](\pi_4) \geq \g[R](\pi_3) \geq \g[R](\pi_2)$. Therefore, sampling arms based on lifetime matters. The intuition is that it is more rewarding to learn the mean reward of an arm that lives longer because we can exploit it for a longer time.
If we define $\kappa = L_2 / L_1 < 1$, then $\kappa$ represents the heterogeneity in lifetime -- a smaller $\kappa$ denotes a larger heterogeneity. As $\kappa \to 1$, $\g[R](\pi_4) \to \g[R](\pi_3)$ and as $\kappa \to 0$, $\g[R](\pi_4) \to \g[R](\pi_1)$. Essentially, $\pi_4$ represents the continuum between a policy that is agnostic to lifetimes (i.e., $\pi_3$) and a policy that greedily chooses arms with longer lifetimes (i.e., $\pi_1$). Thus, a larger heterogeneity results in a larger average reward when we sample by lifetime.

Now, we take the expectation of $\g[R](\cdot)$ with respect to $L_i$. Since $\E \g[R] (\pi_1) = \E \g[R](\pi_2)$, we will formulate $\pi_1$ as $\pi_1^{\prime}$, a policy that chooses the arm with the longer lifetime first, and remove $\pi_2$ from consideration. Simple calculations reveal that $\E \g[R](\pi_1^{\prime}) \geq \E \g[R](\pi_4) \geq \E \g[R](\pi_4)$. And a similar argument regarding heterogeneity as measured by $L_B / L_A$ can be made: $\pi_4$ represents a continuum between $\pi_3$ and $\pi_1^{\prime}$.
\end{example}

Although the analysis, as in Example \ref{example:sample-by-degree}, becomes complicated when we have more arms or allow $\mu$ to be continuous, the underlying intuition remains the same: estimating the mean reward for an arm with a longer lifetime is more rewarding than estimating the mean reward for an arm with shorter lifetime with the same precision. This is because we can play the arm with the longer lifetime for a longer time. Further, the additional reward we gain from sampling based on lifetimes is more pronounced when the heterogeneity in lifetime is larger. This variant is obtained by modifying line \ref{line:ag-sampling} of Algorithm \ref{alg:adaptive-greedy} and line \ref{line:pilot-sampling} of Algorithm \ref{alg:pilot-sampling} to sample by lifetime (or degree) instead of uniformly i.e., $\prob(\text{arm $i$}) \propto L_i$.

We've shown in the example above that the \emph{expected} reward from an arm is higher if the arm's lifetime is longer (i.e. a person has more contacts).  In our previous theoretical results, though, we've been concerned not with the expected performance but with the asymptotic big-O bound, which quantifies the order of the worst-case performance.  In what might seem like a paradoxical result, Corollaries \ref{corollary:ag-br-lifetime} and \ref{corollary:pilot-br-lifetime} say that the previously established asymptotics are unaffected.
This insight can be very useful in settings where we do not have access to the lifetimes of arms as it says that, asymptotically, one method does not outperform the other in the worst case. We provide a technical discussion of this argument in Appendix \ref{section:proof-ag-BR} and \ref{section:proof-pilot-BR}.

\begin{corollary}[Adaptive Greedy sampling by lifetime]
\label{corollary:ag-br-lifetime}
Under the assumptions \ref{assumption:prior}-\ref{assumption:bern}, for any $\epsilon \in (0, 1/3)$, the Bayesian regret of the adaptive greedy algorithm where we sample by lifetime obeys the same asymptotics in Theorem \ref{thm:ag-br}.
\end{corollary}

\begin{corollary}[Pilot sampling by lifetime]
\label{corollary:pilot-br-lifetime}
Suppose that assumptions \ref{assumption:minimum-arms}, and \ref{assumption:min-lifetime} hold. Then the Bayesian regret of the pilot sampling algorithm when choosing arms by lifetime obeys the same asymptotics in Theorem \ref{thm:bayesian-regret-pilot}.
\end{corollary}

Corollaries \ref{corollary:ag-br-lifetime} and \ref{corollary:pilot-br-lifetime} say that in the worst case, the asymptotic behavior of the Bayesian regret does not depend on whether we sample arms uniformly or by lifetime. This is because, in the worst case, we choose arms with the worst mean reward.  That is, the worst-case scenario is one where we happen to choose people with very low infectivity (and thus few infections). The result is still finding few infections, regardless of whether the person has many contacts or few.  Rather than focusing on the number of contacts (which could be logistically challenging to obtain in practice anyway), the crucial idea is to learn the mean reward of an arm quickly, regardless of how it is chosen, to help us decide whether to exploit that arm or explore other arms.\footnote{It is worth noting that focusing search on arms that live longer has been brought up in earlier research. \citet{traca2020reducing} describe a similar idea in mortal bandits by exploring arms that live longer. They restrict the exploration phase in Adaptive Greedy to arms that are in the top $k$\% of the distribution of the remaining lifetimes and they tune $k$ using a subset of available data. In contrast, our approach is free of hyperparameters and also works for Pilot sampling.}

\subsection{Summary of theoretical results}

We first showed that the theoretical lower bound on the Bayesian regret of standard bandits, established by \citet{bayati2020unreasonable}, extends to mortal bandits in Theorem \ref{thm:lower-bound}. Theorem \ref{thm:ag-br} then showed that the upper bound of Adaptive Greedy sampling achieves this lower bound up to some logarithmic factors in big $\g[O]$. Theorem \ref{thm:bayesian-regret-pilot} showed that Pilot sampling achieves the lower bound in big $\g[O]$. In Example \ref{example:sample-by-degree}, we demonstrated that sampling by lifetime can lead to a higher expected reward than when sampling arms uniformly. However, Corollaries \ref{corollary:ag-br-lifetime} and \ref{corollary:pilot-br-lifetime} show that sampling based on lifetime does not change the asymptotic behavior.

Next, we focus on empirical simulations of algorithmic performance for different distributions of mean rewards and lifetimes. This addresses three issues. First, as is well understood, matching orders in big $\g[O]$ does not mean that one algorithm is equally efficient as the other. This is evident from Example \ref{example:sample-by-degree}. Second, finite sample behavior (in terms of $N$) can be very different from asymptotic behavior (see Appendix \ref{section:asymp-regret}). Third, from a perspective based purely on asymptotic behavior, it is unclear if there is value in learning the distribution of the mean rewards i.e., $\gamma$, and the lifetimes.

%% file: sections/simulations.tex
\section{Simulations}
\label{section:sims}
\label{section:nobranching-sims}

We perform a series of numerical experiments to evaluate the performance of the algorithms described in Section \ref{section:methods}. We also compare these algorithms with a baseline naive sampler that picks an arm uniformly at random and pulls it until it dies, effectively not performing any learning, and with the widely used Thompson sampling algorithm, which we discuss in Appendix \ref{section:comparison-algorithms}. In particular, Thompson sampling has been shown to enjoy the best Bayesian regret bounds for a variety of model classes in the stochastic (non-mortal) setting \citep{russo2014learning}.

\subsection{Simulation setup}

For generating synthetic data, we fix $N$ arms with prior parameters for mean reward $(\alpha, \beta)$. For each arm $i$, we draw mean reward $\mu_i \sim \text{Beta}(\alpha, \beta)$ and lifetime $L_i \sim F$, where the choice of $F$ is described below. We randomly choose $X_{i} \sim \text{Binomial}(L_i, \mu_i)$ pulls as the rewards. We set a total budget of $T$ pulls.\footnote{ If we have $\sum_{i=1}^N L_i < T$, then we add new arms to the data until $\sum_{i=1}^N L_i > T$.} For prior distribution on the mean reward, we choose $(\alpha, \beta) \in \{(0.09, 0.6)$, $(1, 3)$, $(1, 1)$, $(10, 10)$, $(3, 1)$, $(0.6, 0.09\}$. This allows us to vary the skew of the mean rewards, which range from distributions that have some highly infective people (first and fifth) to those with a uniform or normal PCI distribution (third and fourth) and those where most people are not infective (the last). In Figure \ref{fig:beta-densities}, we visualize the densities of these distributions. Irrespective of the choice of $(\alpha, \beta)$, the priors for all Thompson sampling simulations were initialized at $\text{Beta}(1, 1)$, a uniform prior. For the average lifetime, we chose two different families of distribution, zero-truncated Poisson and Pareto. We chose Poisson since it is a relatively homogeneous distribution with the same mean and variance. We chose the mean to be 500. We chose Pareto as the other distribution as it is a heavy-tailed distribution. For Pareto, we fixed the location as 1 and chose the shape = 0.6. Based on the literature on social networks, we expect that the degree distribution across the population will be heavily right-skewed \citep[see][]{newman2003social, mccormick2010many, diprete2011segregation}.

The key result we will show is that when the degree distribution is heavy-tailed or when the PCI distribution is right-skewed, Pilot Sampling yields significant advantages over either the Adaptive Greedy or Thompson sampling algorithms.

\subsubsection{Simulation results}

\begin{figure*}[!p]

    \centering
    \begin{subfigure}[t]{0.5\textwidth}
        \centering
        \includegraphics[height=2in]{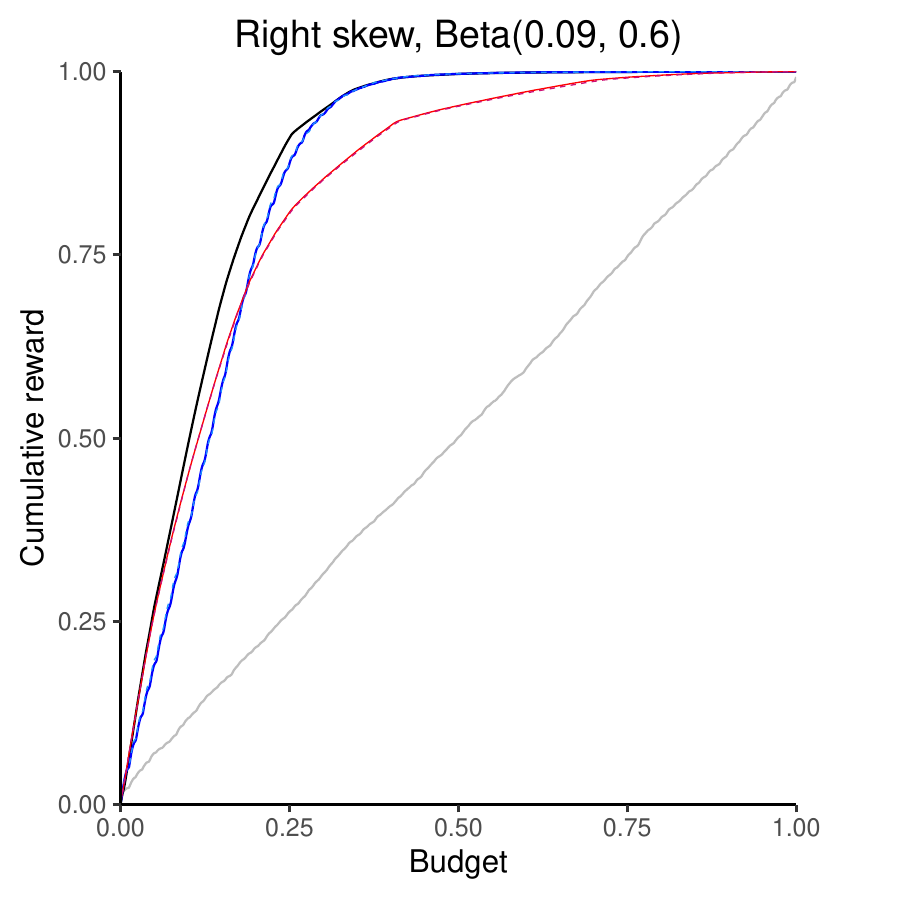}
    \end{subfigure}%
    ~
    \begin{subfigure}[t]{0.5\textwidth}
        \centering
        \includegraphics[height=2in]{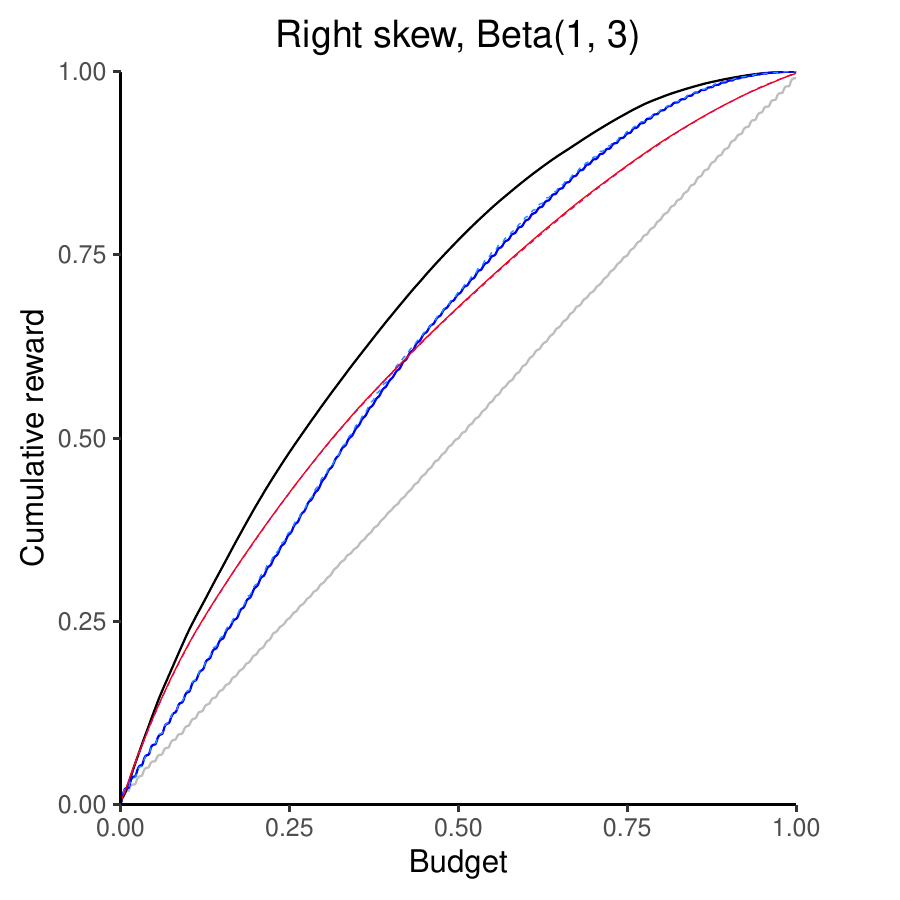}
    \end{subfigure}
    \\
    
    \begin{subfigure}[t]{0.5\textwidth}
        \centering
        \includegraphics[height=2in]{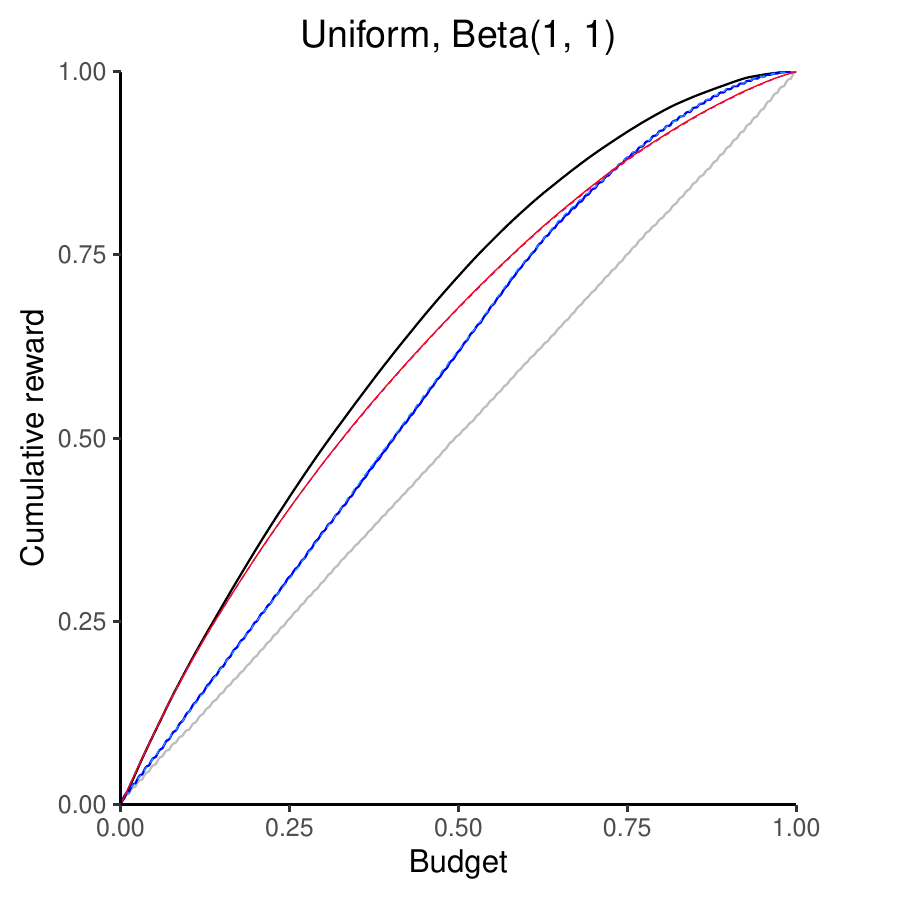}
    \end{subfigure}%
    ~
    \begin{subfigure}[t]{0.5\textwidth}
        \centering
        \includegraphics[height=2in]{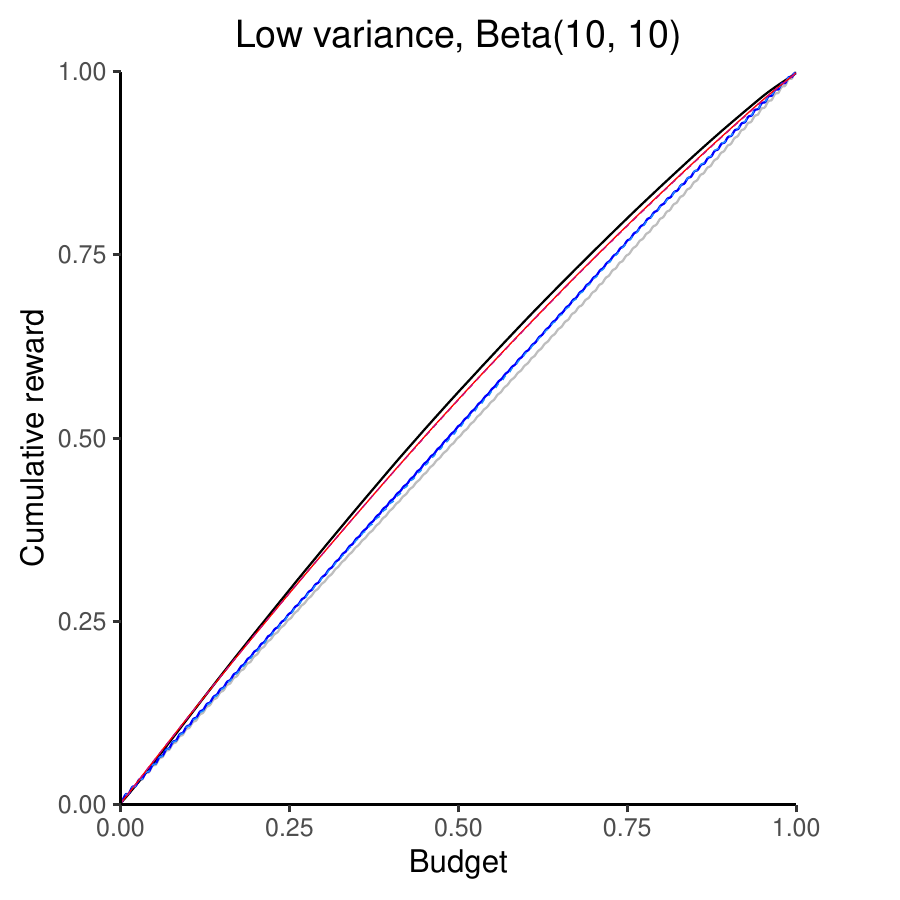}
    \end{subfigure}
    
    \begin{subfigure}[t]{0.5\textwidth}
        \centering
        \includegraphics[height=2in]{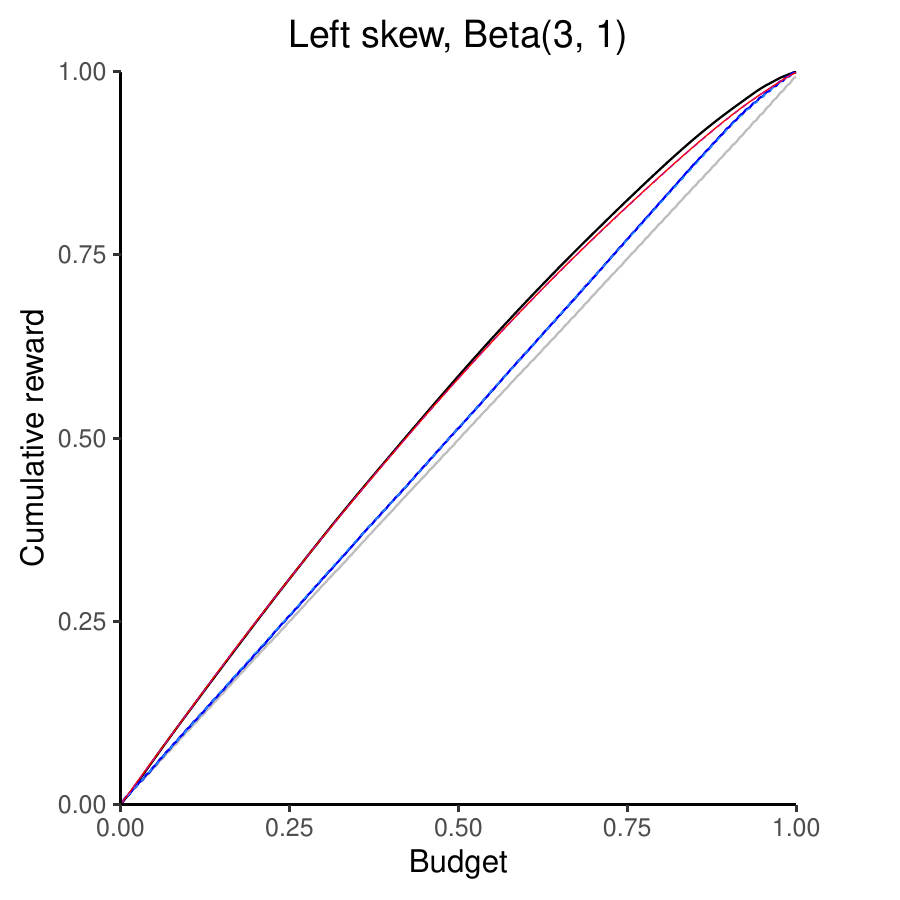}
    \end{subfigure}%
    ~
    \begin{subfigure}[t]{0.5\textwidth}
        \centering
        \includegraphics[height=2in]{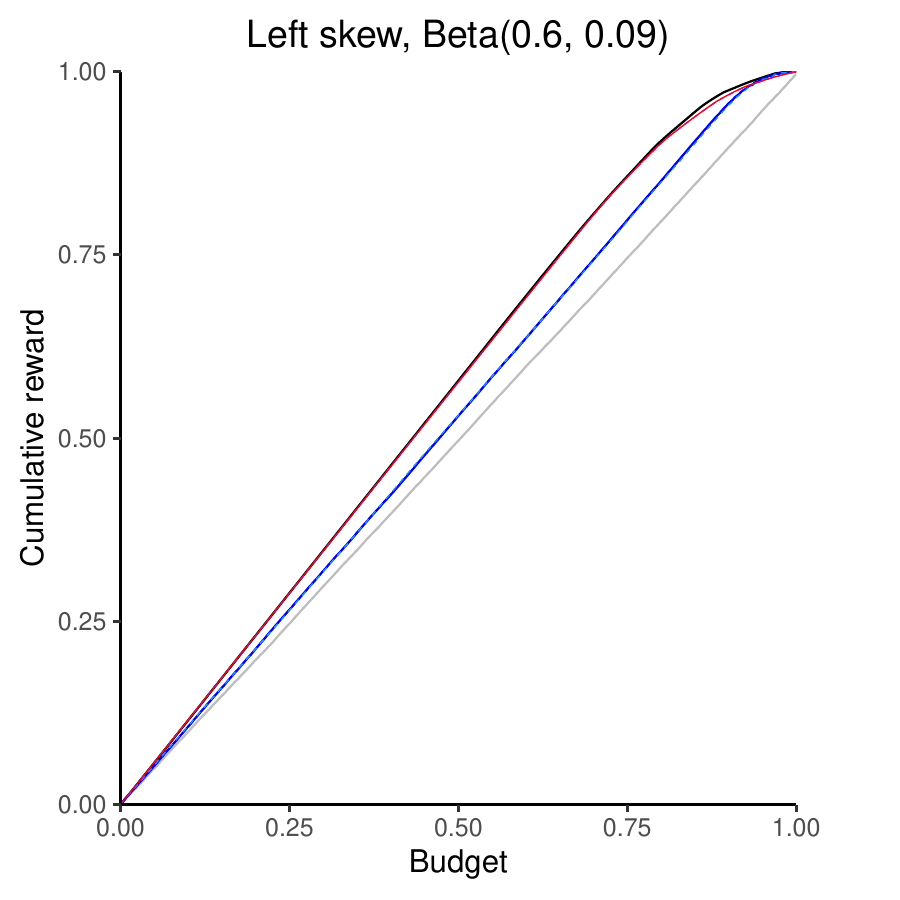}
    \end{subfigure}
    
    \caption{Cumulative reward over the total time horizon for different policies and various reward distributions based on data simulated without branching and Poisson(500) lifetime.  The axes are normalized to facilitate visual comparison. Thompson sampling is in black; pilot sampling with uniform sampling and lifetime sampling are in dark blue and light blue respectively; adaptive greedy with uniform sampling and sampling by lifetime are in red and pink respectively, and; naive sampling is in grey. Pilot sampling performs better when the rewards are heavily right skewed while adaptive greedy performs better in other scenarios. Thompson sampling appears to perform consistently well in all scenarios. In this setup, sampling by degree seems identical to sampling uniformly.}
    \label{fig:comparisons-large-degree}
\end{figure*}

First, consider the case where lifetimes are Poisson distributed. We set the initial number of arms $N = 10$, the average lifetime to $\lambda = 500$, and the budget to $T = 50,000$. The results from these simulations are shown in Figure \ref{fig:comparisons-large-degree}. For a heavily right-skewed distribution of the mean reward ($\text{Beta}(0.09, 0.6)$, Thompson and pilot sampling perform the best. For a distribution with low variance ($\text{Beta}(10, 10)$), Thompson and adaptive greedy perform the best. For a distribution that is not heavily skewed or does not have a small variance, all three methods perform roughly the same. Finally, for a left skewed distribution ($\text{Beta}(3, 1)$, $\text{Beta}(0.6, 0.09)$), Thompson and adaptive greedy perform the best. Overall, it seems like Thompson is consistently doing well in all situations while the other two algorithms fail in some extreme cases. While some of these conclusions agree with the bounds presented in Figure \ref{fig:regret-bounds}, it is clear that the bounds do not tell the full story. We perform similar simulations with $\lambda = 10$ and $T = 1000$ in Appendix \ref{section:sims-poisson-10}.

%=================
% Figure 1
%=================

\begin{figure*}[!p]

    \centering
    \begin{subfigure}[t]{0.5\textwidth}
        \centering
        \includegraphics[height=2in]{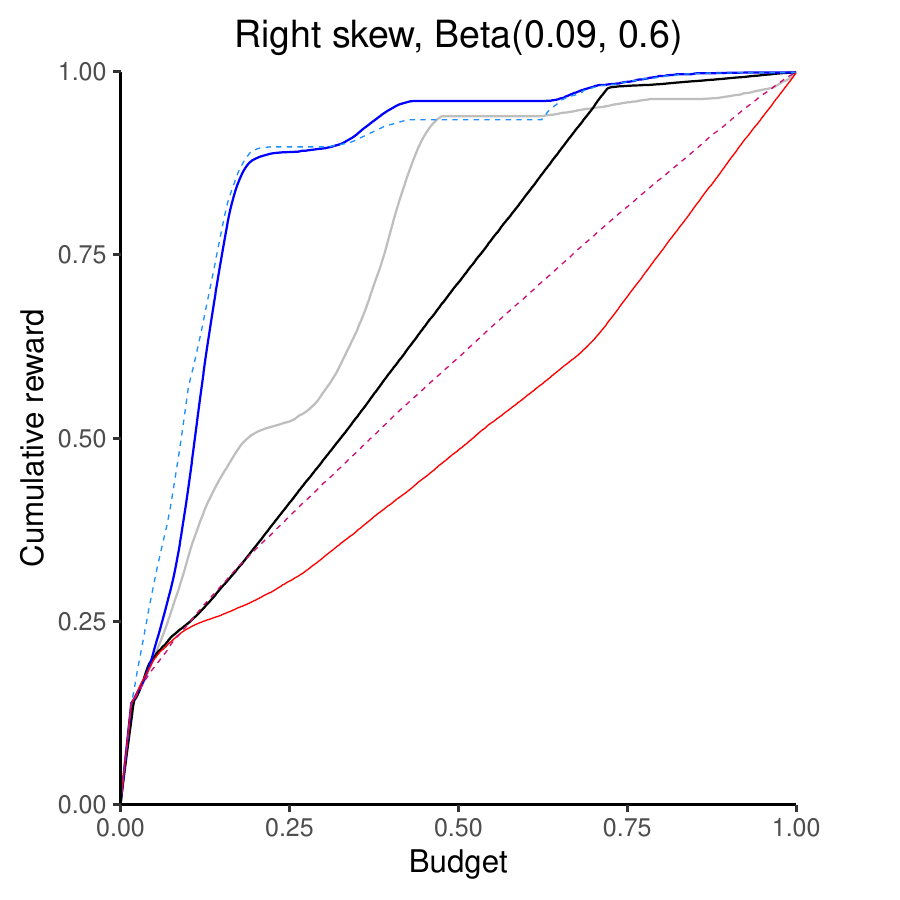}
    \end{subfigure}%
    ~
    \begin{subfigure}[t]{0.5\textwidth}
        \centering
        \includegraphics[height=2in]{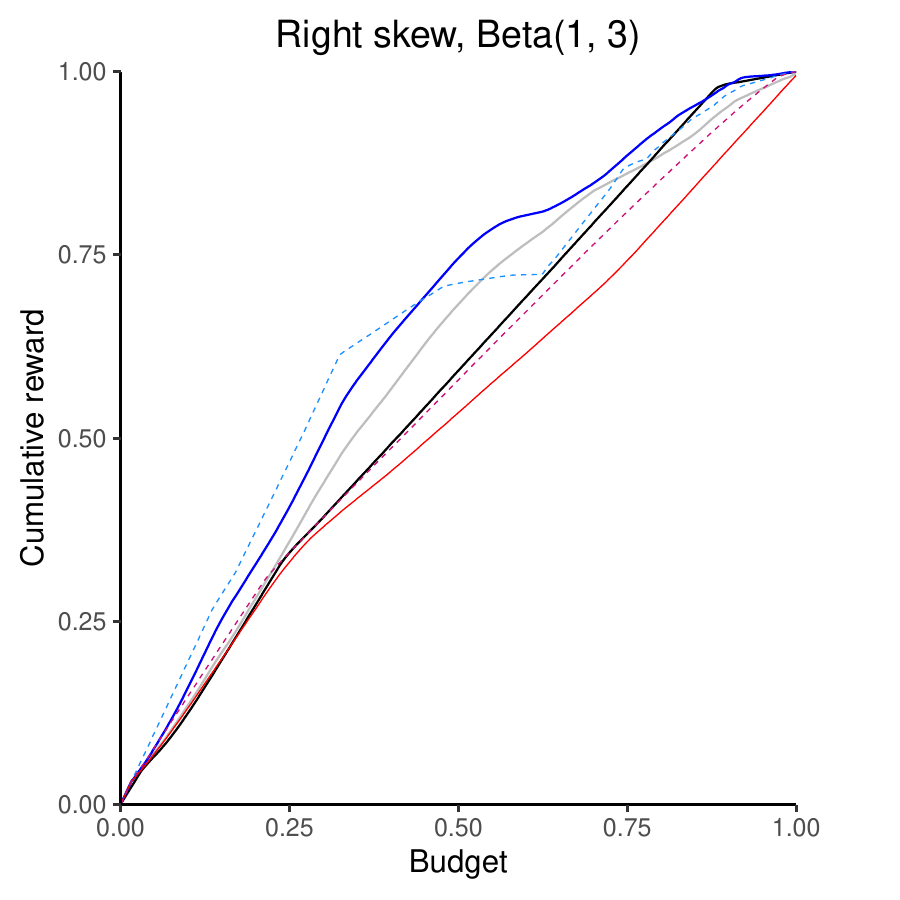}
    \end{subfigure}
    \\
    \begin{subfigure}[t]{0.5\textwidth}
        \centering
        \includegraphics[height=2in]{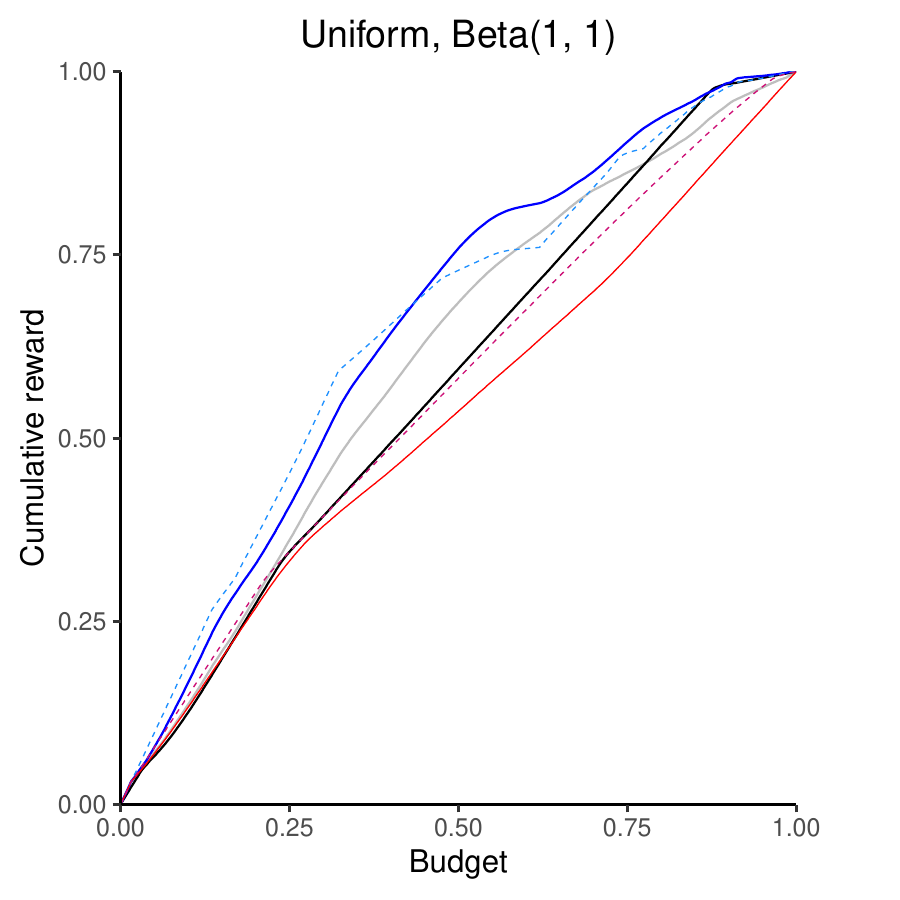}
    \end{subfigure}%
    ~
    \begin{subfigure}[t]{0.5\textwidth}
        \centering
        \includegraphics[height=2in]{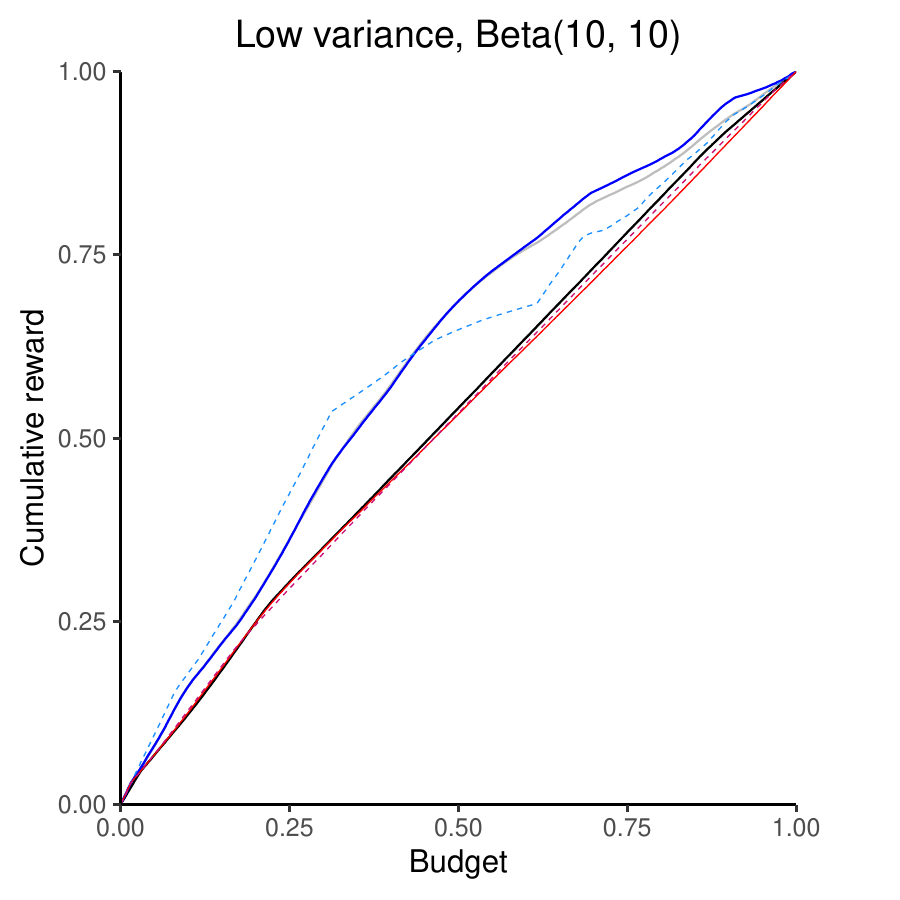}
    \end{subfigure}
    \\
    \begin{subfigure}[t]{0.5\textwidth}
        \centering
        \includegraphics[height=2in]{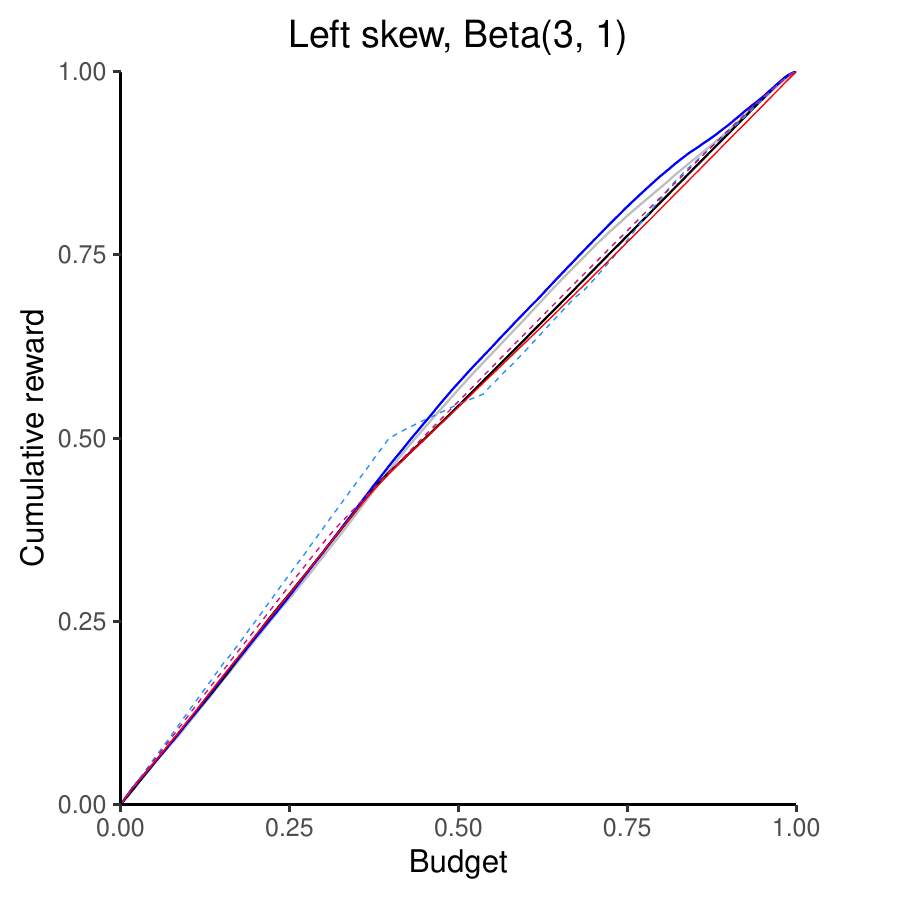}
    \end{subfigure}%
    ~
    \begin{subfigure}[t]{0.5\textwidth}
        \centering
        \includegraphics[height=2in]{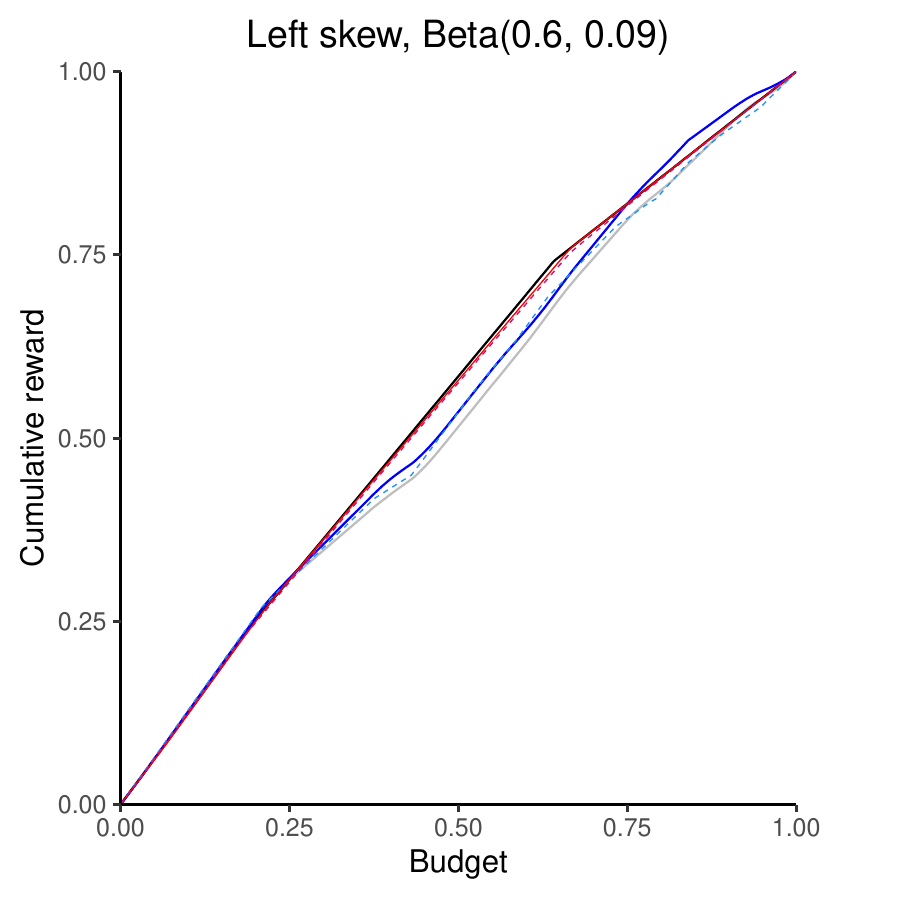}
    \end{subfigure}
    
    \caption{Cumulative reward over the total time horizon for different policies and various reward distributions based on data simulated without branching and Pareto(1, 0.6) lifetime.  The axes are normalized to facilitate visual comparison. Thompson sampling is in black; pilot sampling with uniform sampling and lifetime sampling are in dark blue and light blue respectively; adaptive greedy with uniform sampling and sampling by lifetime are in red and pink respectively, and; naive sampling is in grey. Pilot sampling outperforms all other methods in all scenarios except when rewards are heavily left-skewed. In pilot sampling, it is unclear whether sampling uniformly or by lifetime is better. For adaptive greedy, it is clear that sampling by lifetime dominates sampling arms uniformly.}
    \label{fig:heavy-tail}
\end{figure*}

Next, we used a heavy-tailed lifetime distribution. The lifetime was drawn from a Pareto distribution with shape 0.6 and location 1 (so the expectation diverges). Here, we fix the number of arms $N = 5000$. The results are shown in Figure \ref{fig:heavy-tail}. Note that pilot sampling consistently dominates the other algorithms. This is because pilot sampling commits to exhaustive testing when it finds a positive in the pilot group. In heavy-tailed scenarios, the regret in committing to a bad arm is small: (i) we choose an arm with an average lifetime and are done quickly or (ii) we choose an arm with a very high lifetime and are proportionally given a larger reward. In the Poisson case with no heavy tail, the loss in committing to a bad arm is relatively the same for all arms. Meanwhile, other algorithms are afraid of committing to exploit unless they are very sure. So they continue searching for a good arm for a long time. This is also the same reason why we see Naive sampling beat other methods in some cases. But since it does not intelligently choose to commit arms, Pilot sampling performs better. Also, observe that the sampling-by-lifetime variant of Adaptive Greedy dominates the lifetime-agnostic variant.

We see two main patterns in these simulations. First, as the mean rewards become more and more right-skewed, there is little difference in how all the methods compare against each other. This is due to the fact for a right-skewed distribution, the probability that we choose an arm with a high reward is larger than in the left-skewed case. So all methods perform relatively similarly in this case.

Second, Adaptive Greedy performs better when the mean rewards have low variability. For example, compare Beta(1, 1) And Beta(10, 10) in Figures \ref{fig:comparisons-large-degree} and \ref{fig:heavy-tail}. The reason is that Adaptive Greedy learns the mean reward of an arm by pulling it once. Thus, it takes more pulls in a high-variance setting than in a low-variance setting to learn the mean reward of the arm with the same precision. Compare this to Pilot sampling where we pull an arm multiple times to get a more precise estimate of its mean reward. In a high variance setting, Pilot sampling has an edge over Adaptive Greedy which can spend too much time exploring. In a low variance setting, Pilot sampling's gain in this precision by pulling an arm more than once is not substantial. Therefore, Pilot sampling can waste resources by making these unnecessary additional pulls that Adaptive Greedy does not. This difference does not seem to matter in cases where the lifetimes have a heavy tail but does become important when the distribution of the lifetime does not have a heavy tail.

%=================
% Figure 3
%=================

At first blush, these results appear to contradict our regret bounds in Theorems \ref{thm:ag-br} and \ref{thm:bayesian-regret-pilot}. Based on our discussion in Section \ref{section:methods}, most of our simulations, we appear to be in the regime where asymptotic behavior holds (also see Appendix \ref{section:asymp-regret}) and the asymptotic behavior says that Adaptive Greedy should perform worse than Pilot sampling which, based on simulations, is generally not true. In fact, this is only true in the heavy-tailed or heavily right-skewed cases. The reason this is not contradictory is that the asymptotic behavior only reflects the upper bounds of the Bayesian regret. This is a scenario where upper bounds are not indicative of the average-case behavior. We note that \citep{bayati2020unreasonable} also report very similar results for the standard bandit. They show that even though a greedy algorithm does not achieve universal rate optimality, it performs extremely well in practice.

\paragraph{Guidelines for choosing a sampler}
% \label{section:sims-summary}

The insights from these simulations are summarized in these guidelines: If the degree distribution is heavy-tailed, we should use Pilot sampling. If the degree distribution is not heavy-tailed, then we should use Pilot Sampling if the rewards are heavily right-skewed; use Adaptive Greedy or Thompson Sampling if rewards have a small variance or are heavily left-skewed, and; the choice does not matter in other cases. If the degree distribution is not heavy-tailed and we do not know how the rewards are distributed, we should default to Thompson Sampling as it consistently performs well. In Appendix \ref{section:branching-sims}, we show that these findings hold true in scenarios where new arms can appear. In Appendix \ref{section:asymp-regret}, we provide additional discussion of the asymptotic behavior and empirical simulations.

%% file: sections/results_real_data.tex
\section{Results on COVID-19 contact tracing}
\label{section:real-data}

A fundamental insight from the previous sections is that the algorithms we have proposed have similar asymptotic bounds, but the mean performance depends critically on the shape of the degree and PCI distribution, which are both empirical quantities. In the last section of the paper, we therefore turn to the data to estimate these distributions. Using these data, we implemented the different sampling policies described in Section \ref{section:methods}. These datasets were collected as a part of administrative contact tracing efforts during the outbreak of COVID-19 in India and Pakistan. The first dataset was collected from Punjab (Pakistan), the second dataset was collected from Punjab (India), and the third dataset was collected from southern India (from parts of Andhra Pradesh and Tamil Nadu). In Appendix \ref{section:real-data-appendix}, we summarize the properties of these datasets.

Prior to discussing our estimations, we must acknowledge that our datasets are not perfect, and there are a number of data gaps that we cannot address. For instance, the data collected from Punjab, Pakistan exhibits low infectivity because at the time of collection, 485,853 people were still awaiting test results and, given low infectivity, we treat these individuals as healthy in our simulations. Similarly, the data collected from southern India contains only summary-level information i.e., total counts of traced and infected people, rather than a `line-listing' of each individual contact and whether they were infected. Therefore, it was not possible to trace the infected individuals i.e., contact tracing does not proceed as a branching process. Finally, none of these data come from prospective studies with careful lab studies that can help establish the progeny of an infection. This implies, for instance, that if $B$ is a contact of $A$ and $B$ is now infected, we will follow the contact-tracing line and assume that the causal infection link went from $A$ to $B$, and not because $A$ and $B$ both were infected from a different source, or because $B$ infected $A$. This is a strong assumption, but it is likely to hold in the dataset from Punjab, India, that was carried out through the period of a stringent lockdown with very limited outside contact. 

Our idea then is not necessarily to demonstrate the value of our methods in a carefully prospective study, but rather to assess whether, in very different datasets from different settings and policies, we obtain similar results in terms of the shape of the degree and PCI distribution. To the extent that we do, it increases our confidence in the underlying estimates.

\subsection{Estimating individual heterogeneity in infectivity}

We estimated the parameters of the Beta model for the infectivity of the population using the full dataset with a Bayesian shrinkage estimator as in \citet{arinaminpathy2020quantifying}.

An immediate estimator of PCI for person $i$,  denoted by $\mu_i$, would be the ratio of infected, $z_i$, to the total number of people, $d_i$, that $i$ came into contact with i.e., $\widehat{\theta}_i = z_i / d_i$. However, since the distribution of degree is skewed, this naive estimator would have a different variance for each individual as the total number of contacts changes. Instead, we will use a Bayesian shrinkage estimator following \citet{arinaminpathy2020quantifying}. Therefore, the individual PCI estimates for high-contact individuals will remain mostly unchanged while those of low-contact individuals will be shrunken towards the overall mean.

In particular, we model the log odds of the individual PCI as following a normal distribution with a common mean and variance,
\begin{align}
    \mu_i &= \text{logit}^{-1}(\theta_i) = \frac{1}{1 + e^{-\theta_i}} \nonumber \\
    \theta_i &\sim N(\overline{\theta}, \sigma^2_{\theta}) \nonumber
\end{align}
where $\theta_i$ is the log-odds of $\mu_i$ and $\overline{\theta}$ is the overall mean. The variance $\sigma^2_{\theta}$ is inversely proportional to the shrinkage. As $\sigma^2_{\theta} \to 0$, $\mu_i \to \overline{\theta}$ and as $\sigma^2_{\theta} \to \infty$, $\mu_i \to z_i / d_i$. These hyperparameters are estimated using Monte Carlo Markov Chain methods with diffuse priors. Since $\overline{\theta}, \sigma^2_{\theta} > 0$, $\mu_i$ is guaranteed to be between 0 and 1. We refer the reader to the \citet[Supplementary]{arinaminpathy2020quantifying} for more details.

\begin{figure*}[!p]
    \centering
    \begin{subfigure}[t]{0.5\textwidth}
        \centering
        \includegraphics[width=3.5in]{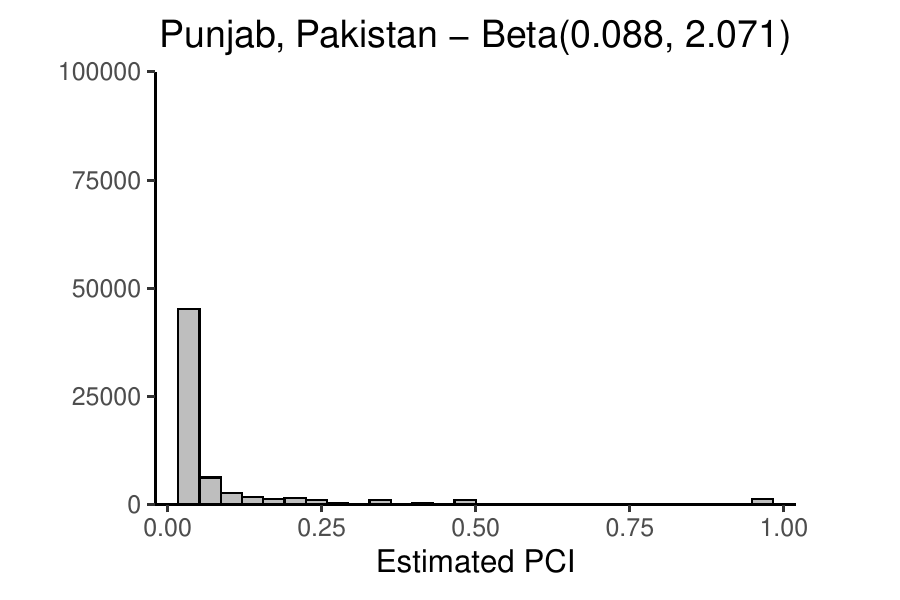}
    \end{subfigure}%
    ~ 
    \begin{subfigure}[t]{0.5\textwidth}
        \centering
        \includegraphics[width=3.5in]{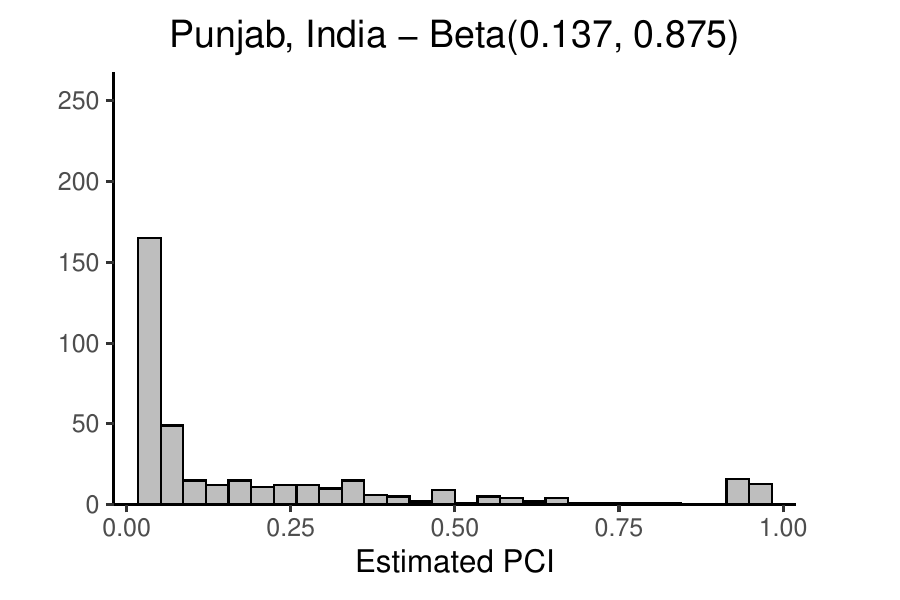}
    \end{subfigure}%
    \
    \begin{subfigure}[t]{0.5\textwidth}
        \centering
        \includegraphics[width=3.5in]{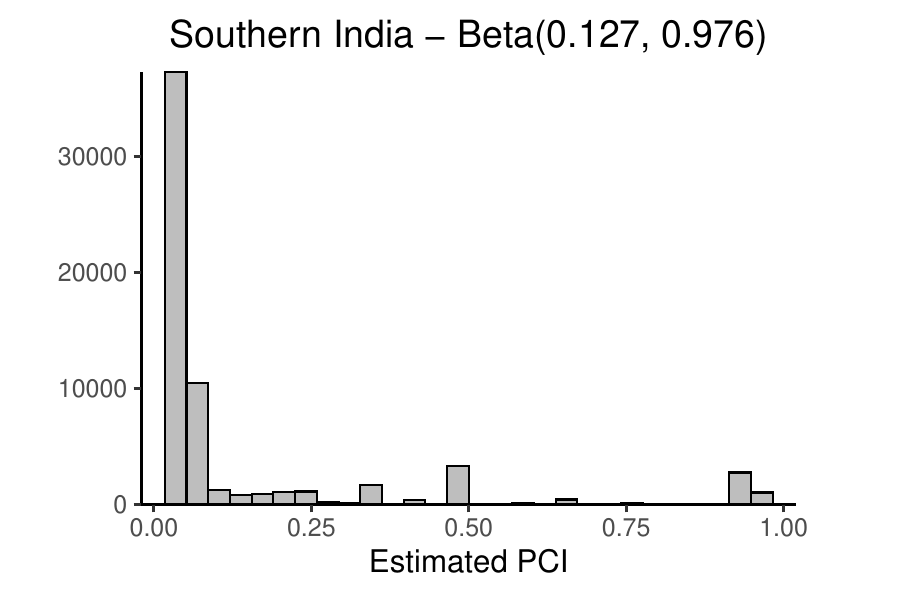}
    \end{subfigure}
    \caption{Distribution of estimated PCI from the three different datasets. PCI was estimated using a Bayes shrinkage estimator. The histogram of the PCI is shown in grey. Using the estimated PCI, parameters of a Beta distribution were fit using the method of moments. As we can see, the Beta distribution is heavily right-skewed.}
    \label{fig:estimated-pci}
\end{figure*}

 We visualize the distribution of infectivity as estimated using Bayes shrinkage in Figure \ref{fig:estimated-pci}.

\subsection{Results on contact tracing}

%=================
% Figure 4
%=================
\begin{figure*}[!p]
    \centering
    \begin{subfigure}[t]{0.5\textwidth}
        \centering
        \includegraphics[width=2.5in]{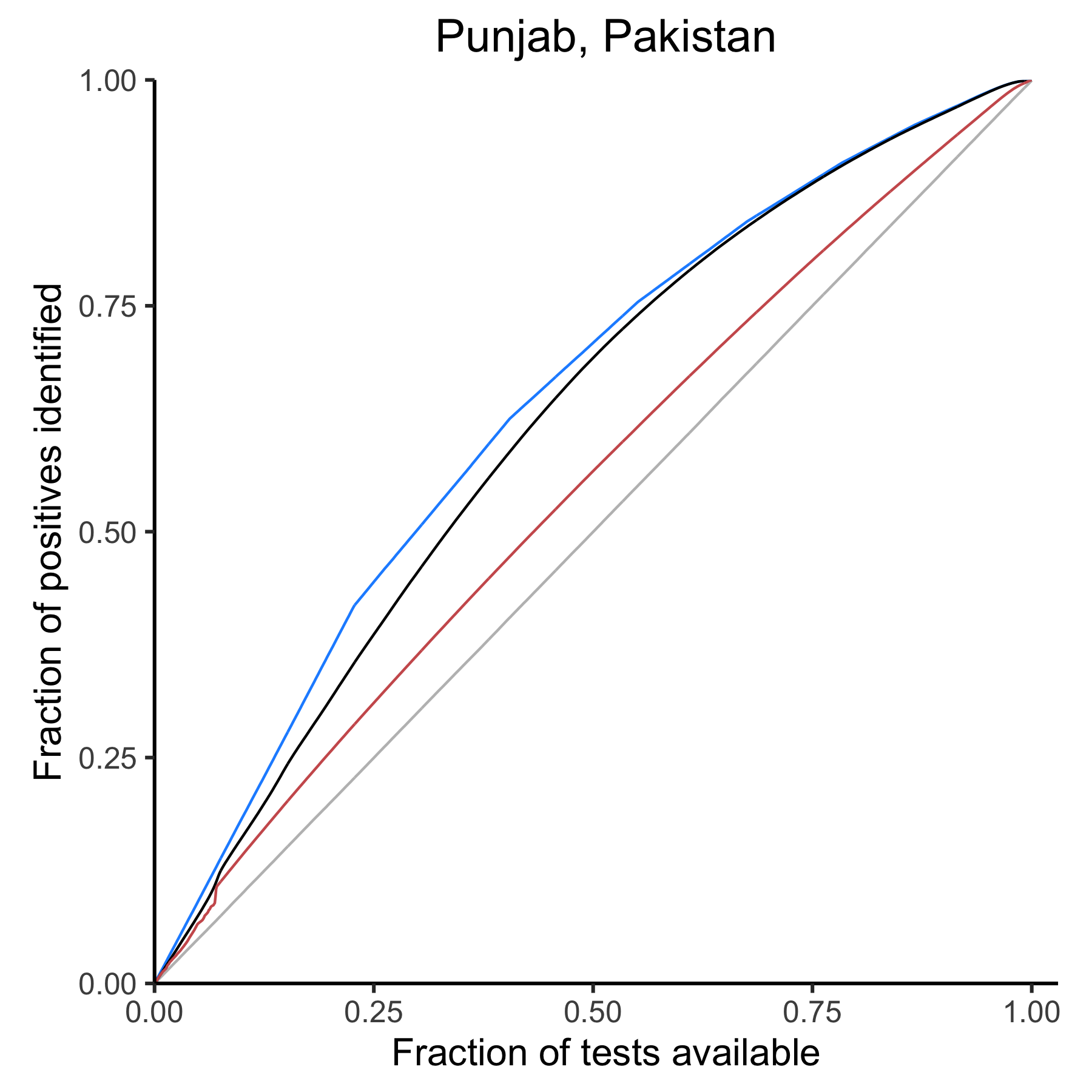}
    \end{subfigure}%
    ~ 
    \begin{subfigure}[t]{0.5\textwidth}
        \centering
        \includegraphics[width=2.5in]{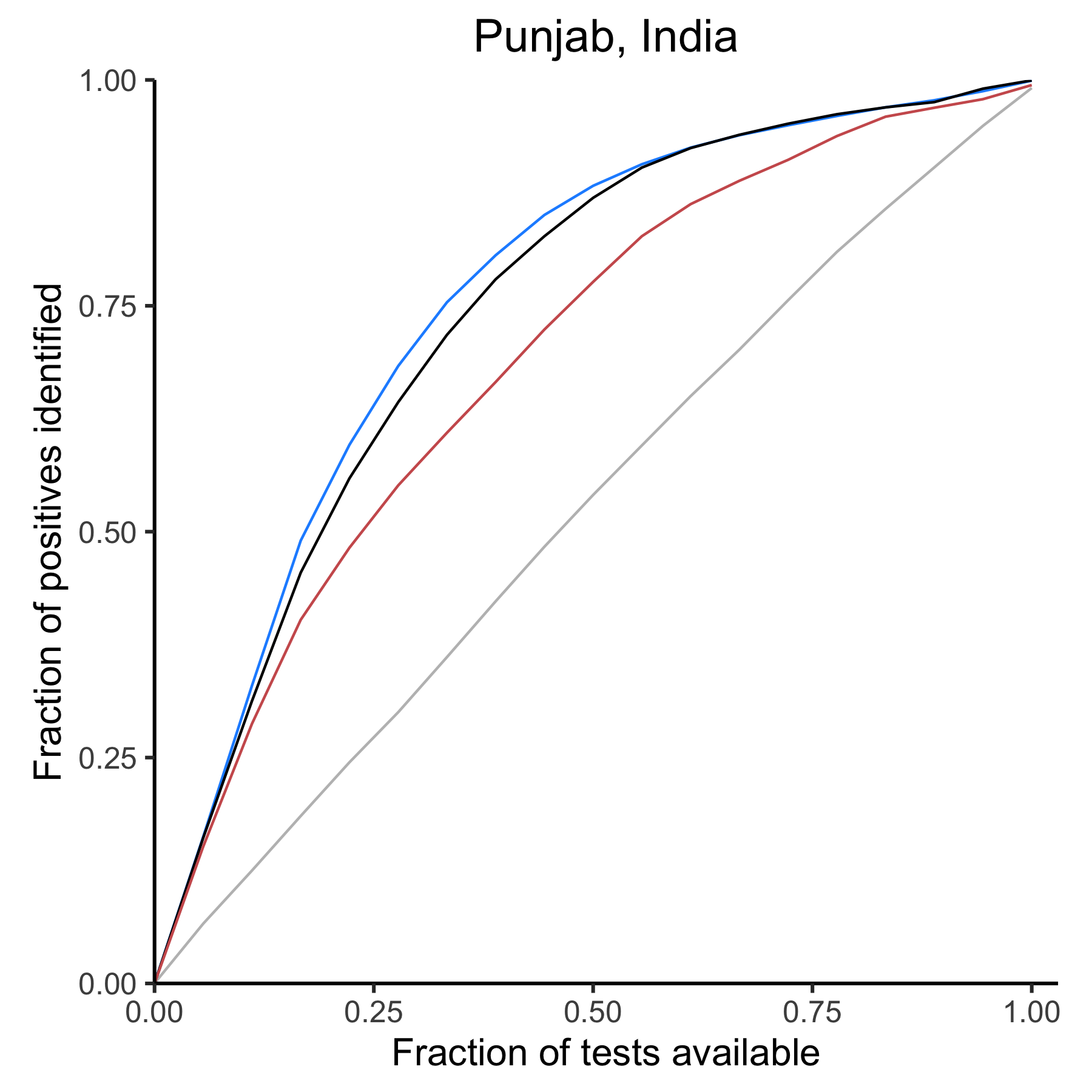}
    \end{subfigure}%
    \
    \begin{subfigure}[t]{0.4\textwidth}
        \centering
        \includegraphics[width=3.5in]{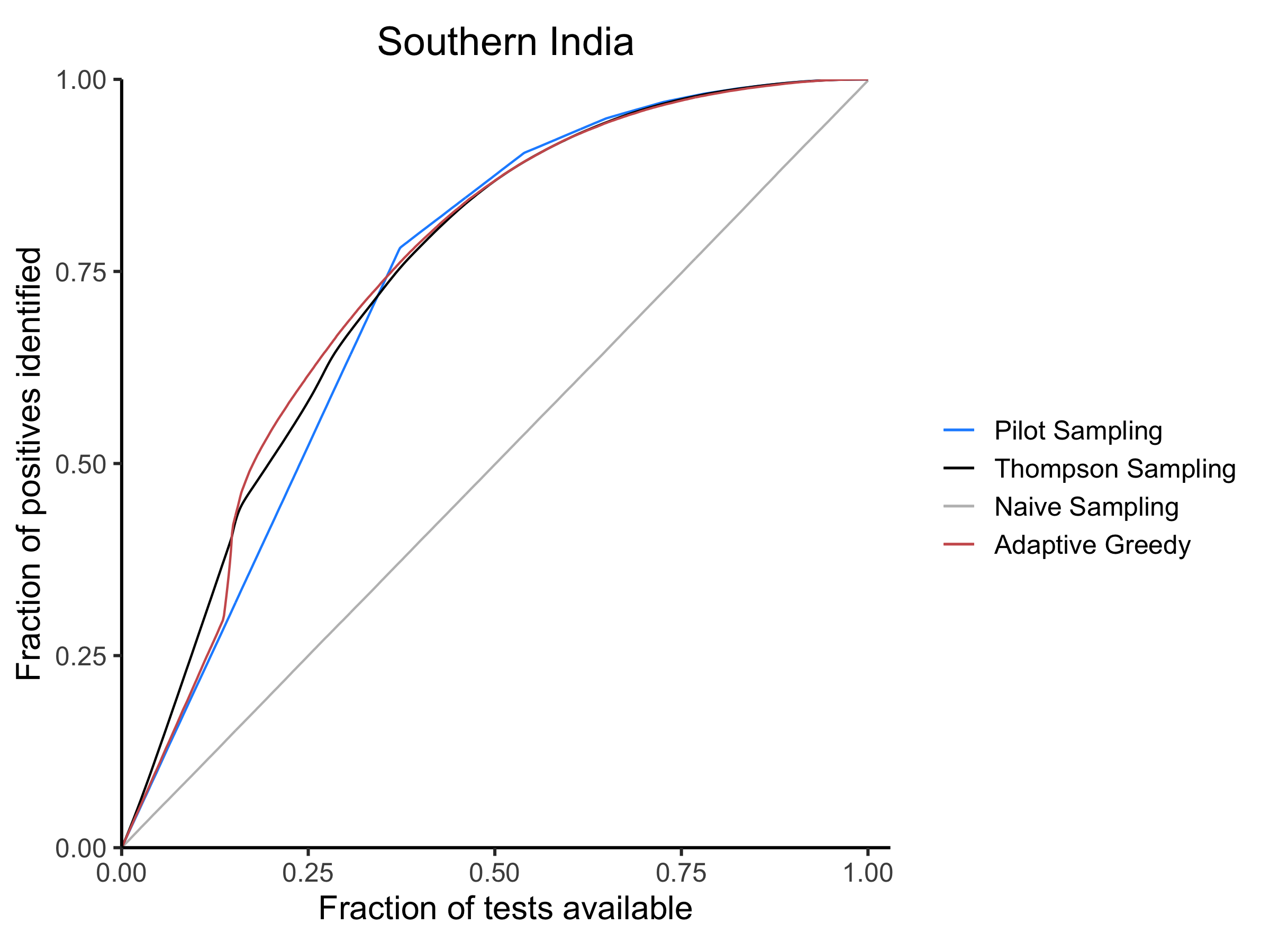}
    \end{subfigure}
    \caption{The top left figure shows results on Punjab, Pakistan dataset. The top right figure shows results on the Punjab, India dataset. The bottom figure shows results on the southern India dataset. Pilot sampling and Thompson sampling are clearly doing better when we have branching data (both Punjab datasets). However, Thompson sampling may be logistically difficult making pilot sampling favorable when implementing contact tracing. Figure \ref{fig:real-data-absolute} shows the same plot with absolute numbers of tests and infections.}
    \label{fig:real-data}
\end{figure*}

We used the PCI distribution estimated previously to initialize Thompson sampling and determine the pilot group size for mortal bandits. Although we used the full dataset to estimate the pilot group size, we note that the group size is not very sensitive to the prior distribution, especially in the right-skewed cases. Based on these parameters and sizes of the datasets, we are in the regime where the asymptotic behavior described in Theorems \ref{thm:ag-br} and \ref{thm:bayesian-regret-pilot} hold (see Appendix \ref{section:asymp-regret}).

The results are shown in Figure \ref{fig:real-data}. The results were averaged over 100 simulations to account for randomly drawing infected people and their contacts. Here, Naive sampling is a straight line because it is equivalent to testing every single person and the rate of uncovering infections reflects the proportion of infections in the full population. As expected, Pilot sampling, Thompson sampling, and Adaptive Greedy vastly outperform Naive sampling. Pilot sampling and Thompson sampling perform at similar levels with Pilot sampling dominating both the Punjab datasets. Adaptive Greedy does not perform well here. On the other hand, in the dataset from southern India, all three algorithms perform similarly. In Table \ref{tab:infections-tests-results}, we look at how many people need to be tested to identify 80\% of all infections. The advantages are quite dramatic -- the use of Pilot sampling would have allowed teams to pick up 80\% of the infected individuals in the samples with only 40\% of the tests in Pakistan and South India and 62\% of the tests in Punjab, India compared to 80\% of tests with Naive sampling.

\begin{table}[!tb]
    \centering
    \renewcommand{\arraystretch}{1.2}
    \begin{tabular}{c | c | c | c}
        \hline
        Algorithm & Punjab, Pakistan & Punjab, India & Southern India  \\
        \hline
        Naive Sampling & 80\% & 78\% & 80\% \\
        Thompson Sampling & \textbf{62\%} & \textbf{39\%} & 42\%  \\
        Adaptive Greedy & 75\% & 50\% & 41\%  \\
        Pilot Sampling & \textbf{62\%} & \textbf{39\%} & \textbf{40\%}  \\
         \hline
    \end{tabular}
    \caption{Fraction of tests performed in order to identify 80\% of all infections.}
    \label{tab:infections-tests-results}
\end{table}

This demonstrates that the conclusions we drew from controlled simulations do in fact hold up when confronted with data collected during the COVID-19 pandemic. Clearly, Thompson sampling and Pilot sampling have an edge over Adaptive Greedy. These findings along with simulations in Section \ref{section:sims} strongly suggest that we should abandon Naive sampling in favor of more efficient methods. 

One reason why governments may be reluctant to try active learning algorithms is because of their complexity and other logical constraints. For instance, door-to-door testing makes Thompson sampling and Adaptive Greedy approaches hard to implement in the field but may be more amenable to phone call testing. What makes our results particularly appealing is that in all three datasets, the distribution of PCI and degree is such that pilot sampling either performs better or very similarly to other more complex algorithms. Pilot sampling can be easily implemented in the field and offers an opportunity for governments to engage in contact tracing that they can actually do.

%% file: sections/conclusion.tex
\section{Discussion}
\label{section:conclusion}

In this paper, we address the problem of efficient contact tracing by framing it as a mortal bandit problem. We showed that the lower bound for the Bayesian regret in standard and mortal bandits are identical, and presented new Bayesian regret bounds for the Adaptive Greedy and Pilot Sampling algorithms. Through empirical simulations, we provide guidelines for choosing appropriate policies. If the distribution of the lifetime of arms is heavy-tailed, then we should use Pilot Sampling. If the distribution of lifetime is not heavy-tailed, then we should use Pilot Sampling if the rewards are heavily right-skewed; use Adaptive Greedy or Thompson Sampling if rewards have a small variance or are heavily left-skewed, and; the choice does not matter in other cases. If the distribution of lifetimes is not heavy-tailed and we do not know how the rewards are distributed, we should default to Thompson Sampling. We use our theoretical results and findings from empirical simulations on data from COVID-19 contact tracing in three different regions -- Punjab (Pakistan), Punjab (India), and South India. We show that Pilot Sampling outperforms Adaptive Greedy in both of the Punjab datasets and performs similarly to Adaptive Greedy in the dataset from southern India. These results are in line with the results from the empirical simulations.

We outline three possible extensions along with their relevance for contact tracing. First, we assumed that the distribution of mean rewards (PCI) is known. However, the distribution of PCI is unlikely to be known at the beginning of the epidemic. Therefore, contact tracing serves the dual purpose of controlling the epidemic through contact tracing and estimating the distribution of PCI in the population. Thus, the value of information regarding the distribution of PCI is very high in the initial stages of the epidemic -- performance may be higher if all contacts are tested until the PCI distribution is known with low uncertainty.

Relatedly, we claimed that Pilot Sampling and Adaptive Greedy are able to match the order lower bounds when we run the policies on an appropriately sized subset of arms. The size of the arms depends on the $\gamma$ parameter. In practice, this may be unavailable. For example, during a new outbreak, we may not know the behavior of the disease. There are a variety of estimators with desirable convergence properties available from extreme value theory that help estimate $\beta$ \citep[for example, see][]{hill1975simple, pickands1975statistical, haan2006extreme}. The situation with bandits is slightly different as we never really observe the true mean reward of each arm. \citet{carpentier2015simple} tackle this case for the standard bandit. They propose a modification to the estimator of \citet{carpentier2015adaptive} and describe a two-phase algorithm that first estimates $\gamma$ and then calls a different policy.

Second, we have disregarded any potential correlation between the number of contacts and PCI. If PCI is purely biological, this may not be a poor assumption, but if PCI is, in part behavioral, both positive and negative correlations may occur.
Estimating this correlation with any precision requires considerable data. It will be useful to compute the sample size requirements in order to estimate this with sufficiently high power. After this correlation is estimated (or is known a priori), a natural question is the effect of this correlation on choosing an optimal contact tracing policy.

Third, we have abstracted from the full network structure thus far, in part because we just don’t see how such a structure can be learned in the throes of an epidemic. However, it will clearly matter. Suppose one person’s PCI based on data accumulated thus far is 20\% and another’s is 100\%. Each has 5 remaining contacts to trace. However, the person whose PCI is 20\% has one contact who regularly meets thousands of people, while the contacts of the person whose PCI is 100\% each have no further contacts. Clearly, tracing and quarantining the contacts of the 20\% person will still be the most effective action in this case. It will be beneficial to assess the extent to which a lack of knowledge of the full structure of the network graph affects our results.

%% file: sections/appendix_ag.tex
\section{Technical details of Adaptive Greedy sampling}
\label{section:proof-ag-BR}

\subsection{Proof of Theorem \ref{thm:ag-br}}

Before we prove Theorem \ref{thm:ag-br}, we first present a useful result.

%%%
%%% Theorem
%%%
\begin{theorem}[Bayesian Regret for Adaptive Greedy]
\label{thm:br-adaptive-greedy}
Under the assumptions \ref{assumption:prior}-\ref{assumption:independence}, for any $\epsilon \in (0, 1/3)$, the Bayesian regret of the adaptive greedy algorithm is given by
\begin{align}
    BR_{T, N}(AG) &\leq T \left(\E_{\Gamma} \left[\mathbb{I}(1 - \mu \leq \epsilon) \prob( \exists t : \widehat{\mu}^t < 1 - 2 \epsilon ) \right] \right)^N + 3 \epsilon T + \nonumber \\
    N \E_{\Gamma} \Big[ \mathbb{I}(1 - \mu > 3 \epsilon) &  \min \left\{ 1 + \min\{T/N_m, L\} (1 - \mu) (1 - \epsilon) + \frac{3}{C_1 (1 - \mu - 2 \epsilon)}, \min\{L, T\} (1-\mu) \right\} \Big]
\end{align}
\end{theorem}

\begin{proofof}[Theorem \ref{thm:br-adaptive-greedy}]

Let $\muopt = \max_{i \in N} \mu_i$ and $\Delta_i = \muopt - \mu_i$. Fix $\epsilon \in (0, \muopt / 3)$. In the adaptive greedy sampling strategy, we exploit the arm with the largest sample mean with probability $\max_i \widehat{\mu}_i^t$. And with probability $1 - \max_i \widehat{\mu}_i^t$, we randomly choose an arm to explore. Here, $\widehat{\mu}_i^t$ is the sample mean reward of arm $i$ after time $t$. Following the idea from \citet{bayati2020unreasonable}, we will assume that $\muopt = 1$. This will loosen the bound but also make it easier to handle integration over priors.

Let us call all arms $i$ with $\Delta_i \leq \epsilon$ as $\epsilon$-optimal. And all arms $i$ such that $\Delta_i > 3 \epsilon$ are called sub-optimal. Following standard bandit literature, let us define a bad event to distinguish the randomness of the distributions from the sampling. The bad event is
\begin{align*}
    \g^c &= \bigcap_{k : \Delta_{k} < \epsilon} \left\{ \exists t : \widehat{\mu}_k^t < 1 - 2 \epsilon \right\}
\end{align*}
Essentially, the bad event happens when for every $\epsilon$-optimal arm, there is at least one time when its sample mean drops below $1 - 2 \epsilon$. So the good event happens when there is at least one $\epsilon$-optimal arm whose sample mean remains above $1 - 2 \epsilon$. Using the fact that rewards from different arms are independent,
\begin{align*}
    \prob(\g^c) &= \prod_{i : 1 - \mu_i \leq \epsilon} \prob( \exists t : \widehat{\mu}_i^t < 1 - 2 \epsilon ) \\
    &= \prod_{i=1}^N \mathbb{I}(1 - \mu_i \leq \epsilon) \prob( \exists t : \widehat{\mu}_i^t < 1 - 2 \epsilon )
\end{align*}

Next, let us estimate the number of times each arm is pulled under a good event,
\begin{align*}
    \E[N_i(T) \mid \g] &\leq 1 + \sum_{t=1}^{T} \prob(\text{arm $i$ is chosen} \mid \g) \\
    \prob(\text{arm $i$ is chosen} \mid \g) &= \prob(\text{arm $i$ is chosen} \mid \text{explored}, \g) \prob(\text{explored} \mid \g) + \\
    & \qquad \prob(\text{arm $i$ is chosen through exploitation} \mid \g) 
\end{align*}
Since we are in the good event, there must be at least one $\epsilon$-optimal arm $k$ such that $\widehat{\mu}_k^t \geq 1 - 2 \epsilon$ for all $t$. So if we chose arm $i$ through exploitation, it must be that $\widehat{\mu}_i^t \geq 1 - 2 \epsilon$
\begin{align*}
    \prob(\text{arm $i$ is chosen} \mid \g) &\leq \frac{1}{N_t} (1 - \max_j \widehat{\mu}_j^t) + \prob(\widehat{\mu}_i^t \geq 1 - 2 \epsilon)
\end{align*}
Notice that in a good event, $\max_j \widehat{\mu}_j^t \geq 1 - 2 \epsilon$. Further, assume that there are always at least $N_m$ arms left to play. This allows us to write
\begin{align*}
    \prob(\text{arm $i$ is chosen} \mid \g) &\leq \frac{1 - \epsilon}{N_m} + \prob(\widehat{\mu}_i^t \geq 1 - 2 \epsilon) \\
    &= \frac{1 - \epsilon}{N_m} + \prob(\widehat{\mu}_i^t  - \mu_i \geq \Delta_i - 2 \epsilon) \\
    &\leq \frac{1 - \epsilon}{N_m} + \exp \left\{ -t \frac{(\Delta_i - 2 \epsilon)^2}{2} \right\}
\end{align*}
where in the last inequality, we used the fact that $\widehat{\mu}_i^t$ is $1/t$-subgaussian. Going back to the counts,
\begin{align*}
    \E[N_i(T) \mid \g] &\leq 1 + \sum_{t=1}^{T} \frac{1 - \epsilon}{N_m} + \exp \left\{ -t \frac{(\Delta_i - 2 \epsilon)^2}{2} \right\} \\
    &\leq 1 + T \frac{1 - \epsilon}{N_m} + \sum_{t=1}^{\infty} \exp \left\{ -t \frac{(\Delta_i - 2 \epsilon)^2}{2} \right\} \\
    &\leq 1 + \min\{T/N_m, L_i\} (1 - \epsilon) + \frac{1}{1 - \exp \left\{ \frac{(\Delta_i - 2 \epsilon)^2}{2} \right\}} \\
    &\leq 1 + \min\{T/N_m, L_i\} (1 - \epsilon)  + \frac{1}{C_1 (\Delta_i - 2 \epsilon)^2}
\end{align*}
where in the third inequality we bounded the number of times an arm $i$ can be explored by $L_i$ and in the last line we used the fact that $\exp(-x) \leq 1 - 2C_1x, C_1 = (1 - \exp(-1))/2$ for $x \in [0, 1]$.

This implies,
\begin{align*}
    (1 - \mu_i) \E[N_i(T) \mid \g] &\leq 1 - \mu_i + \min\{T/N_m, L_i\} (1 - \mu_i) (1 - \epsilon) + \frac{1 - \mu_i}{C_1 (\Delta_i - 2 \epsilon)^2} \\
    &\leq 1 + \min\{T/N_m, L_i\} (1 - \mu_i) (1 - \epsilon) + \frac{3}{C_1 (\Delta_i - 2 \epsilon)}
\end{align*}
as $1 - \mu_i > 3 \epsilon \implies 1 - \mu_i \leq 3(1 - 2 \epsilon - \mu_i)$. Finally, we will bound the number of pulls on the arm by $\min\{L_i, T\}$,
\begin{align*}
    (1 - \mu_i) \E[N_i(T) \mid \g] &\leq \min \left\{ 1 + \min\{T/N_m, L_i\}(1 - \mu_i) (1 - \epsilon) + \frac{3}{C_1 (\Delta_i - 2 \epsilon)}, \min\{L_i, T\} (1-\mu_i) \right\}
\end{align*}

Now,
\begin{align*}
    \E[R_T \mid \g] &= \sum_{i : \Delta_i \leq 3 \epsilon} \Delta_i \E[N_i(T) \mid \g] + \sum_{i : \Delta_i > 3 \epsilon} \Delta_i \E[N_i(T) \mid \g] \\
    &\leq 3 \epsilon T + \sum_{i : \Delta_i > 3 \epsilon} \min \left\{ 1 + \min\{T/N_m, L_i\} (1 - \mu_i) (1 - \epsilon) + \frac{3}{C_1 (\Delta_i - 2 \epsilon)}, \min\{L_i, T\} (1-\mu_i) \right\} \\
    &= 3 \epsilon T + \sum_{i=1}^N \mathbb{I}(1 - \mu_i > 3 \epsilon)   \min \Bigg\{ 1 + \min\{T/N_m, L_i\} (1 - \mu_i) (1 - \epsilon) + \frac{3}{C_1 (\Delta_i - 2 \epsilon)},\\
    & \qquad \min\{L_i, T\} (1-\mu_i) \Bigg\}
\end{align*}

Therefore, the regret is
\begin{align*}
    \E[R_T] &= \E[R_T \mid \g^c] \prob(\g^c) + \E[R_T \mid \g] \prob(\g) \\
    &\leq T \prob(\g^c) + \E[R_T \mid \g] \\
    \implies BR_T &= \E_{\Gamma} \E[R_T] \\
    &\leq T \left(\E_{\Gamma} \left[\mathbb{I}(1 - \mu \leq \epsilon) \prob( \exists t : \widehat{\mu}^t < 1 - 2 \epsilon ) \right] \right)^N + 3 \epsilon T + \\
     N \E_{\Gamma} \Big[ \mathbb{I}(1 - \mu > 3 \epsilon) &  \min \left\{ 1 + \min\{T/N_m, L\} (1 - \mu)(1 - \epsilon) + \frac{3}{C_1 (1 - \mu - 2 \epsilon)}, \min\{L, T\} (1-\mu) \right\} \Big]
\end{align*}
\end{proofof}

%%%
%%% Theorem
%%%

\begin{proofof}[Theorem \ref{thm:ag-br}]
The general strategy is similar to the one used by \citet{bayati2020unreasonable}. The regret is,
\begin{align*}
    & BR_{T, N}(AG) \leq T \left(\E_{\Gamma} \left[\mathbb{I}(1 - \mu \leq \epsilon) \prob( \exists t : \widehat{\mu}^t < 1 - 2 \epsilon ) \right] \right)^N + 3 \epsilon T + \\
    &N \E_{\Gamma} \left[ \mathbb{I}(1 - \mu > 3 \epsilon)  \min \left\{ 1 + \min\{T/N_m, L\} (1 - \mu)(1 - \epsilon) + \frac{3}{C_1 (1 - \mu - 2 \epsilon)}, \min\{L, T\} (1-\mu) \right\} \right]
\end{align*}
Consider the first term. Since $\mu$ comes from a $\gamma$-regular prior, $\prob(\mu > 1 - \epsilon) \geq c_{\min} \epsilon^{\gamma}$ for some absolute constant $c_{\min}$. From Lemma \ref{lemma:bound-bernoulli-bad-event},
\begin{align*}
    \E_{\Gamma} \left[\mathbb{I}(1 - \mu \leq \epsilon) \prob( \exists t : \widehat{\mu}^t < 1 - 2 \epsilon ) \right] &\leq \left(1 - \exp(-0.5)/3\right) \E_{\Gamma} [\mathbb{I}(1 - \mu \leq \epsilon)] \\
    &\leq c_{\min} \epsilon^{\gamma} \left(1 - \exp(-0.5)/3\right) \\
    &\leq 1 - c_{\min} \epsilon^{\gamma} \frac{\exp(-0.5)}{3} \\
    &= 1 - c_0 \epsilon^{\gamma} \\
    &\leq \exp\left\{ - c_0 \epsilon^{\gamma} \right\} \\
    \implies \text{Term 1} &= T \exp\left\{ - N c_0 \epsilon^{\gamma} \right\}
\end{align*}
where $c_0 = c_{\min} \exp(-0.5)/3$ and we used $1 - x \leq \exp(-x)$. Our strategy now is to control the first term as $\g[O](1)$ and treat the third term using Lemma \ref{lemma:bound-good-event}.

To analyze the third term, we will use the fact that $\min\{f + g, h\} \leq \min\{f, h\} + g$ to pull out the $\min\{T/L_m, L\}(1-\mu)(1-\epsilon)$ outside the outer min operator. Bounding $\min\{T/N_m, L\} \leq L$, we have
\begin{align*}
    \E_{\Gamma} \left[ \mathbb{I}(1 - \mu > 3 \epsilon) L (1-\mu)(1-\epsilon)\right] &\leq L (1-\epsilon)(1 - c_{\min}3^{\gamma} \epsilon^{\gamma})
\end{align*}
We will also upper bound $\min\{T, L\} \leq T$. So, an application of Lemma \ref{lemma:bound-good-event} takes care of the remaining pieces of the third term.

In particular, if $\gamma = 1$, then the Bayesian regret is upper bounded by
\begin{align*}
    T \exp\left\{ - N c_0 \epsilon^{\gamma} \right\} + 3 \epsilon T + N L (1-\epsilon)(1 - c_{\min}3^{\gamma} \epsilon^{\gamma}) + \frac{3 C_0}{C_1} N (5 + \log(1/\epsilon))
\end{align*}
where $C_0$ is the constant in Lemma \ref{lemma:bound-good-event}. Now, if we choose $\epsilon^{\gamma} = C_2 \log T / (N c_0)$ where $C_2 \geq 1$, then
\begin{align*}
    % T^{1 - C_2} + 3 C_2 c_0^{-1} N^{-1} \log T + \frac{NT}{N_m} (1 - C_2 c_0^{-1} N^{-1}) ( 1- c_{\min} 3 \frac{C_2}{c_0} N^{-1} \log T) + \frac{3 C_0}{C_1} N (5 - \log C_2 - \log \log T + \log N + \log c_0)
    BR_{N, T} &= \g[O] \left( T^{1 - C_2} + N^{-1} T \log T + N + \log T + N^{-1} (\log T)^2 + N \log \log T + N \log N \right) \\
    &= \g[O] \left(N^{-1} T \log T + N (\log \log T + \log N) + \log T \right) \\
    &= \tilde{\g[O]} \left(T N^{-1} + N \right)
\end{align*}
And when $\gamma > 1$, if we choose the same $\epsilon$ as above,
\begin{align*}
    BR_{N, T} &= \g[O] \left( T^{1 - C_2} + N^{-1/\gamma} T (\log T)^{1/\gamma} + N + N^{1 - 1/\gamma} (\log T) ^{1/\gamma} + \log T + N^{-1/\gamma} (\log T)^{1 + 1/\gamma} + N \right) \\
    &= \g[O] (N^{-1/\gamma} (\log T)^{1/\gamma} (T + N) + N + \log T) \\
    &= \tilde{\g[O]} (T N^{-1/\gamma} + N)
\end{align*}
Finally, when $\gamma < 1$ and choosing the same $\epsilon$,
\begin{align*}
    BR_{N, T} &= \g[O] \left( N^{-1/\gamma} (\log T)^{1/\gamma} (T + N) + N + \log T + N \min (\sqrt{T}, N^{1/\gamma}(\log T)^{-1/\gamma} )^{1 - \gamma}\right) \\
    &= \tilde{\g[O]} \left( T N^{-1/\gamma} + N \min (\sqrt{T}, N^{1/\gamma} )^{1 - \gamma}\right)
\end{align*}

Thus,
\begin{align*}
    BR_{T, N} &=  \begin{cases}
        \g[O] \left( N^{-1/\gamma} (\log T)^{1/\gamma} (T + N) + N + \log T + N \min (\sqrt{T}, N^{1/\gamma}(\log T)^{-1/\gamma} )^{1 - \gamma}\right), & \gamma < 1 \\
        \g[O] \left(N^{-1} T \log T + N (\log \log T + \log N) + \log T \right), & \gamma = 1 \\
        \g[O] (N^{-1/\gamma} (\log T)^{1/\gamma} (T + N) + N + \log T), & \gamma > 1
    \end{cases} \\
    &= \begin{cases}
        \Tilde{\g[O]} \left( TN^{-1/\gamma} + N \min(\sqrt{T}, N^{1 / \gamma})^{1 - \gamma} \right), & \gamma < 1 \\
        \Tilde{\g[O]} \left( T N^{-1/\gamma} + N \right), & \gamma \geq 1
    \end{cases}
\end{align*}
\end{proofof}

\subsection{Mean rewards from a beta prior}

Under assumption \ref{assumption:beta}, we have $\gamma = \alpha + \beta - 1$ for $\beta > 1$ and $\gamma = \alpha$ otherwise. Corollary \ref{thm:br-adaptive-greedy-beta} gives an explicit bound under \ref{assumption:bern} and \ref{assumption:beta}.

\begin{corollary}[Adaptive Greedy - Bayesian Regret with Beta priors]
\label{thm:br-adaptive-greedy-beta}

Under the setup of Theorem \ref{thm:ag-br} with an additional assumption \ref{assumption:beta}, the Bayesian regret of the adaptive greedy algorithm is given by
\begin{align*}
    BR_T &\leq T \exp\left\{ - N c_0 \epsilon^{\gamma} \right\} + 3 \epsilon T + N \min \Bigg\{L \frac{\beta + 1}{\alpha + \beta + 1} (1 - F_{\alpha, \beta + 1}(3 \epsilon)), \nonumber \\
    & \quad \left(1 + \frac{3}{C_1 \epsilon} \right) ( 1 - F_{\alpha, \beta}(3 \epsilon)) + \min\{T/N_m, L\} (1 - \epsilon) \frac{\beta + 1}{\alpha + \beta + 1} (1 - F_{\alpha, \beta + 1}(3 \epsilon))  \Bigg\}
\end{align*}
where $\gamma = \alpha + \beta - 1$ for $\beta > 1$ and $\gamma = \alpha$ otherwise, and $F_{a, b}$ is the distribution of $\text{Beta}(a, b)$ and $L := \E[ L_i]$.
\end{corollary}

\begin{proofof}[Corollary \ref{thm:br-adaptive-greedy-beta}]
Following the same idea as the proof of Theorem \ref{thm:br-adaptive-greedy}, the first term is,
\begin{align*}
    \E_{\Gamma} \left[\mathbb{I}(1 - \mu \leq \epsilon) \prob( \exists t : \widehat{\mu}^t < 1 - 2 \epsilon ) \right] &\leq \exp\left\{ - c_0 \epsilon^{\gamma} \right\}
\end{align*}
where $\gamma = \alpha + \beta -1$.
When analyzing the second expectation, we will bring the $\min$ operator outside the expectation and treat each inner expectation separately. The first expectation is,
\begin{align*}
    & \E_{\Gamma} \left[ \mathbb{I}(1 - \mu > 3 \epsilon) \left( 1 + \min\{T/N_m, L\}(1 - \mu) \frac{1 - \epsilon}{N_m} + \frac{3}{C_1 (1 - \mu - 2 \epsilon)} \right) \right] \\
    &\leq \E_{\Gamma} \left[ \mathbb{I}(1 - \mu > 3 \epsilon) \left( 1 + \min\{T/N_m, L\}(1 - \mu) \frac{1 - \epsilon}{N_m} + \frac{3}{C_1 \epsilon} \right) \right] \\
    &= \int_{3 \epsilon}^1 \left( 1 + \min\{T/N_m, L\}(1 - \mu) \frac{1 - \epsilon}{N_m} + \frac{3}{C_1 \epsilon} \right) \frac{\mu^{\alpha - 1} (1 - \mu)^{\beta - 1}}{B(\alpha, \beta)} d\mu \\
    &= \left(1 + \frac{3}{C_1 \epsilon} \right) ( 1 - F_{\alpha, \beta}(3 \epsilon)) + \min\{T/N_m, L\} (1 - \epsilon) \frac{\beta + 1}{\alpha + \beta + 1} (1 - F_{\alpha, \beta + 1}(3 \epsilon))
\end{align*}
And the second is,
\begin{align*}
    \E_{\Gamma} \left[ \mathbb{I}(1 - \mu > 3 \epsilon) \min\{L, T\} (1-\mu) \right] &= \min\{L, T\} \frac{\beta + 1}{\alpha + \beta + 1} (1 - F_{\alpha, \beta + 1}(3 \epsilon))
\end{align*}
Assuming that $\min\{T, L\} = L$ gives us the desired result.
\end{proofof}

\subsection{Sampling by lifetime}

Corollary \ref{corollary:ag-lifetime} describes the upper bounds of the Bayesian regret of Adaptive Greedy when sampling by lifetimes. Essentially this allows us to conclude that previously established asymptotics of the upper bounds are not affected thereby proving Corollary \ref{corollary:ag-br-lifetime}. 

\begin{corollary}[Adaptive Greedy sampling by lifetime]
\label{corollary:ag-lifetime}
Under the assumptions \ref{assumption:prior}-\ref{assumption:independence}, for any $\epsilon \in (0, 1/3)$, the Bayesian regret of the adaptive greedy algorithm where we sample by lifetime is given by
\begin{align*}
    BR_{T, N}(AG) &\leq T \left(\E_{\Gamma} \left[\mathbb{I}(1 - \mu \leq \epsilon) \prob( \exists t : \widehat{\mu}^t < 1 - 2 \epsilon ) \right] \right)^N + 3 \epsilon T + \nonumber \\
    N \E_{\Gamma} \mathbb{I}(1 - \mu > 3 \epsilon) &  \min \left\{ 1 + \min\{\frac{T}{\sum_{i=1}^{N_m} L_{(i)}}, 1\} L (1 - \mu) (1 - \epsilon) + \frac{3}{C_1 (1 - \mu - 2 \epsilon)}, \min\{L, T\} (1-\mu) \right\}
\end{align*}
where $\{L_{(i)}\}_{i=1}^N$ are the order statistics of $\{L_i\}_{i=1}^N$.

If assumption \ref{assumption:bern} also holds, then the asymptotics in Theorem \ref{thm:ag-br} still hold. Further, if we assume \ref{assumption:bern} and \ref{assumption:beta} hold, then the result of Corollary \ref{thm:br-adaptive-greedy-beta} naturally extends to
    \begin{align*}
    & BR_T \leq T \exp\left\{ - N c_0 \epsilon^{\gamma} \right\} + 3 \epsilon T + N \min \Bigg\{L \frac{\beta + 1}{\alpha + \beta + 1} (1 - F_{\alpha, \beta + 1}(3 \epsilon)), \nonumber \\
    & \left(1 + \frac{3}{C_1 \epsilon} \right) ( 1 - F_{\alpha, \beta}(3 \epsilon)) + \E \left[ \min\{\frac{T}{\sum_{i=1}^{N_m} L_{(i)}}, 1\} L \right] (1 - \epsilon) \frac{\beta + 1}{\alpha + \beta + 1} (1 - F_{\alpha, \beta + 1}(3 \epsilon))  \Bigg\}
\end{align*}
where $\gamma = \alpha + \beta - 1$ and $F_{a, b}$ is the distribution of $\text{Beta}(a, b)$ and the expectation is over $\{L_i\}_{i=1}^N$.
\end{corollary}

\begin{proofof}[Corollary \ref{corollary:ag-lifetime}]
The key difference from the previous proofs lies in the exploration phase. Instead of sampling uniformly, we are sampling proportional to lifetimes. Therefore,
\begin{align*}
    \prob(\text{arm $i$ is chosen} \mid \g) &\leq \frac{L_i}{\sum_{j=1}^{N_t} L_j} (1 - \max_j \widehat{\mu}_j^t) + \prob(\widehat{\mu}_i^t \geq 1 - 2 \epsilon)
\end{align*}
Since there are at least $N_m$ arms to play and $\sum_{j=1}^{N_m} L_j \geq \sum_{j=1}^{N_m} L_{(j)}$ where $L_{(1)}, \dots, L_{(N)}$ are the order statistics of $L_1, \dots, L_N$, we have
\begin{align*}
    \prob(\text{arm $i$ is chosen} \mid \g) &\leq \frac{L_i}{\sum_{j=1}^{N_m} L_{(j)}} (1 - \max_j \widehat{\mu}_j^t) + \prob(\widehat{\mu}_i^t \geq 1 - 2 \epsilon)
\end{align*}
Pushing this through the remainder of the steps gives the desired result and the extensions of Theorems \ref{thm:ag-br} and \ref{thm:br-adaptive-greedy-beta}.
\end{proofof}

%% file: sections/appendix_pilot.tex
\section{Technical details of Pilot sampling}
\label{section:proof-pilot-BR}

\subsection{Proof of Theorem \ref{thm:bayesian-regret-pilot}}

\begin{proofof}[Theorem \ref{thm:bayesian-regret-pilot}]
Let $\muopt = \max_{i \in N} \mu_i$ and $\Delta_i = \muopt - \mu_i$. In the pilot strategy, we get to pull an arm $\min\{K, L_i\}$ times. For simplicity of analysis, we will assume that $\min\{K, L_i\} = K$. If there is at least one positive in the $K$ pulls, then we pull until the arm dies (or we run out of budget). Again, we will assume that $\muopt = 1$. This will loosen the bound but also make it easier to handle integration over priors.

Let $\E[R_{N, T}]$ be the mean regret when we have $N$ arms and a budget of $T$. Immediately, we have $\E R_{N, T} \leq T$. Therefore, for $T_1 \geq T_2$, $\E R_{N, T_1} - \E R_{N, T_2} \leq T_1 - T_2$.
And if we have $\{L_i\}_{i=1}^N$ arms initially and we remove arm $L_j$ to get $N_j = N - 1$ arms, then for the same budget $T$, then $\E  R_{N, T} - \E R_{N_j, T} \leq L_j$.

Suppose we pick arm $i$ first. Then the conditional mean regret is,
\begin{align*}
    \E \left[ R_{N, T} \mid \text{choose $i$} \right] &\leq \left[ L_i(1 - \mu_i) + \E R_{N_i, T-L_i}\right](1 - (1 - \mu_i)^K) + \\
    & \qquad \left[K(1-\mu_i) + \E R_{N_i, T-K} \right](1-\mu_i)^K \\
    &\leq L_i(1-\mu_i) - (L_i - K) (1-\mu_i)^{K+1} + \E R_{N_i, T-L_i} + \\
    & \qquad (\E R_{N_i, T-K} - \E R_{N_i, T-L_i}) (1-\mu_i)^K \\
    &\leq L_i(1-\mu_i) + (L_i - K) \left[ (1-\mu_i)^{K} - (1-\mu_i)^{K+1} \right] + \E R_{N_i, T-L_i}
\end{align*}

Therefore,
\begin{align*}
    \E R_{N, T} &= \sum_{i=1}^N \prob (\text{choose $i$}) \E \left[ R_{N, T} \mid \text{choose $i$} \right] \\
    &\leq \sum_{i=1}^N \prob(i)  \left(L_i(1-\mu_i) + (L_i - K) \left[ (1-\mu_i)^{K} - (1-\mu_i)^{K+1} \right]\right) + \sum_{i=1}^N \prob(i) \E R_{N_i, T-L_i}
\end{align*}
If we sample arms uniformly, then $\prob(i) = 1/N$. Therefore,
\begin{align*}
    \E R_{N, T} &\leq \frac{1}{N} \sum_{i=1}^N \left(L_i(1-\mu_i) + (L_i - K) \left[ (1-\mu_i)^{K} - (1-\mu_i)^{K+1} \right]\right) + \frac{1}{N} \sum_{i=1}^N \E R_{N_i, T-L_i} \\
    \implies \E_{\Gamma} \E R_{N, T} &\leq \E_{\Gamma} \left(L_i(1-\mu_i) + (L_i - K) \left[ (1-\mu_i)^{K} - (1-\mu_i)^{K+1} \right]\right) + \frac{1}{N} \sum_{i=1}^N \E_{\Gamma} \E R_{N_i, T-L_i}
\end{align*}
Notice that the first term is independent of $T$ and $N$ (technically, we should apply a min operator to account for $T < L_i$, but we will come back to that later). And the second term can be recursively expanded at most $\E \lfloor T / L_{(1)} \rfloor$ times on average where $L_{(1)} = \min_{i = 1, \dots, N} L_i$ is the smallest lifetime of $N$ samples. Thus, we have
\begin{align*}
    BR_T &= \E_{\Gamma} \E R_{N, T} \\
    &\leq \E \left\lfloor \frac{T}{L_{(1)}} \right\rfloor \E_{\Gamma} \left(L(1-\mu) + (L - K) \left[ (1-\mu)^{K} - (1-\mu)^{K+1} \right]\right) + \E_{\Gamma} L(1 - \mu)
\end{align*}
where the last term captures any remaining available pulls after exhausting $\lfloor T/L \rfloor$ arms. Notice that this naturally captures the case where $T < L$, which we ignored above. And to finally account for $N < T / L$,
\begin{align*}
    BR_T &\leq \min \left\{N-1, \E \lfloor T/L_{(1)} \rfloor \right\} \E_{\Gamma} \left(L(1-\mu) + (L - K) \left[ (1-\mu)^{K} - (1-\mu)^{K+1} \right]\right) + \E_{\Gamma} L(1 - \mu)
\end{align*}

This gives us that $BR_{T, N}(\text{Pilot}) = \g[O](\min\{N, T\})$.
\end{proofof}

\subsection{Mean rewards from a beta prior}

Theorem \ref{thm:bayesian-regret-pilot} immediately allows us to apply Beta priors described in Assumption \ref{assumption:beta}. This is presented in Corollary \ref{thm:bayesian-regret-pilot-beta}.

\begin{corollary}[Pilot Sampling - Bayesian Regret under Beta Priors]
\label{thm:bayesian-regret-pilot-beta}
Consider the setup of Theorem \ref{thm:bayesian-regret-pilot}. Additionally assume that \ref{assumption:independence} and \ref{assumption:beta} hold. Then, the Bayesian regret of the pilot sampling algorithm is given by
\begin{align*}
    BR_{T, N}(Pilot) &\leq \min \left\{N-1, \E\lfloor T/L_{(1)} \rfloor \right\} \left[ L \frac{\beta}{\alpha + \beta} + (L - K) \frac{\alpha}{\beta + \alpha + K + 1} \prod_{r=0}^{K} \frac{\beta + r}{\beta + \alpha + r} \right] + \\
    & \qquad L \frac{\beta}{\alpha + \beta}
\end{align*}
where $L := \E[L_i]$.
\end{corollary}

\begin{proofof}[Corollary \ref{thm:bayesian-regret-pilot-beta}]
Suppose that $\mu_i \sim \text{Beta}(\alpha, \beta)$. Therefore, $1 - \mu_i \sim \text{Beta}(\beta, \alpha)$. Then the $m$-th moment of $1 - \mu_i$ is given by
\begin{align*}
    \E[(1 - \mu_i)^m] &= \prod_{r=0}^{m-1} \frac{\beta + r}{\beta + \alpha + r}
\end{align*}
Therefore,
\begin{align*}
    BR_T &\leq \min \left\{N-1, \E \lfloor T/L_{(1)} \rfloor \right\} \left[ L \frac{\beta}{\alpha + \beta} + (L - K) \frac{\alpha}{\beta + \alpha + K + 1} \prod_{r=0}^{K} \frac{\beta + r}{\beta + \alpha + r} \right] + L \frac{\beta}{\alpha + \beta}
\end{align*}
where $L := \E[L_i]$.
\end{proofof}

\subsection{Sampling by lifetime}

Corollary \ref{thm:bayesian-regret-pilot-lifetime} provides an upper bound on the Bayesian regret of pilot sampling when we sample by lifetimes. At first glance, the asymptotic order does not match with Theorem \ref{thm:bayesian-regret-pilot}. Here, we have $BR = \g[O](\min\{N, T\}N/N_m)$ with an inflation factor of $N/N_m$. To gain tractability over the problem, we sacrificed lower-bounded $\sum_{i=1}^N L_i$ by $\sum_{i=1}^{N_m} L_{(i)}$. This is the reason we see the sum of the order statistics in the regret resulting in an apparently larger bound (as $N \geq N_m$). Since $N_m$ is the number of arms available to play at any time $t < T$, it makes sense that $N_m = o(N - T/L)$ (we assume $T < NL$ as otherwise any policy is good). Therefore, $N / N_m = \g[O](N / (N - T/L)) = \g[O](1)$, which implies that $BR = \g[O](\min\{N, T\})$. Thus, the asymptotic behavior matches that of the variant that uniformly samples arms. This is exactly what we state in Corollary \ref{corollary:pilot-br-lifetime}.

\begin{corollary}[Pilot sampling by lifetime]
\label{thm:bayesian-regret-pilot-lifetime}
Suppose that assumptions \ref{assumption:prior}, \ref{assumption:minimum-arms}, and \ref{assumption:min-lifetime} hold. Then the Bayesian regret of the pilot sampling algorithm when choosing arms by lifetime is given by
\begin{align*}
    BR_{T, N}(Pilot) &< \min\{N - 1, \E \lfloor T/ L_{(1)}\rfloor\} \E_{\Gamma} \Bigg[\left(\frac{N L^2}{\sum_{i=1}^{N_m} L_{(i)}} -K \right) \left( (1-\mu)^{K} - (1-\mu)^{K+1} \right) + \\
    & \qquad \frac{N L^2}{\sum_{i=1}^{N_m} L_{(i)}} (1 - \mu)\Bigg] +  \E_{\Gamma} L_{(N)} (1 - \mu),
\end{align*}
where $\{L_{(i)}\}_{i=1}^N$ are the order statistics of $\{L_i\}_{i=1}^N$.

Additionally, if we assume \ref{assumption:independence} and \ref{assumption:beta}, then the Bayesian regret of the pilot sampling algorithm is given by
\begin{align*}
    BR_{T, N}(Pilot) &\leq \min\{N - 1, \E \lfloor T/ L_{(1)}\rfloor\} \E_{\Gamma} \Bigg[\left(\frac{NL^2}{\sum_{i=1}^{N_m} L_{(i)}} -K \right) \frac{\alpha}{\beta + \alpha + K + 1} \prod_{r=0}^{K} \frac{\beta + r}{\beta + \alpha + r} + \\
    &\qquad \frac{NL^2}{\sum_{i=1}^{N_m} L_{(i)}} \frac{\beta}{\alpha + \beta} \Bigg] + \E_{\Gamma} L_{(N)} \frac{\beta}{\alpha + \beta}
\end{align*}
\end{corollary}

\begin{proofof}[Corollary \ref{thm:bayesian-regret-pilot-lifetime}]
From the proof of Theorem \ref{thm:bayesian-regret-pilot}, we have
\begin{align*}
    \E R_{N, T} &\leq \sum_{i=1}^N \prob(i)  \left(L_i(1-\mu_i) + (L_i - K) \left[ (1-\mu_i)^{K} - (1-\mu_i)^{K+1} \right]\right) + \sum_{i=1}^N \prob(i) \E R_{N_i, T-L_i} \\
    &\leq \sum_{i=1}^N \frac{L_i}{\sum_{j=1}^N L_j} \left(L_i(1-\mu_i) + (L_i - K) \left[ (1-\mu_i)^{K} - (1-\mu_i)^{K+1} \right]\right) + \sum_{i=1}^N \frac{L_i}{\sum_{j=1}^N L_j} \E R_{N_i, T-L_i} \\
    &= - K\left( (1-\mu_i)^{K} - (1-\mu_i)^{K+1} \right) + \sum_{i=1}^N \frac{L_i^2}{\sum_{j=1}^N L_j} \left((1-\mu_i) + (1-\mu_i)^{K} - (1-\mu_i)^{K+1} \right) + \\
    & \qquad \sum_{i=1}^N \frac{L_i}{\sum_{j=1}^N L_j} \E R_{N_i, T-L_i}
\end{align*}
Here, we will use the assumption \ref{assumption:minimum-arms} which says that there are always at least $N_m$ arms left to play. Therefore, $\sum_{i=1}^N L_i \geq \sum_{i=1}^{N_m} L_{(i)}$ where $\{L_{(i)}\}_{i=1}^N$ are the order statistics of $\{L_i\}_{i=1}^N$. Let us define $L^{N_m} = \sum_{i=1}^{N_m} L_{(i)}$ for brevity. Therefore,
\begin{align*}
    \E R_{N, T} &\leq - K\left( (1-\mu_i)^{K} - (1-\mu_i)^{K+1} \right) + \sum_{i=1}^N \frac{L_i^2}{L^{N_m}} \left((1-\mu_i) + (1-\mu_i)^{K} - (1-\mu_i)^{K+1} \right) + \\
    & \qquad \sum_{i=1}^N \frac{L_i}{\sum_{i=1}^N L_i} \E R_{N_i, T-L_i}
\end{align*}
Recursively expanding $\E R_{N_i, T-L_i}$ once,
\begin{align*}
    \E R_{N, T} &\leq - 2 K\left( (1-\mu_i)^{K} - (1-\mu_i)^{K+1} \right) + \sum_{i=1}^N \frac{L_i^2}{L^{N_m}} \left((1-\mu_i) + (1-\mu_i)^{K} - (1-\mu_i)^{K+1} \right) + \\
    & \qquad \sum_{i=1}^N \frac{L_i}{\sum_{i=1}^N L_i} \sum_{j \neq i}^{N} \frac{L_j^2}{L^{N_m}} \left((1-\mu_j) + (1-\mu_j)^{K} - (1-\mu_j)^{K+1} \right) + \frac{L_j}{\sum_{k \neq i}^N L_k} \E R_{N_{i,j}, T-L_i-L_j}
\end{align*}
Now, look at the second and third terms. Let us call $\tilde{\mu}_j = (1-\mu_j) + (1-\mu_j)^{K} - (1-\mu_j)^{K+1}$ for brevity. There are $N$ copies each of $L_i^2 \tilde{\mu}_i / L^{N^m}$ for all $i = 1, \dots, N$ where the sum of coefficients on copies is $2 - L_i / \sum_{j=1}^N L_j$. Therefore,
\begin{align*}
    \sum_{i=1}^N \frac{L_i^2}{L^{N_m}} \tilde{\mu}_i + \sum_{i=1}^N \frac{L_i}{\sum_{i=1}^N L_i} \sum_{j \neq i}^{N} \frac{L_j^2}{L^{N_m}} \tilde{\mu}_j &= \sum_{i=1}^N \sum_{j \neq i}^{N} \frac{L_i^2}{L^{N_m}} \tilde{\mu}_i + \frac{L_i}{\sum_{i=1}^N L_i}  \frac{L_j^2}{L^{N_m}} \tilde{\mu}_j \\
    &= \sum_{i=1}^N \left(2 - \frac{L_i}{\sum_{j=1}^N L_j}\right) \frac{L_i^2}{L^{N_m}} \tilde{\mu}_i
\end{align*}
Thus,
\begin{align*}
    \E R_{N, T} &\leq - 2 K\left( (1-\mu_i)^{K} - (1-\mu_i)^{K+1} \right) + \sum_{i=1}^N \left(2 - \frac{L_i}{\sum_{i=1}^N}\right) \frac{L_i^2}{L^{N_m}} \tilde{\mu}_i + \\
    & \qquad \sum_{i=1}^N \sum_{j \neq i}^{N} \frac{L_i}{\sum_{i=1}^N L_i} \frac{L_j}{\sum_{k \neq i}^N L_k} \E R_{N_{i,j}, T-L_i-L_j}
\end{align*}
At this point, it is easy to see where this recursion is going. If we expand $\E R_{N, T}$ $m$ times, then the coefficient on the first term will $m$; the coefficient of the terms within the second sum will be strictly smaller than $m$, and; the final term will have $m$ summations. As in our strategy for Theorem \ref{thm:bayesian-regret-pilot}, we can recursively expand out $\E R_{N_i, T - L_i}$ for at most $\lfloor T / L_{(1)} \rfloor$ times. The behavior of the last term is hard to analyze. There are $\lfloor T / L_{(1)} \rfloor$ summations. The $k$th sum is a weighted average over $N-k+1$ items. To make things concrete, note that the last term corresponds to any leftover budget that is not enough to fully play an arm. Therefore, we can bound the last term over the arm with the longest lifetime. So, applying expectation over $(\mu, L) \sim \Gamma$ before expanding out the recursion, we have
\begin{align*}
    \E_{\Gamma} \E R_{N, T} &< \E_{\Gamma} \left\lfloor \frac{T}{L_{(1)}} \right\rfloor \E_{\Gamma} \left[ \sum_{i=1}^N \frac{L_i^2}{L^{N_m}} \tilde{\mu}_i - K\left( (1-\mu_i)^{K} - (1-\mu_i)^{K+1} \right) \right] + \E_{\Gamma} L_i (1 - \mu_i) \\
    &= \E_{\Gamma} \left\lfloor \frac{T}{L_{(1)}} \right\rfloor E_{\Gamma} \left[ N  \frac{L^2}{L^{N_m}} \tilde{\mu} -K \left( (1-\mu)^{K} - (1-\mu)^{K+1} \right) \right] + \\
    & \qquad \E_{\Gamma} L_{(N)} (1 - \mu)
\end{align*}

If we assume that $\mu \perp L$ as in \ref{assumption:independence} and that $\mu \sim \text{Beta}(\alpha, \beta)$ as in \ref{assumption:beta}, then
\begin{align*}
    BR_{N, T} &\leq \E_{\Gamma} \left\lfloor \frac{T}{L_{(1)}} \right\rfloor E_{\Gamma} \left[\frac{NL^2}{L^{N_m}} \frac{\beta}{\alpha + \beta} + \left(\frac{NL^2}{L^{N_m}} -K \right) \frac{\alpha}{\beta + \alpha + K + 1} \prod_{r=0}^{K} \frac{\beta + r}{\beta + \alpha + r} \right] + \E L_{(N)} \frac{\beta}{\alpha + \beta}
\end{align*}

Finally, we need to account for the case $N < T / L_{(1)}$. We can do this by replacing $\E (T / L)$ with $\min\{N-1, \E (T/L) \}$. This completes the result.
\end{proofof}

\subsection{Choosing a pilot group size}
\label{section:pilot-group-size}

Let $\mu \sim \text{Beta}(\alpha, \beta)$ to denote the mean reward. \citet{chakrabarti2008mortal} impose an assumption that the lifetime of arms is exponentially distributed with mean $L$. They define the following reward function,
\begin{align}
	\g[R](x) &= \frac{\E[\mu] + (1 - F(x)) (L-1) \E[\mu \mid \mu \geq x]}{1 + (1 - F(x))(L-1)} \label{eq:reward}
\end{align}
where $F$ is the distribution of $\mu$.

Equation \ref{eq:reward} captures the trade-off between exploration and exploitation. The first term corresponds to the action where we pull an arm once. Suppose we are given a threshold $x$. The second term corresponds to the action that we pull arm $i$ until its death given that its mean reward is larger than $x$. Together, Equation \ref{eq:reward} describes the average reward per pull when following the strategy of pulling an arm $n \leq L_i$ times and pulling it until its death if the reward from the first $n$ pulls is at least $nx$.

Therefore, choosing $x^{*} = \argmax_x \g[R](x)$ gives us the optimal threshold. When choosing the pilot size $K$, \citet{chakrabarti2008mortal} show that if $K = \mathcal{O}(\log L / \epsilon^2)$ the average reward per step is $\g[R](x^* - \epsilon)$. We choose $K$ (rounded to the nearest integer greater than $K$) such that $K x^* = 1$. In other words, $K$ is the optimal number of pulls before we decide to abandon or exploit the arm.

Under Assumption \ref{assumption:beta}, Lemma \ref{lemma:pilot-reward} gives the exact form of the reward. Equation \ref{eq:our-reward} is concave for $x \in [0, 1]$ and thus has a unique maximizer, $x^*$, that can be calculated using standard optimization algorithms like gradient descent.

\begin{lemma}
\label{lemma:pilot-reward}
For $\mu \sim \text{Beta}(\alpha, \beta)$, Equation \ref{eq:reward} simplifies as
\begin{align}
    \g[R](x) &= \frac{\alpha}{\alpha + \beta} \frac{1 + (1 - F_{\alpha+1, \beta}(x))(L-1)}{1 + (1 - F_{\alpha, \beta}(x)(L-1)} \label{eq:our-reward}
\end{align}
where $F_{a, b}$ is the distribution of $\text{Beta}(a, b)$.
\end{lemma}

\begin{proofof}[Lemma \ref{lemma:pilot-reward}]
We have $\mu \sim \text{Beta}(\alpha, \beta)$. Let $f$ denote the density of $\mu$. Then, Equation \ref{eq:reward} simplifies as
\begin{align*}
	f(\mu \mid \mu \geq x) &= \frac{f(\mu, \mu \geq x )}{1 - F(x)} \\
	\implies \E[\mu \mid \mu \geq x] &= \frac{1}{1 - F(x)} \int_{x}^{1} \frac{1}{B(\alpha, \beta)}t^{\alpha}(1-t)^{\beta - 1} dt \\
	&= \frac{1}{1 - F(x)} \frac{\alpha}{\alpha + \beta} \int_{x}^{1} \frac{1}{B(\alpha+1, \beta)}t^{\alpha}(1-t)^{\beta - 1}dt \\
	&= \frac{\E[\mu]}{1 - F(x)} (1 - F_{\alpha+1, \beta}(x))
\end{align*}
where $B$ is the beta function and $F_{a, b}$ is the distribution of a $\text{Beta}(a, b)$ random variable. Thus,
\begin{align*}
	\g[R](x) &= \E[\mu]\frac{1 + (1 - F_{\alpha+1, \beta}(x))(L-1)}{1 + (1 - F(x)(L-1)}  \\
	&= \frac{\alpha}{\alpha + \beta} \frac{1 + (1 - F_{\alpha+1, \beta}(x))(L-1)}{1 + (1 - F_{\alpha, \beta}(x)(L-1)}
\end{align*}
\end{proofof}

%% file: sections/appendix_asymptotics.tex
\section{Asymptotic behavior of regret bounds}
\label{section:asymp-regret}

\begin{figure*}[!tbh]

    \centering
    \begin{subfigure}[t]{0.5\textwidth}
        \centering
        \includegraphics[height=2in]{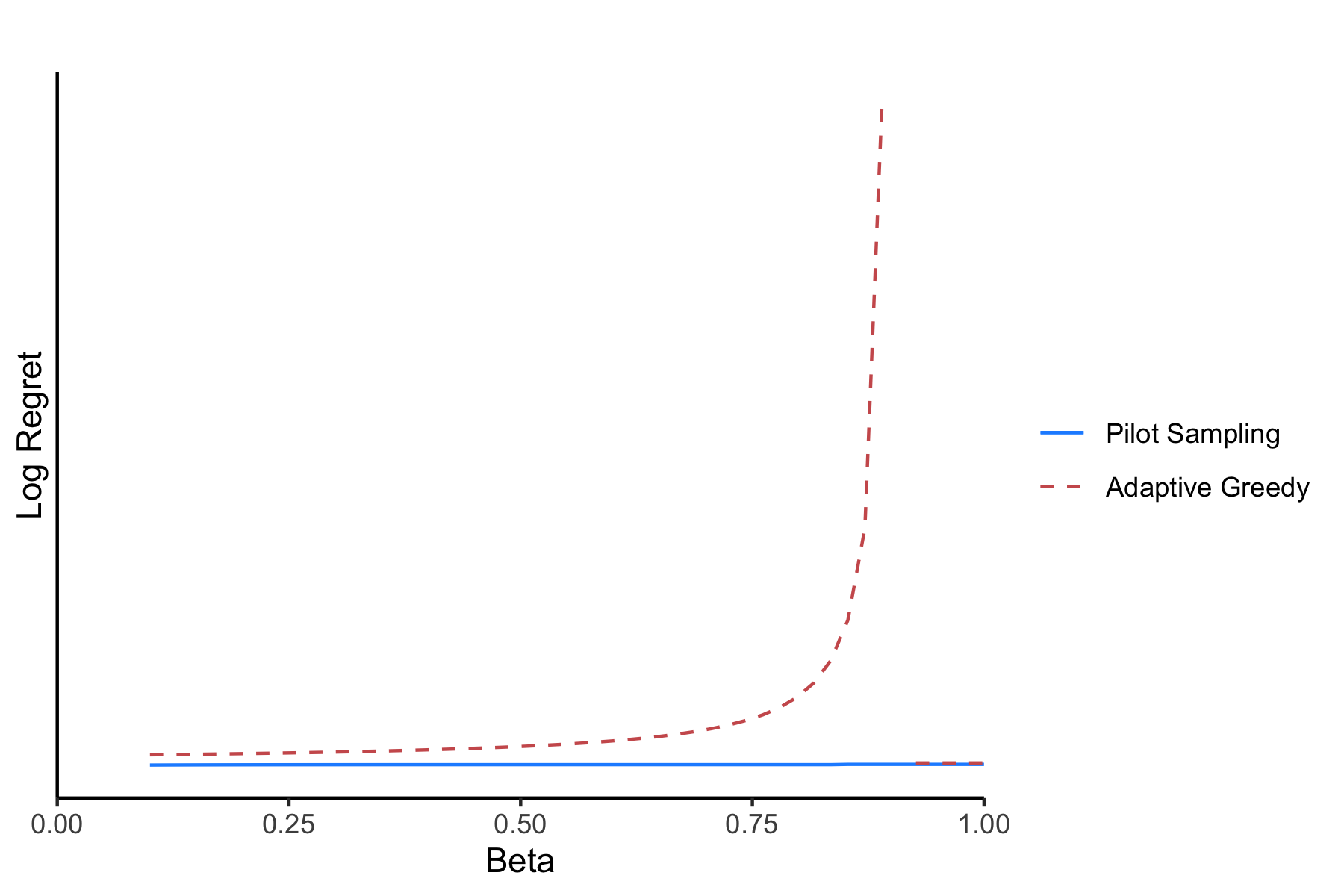}
        \caption{$\alpha = 0.1, \beta < 1$. Regret is presented on a log scale.}
    \end{subfigure}%
    ~
    \begin{subfigure}[t]{0.5\textwidth}
        \centering
        \includegraphics[height=2in]{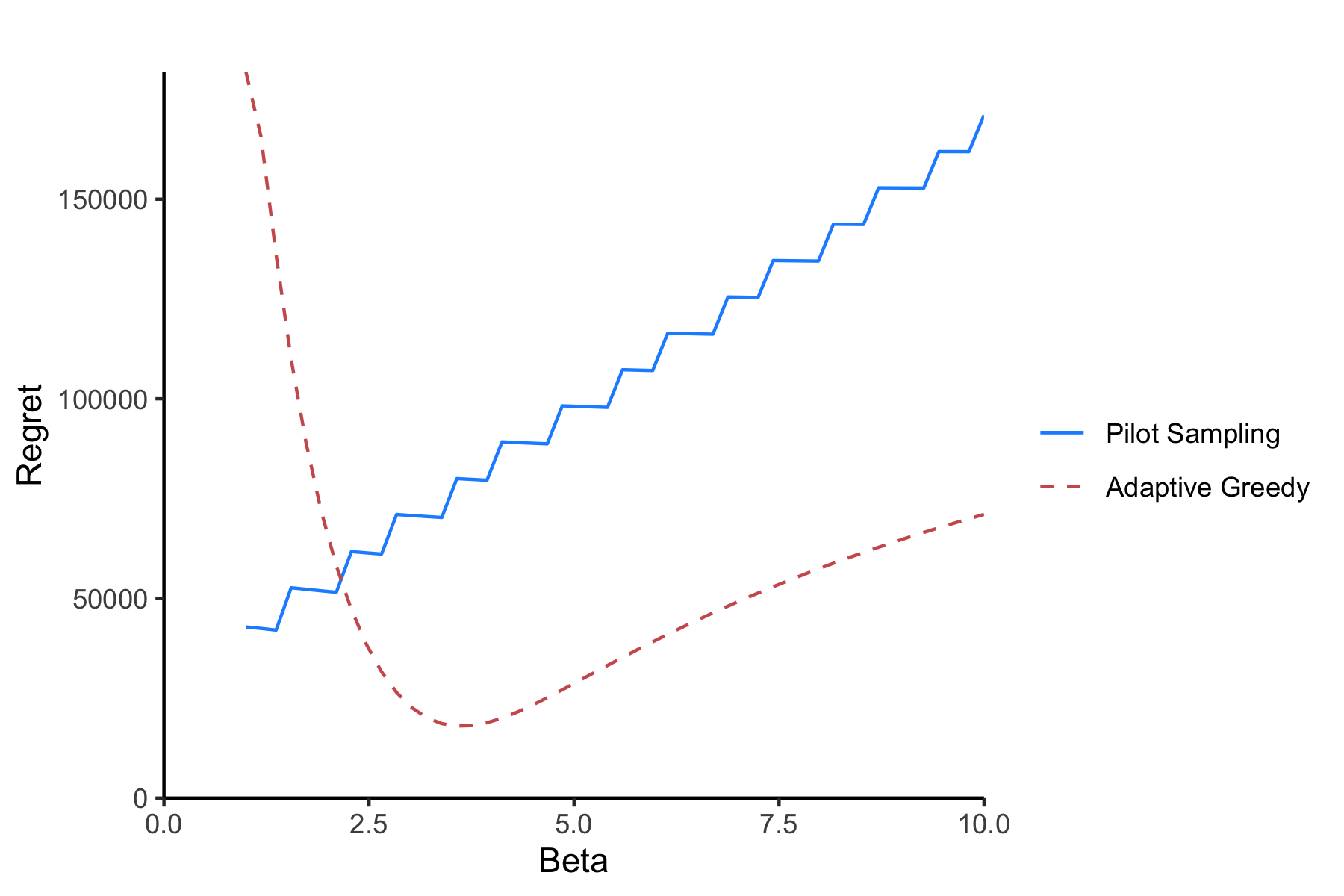}
        \caption{$\alpha = 0.1, \beta > 1$.}
    \end{subfigure}
    \\
    
    \begin{subfigure}[t]{0.5\textwidth}
        \centering
        \includegraphics[height=2in]{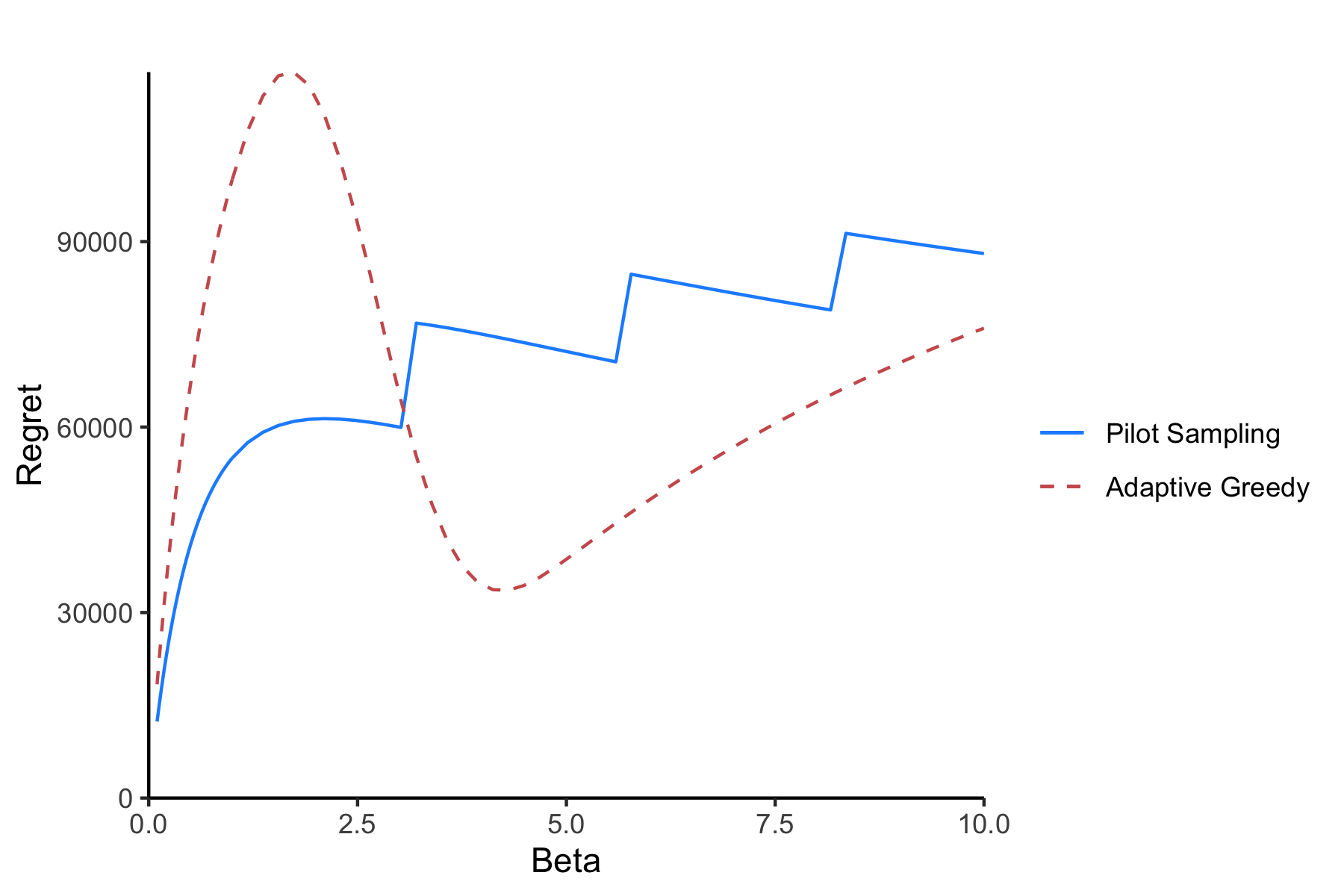}
        \caption{$\alpha = 1$}
    \end{subfigure}%
    ~
    \begin{subfigure}[t]{0.5\textwidth}
        \centering
        \includegraphics[height=2in]{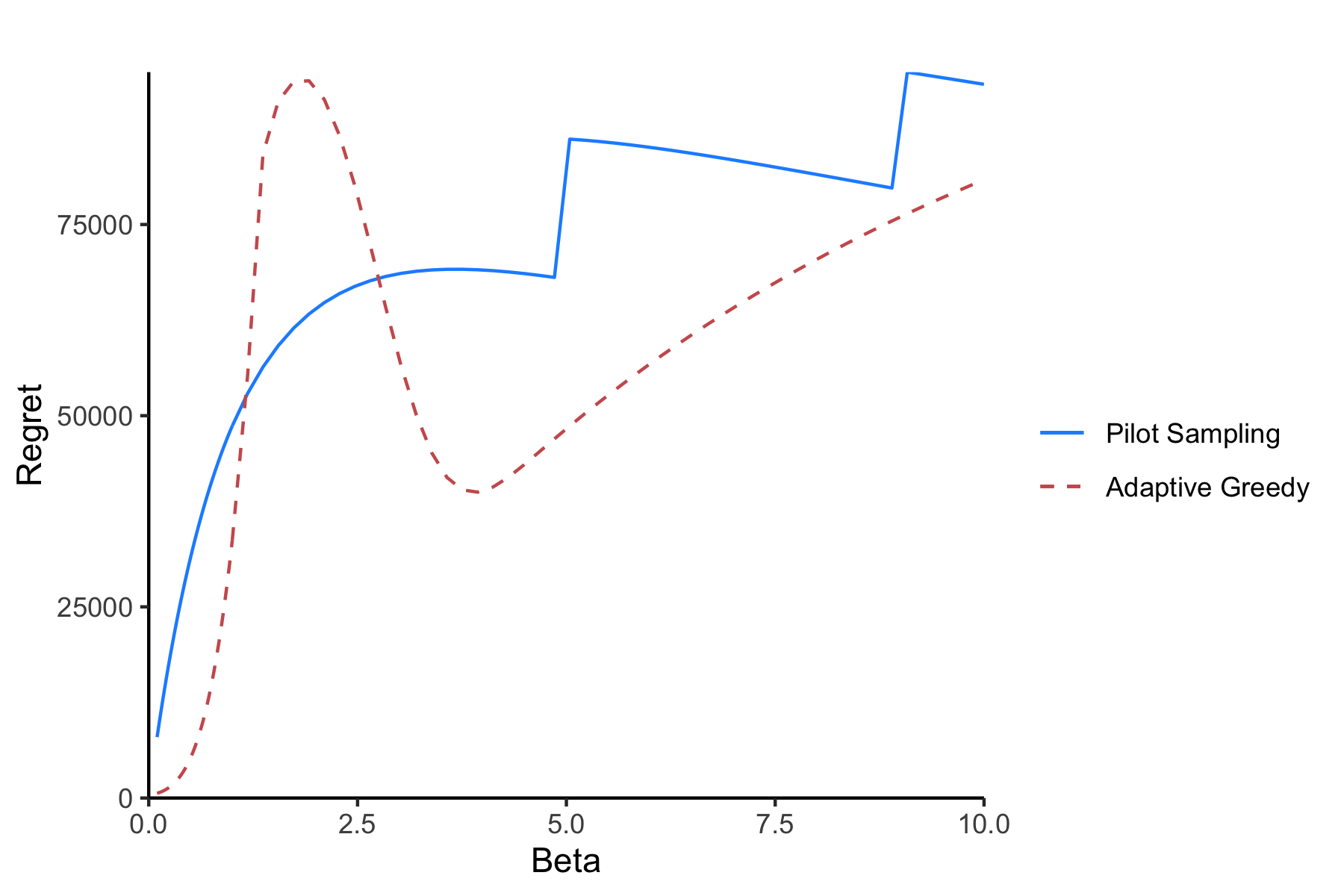}
        \caption{$\alpha = 2$}
    \end{subfigure}
    
    \caption{Bayesian regret bounds for Adaptive Greedy and Pilot Sampling algorithms presented in Theorems \ref{thm:br-adaptive-greedy-beta} and \ref{thm:bayesian-regret-pilot-beta} respectively. Here, we fixed $N = 10,000, N_m = 10, L = 20, T = 50000$. In each panel, we fixed $\alpha \in \{0.1, 1, 2\}$ and varied $\beta \in [0.1, 10]$. When $\alpha = 0.1$, we split the regret bounds into two plots to show the difference in the two algorithms at a meaningful scale. It is clear that when we have a heavily right-skewed distribution, Pilot Sampling has a much smaller regret bound compared to Adaptive Greedy (see panel (a)). This is exactly what we demonstrate in our simulations in Section \ref{section:sims}. When there is a heavy left-skew, Adaptive Greedy has a smaller regret bound which is also in line with our simulations in Section \ref{section:sims}. For other cases, when the regret bounds are comparable, we see little to no difference in performance in our simulations.}
    \label{fig:regret-bounds}
\end{figure*}

In Figure \ref{fig:regret-bounds}, we compare the regret bounds between Adaptive Greedy and Pilot Sampling algorithms for Beta priors.

\begin{figure*}[!tbh]

    \centering
    \begin{subfigure}[t]{0.5\textwidth}
        \centering
        \includegraphics[height=2in]{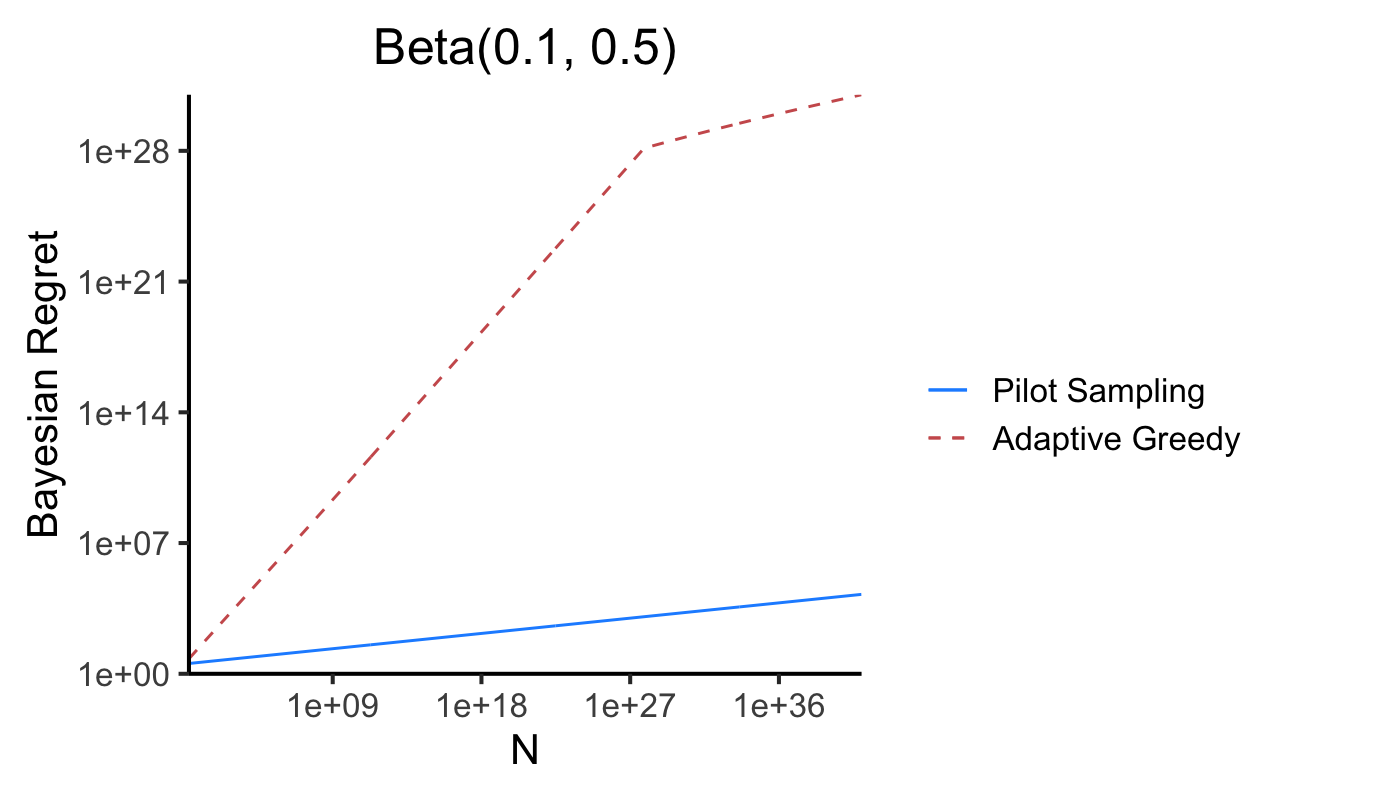}
    \end{subfigure}%
    ~
    \begin{subfigure}[t]{0.5\textwidth}
        \centering
        \includegraphics[height=2in]{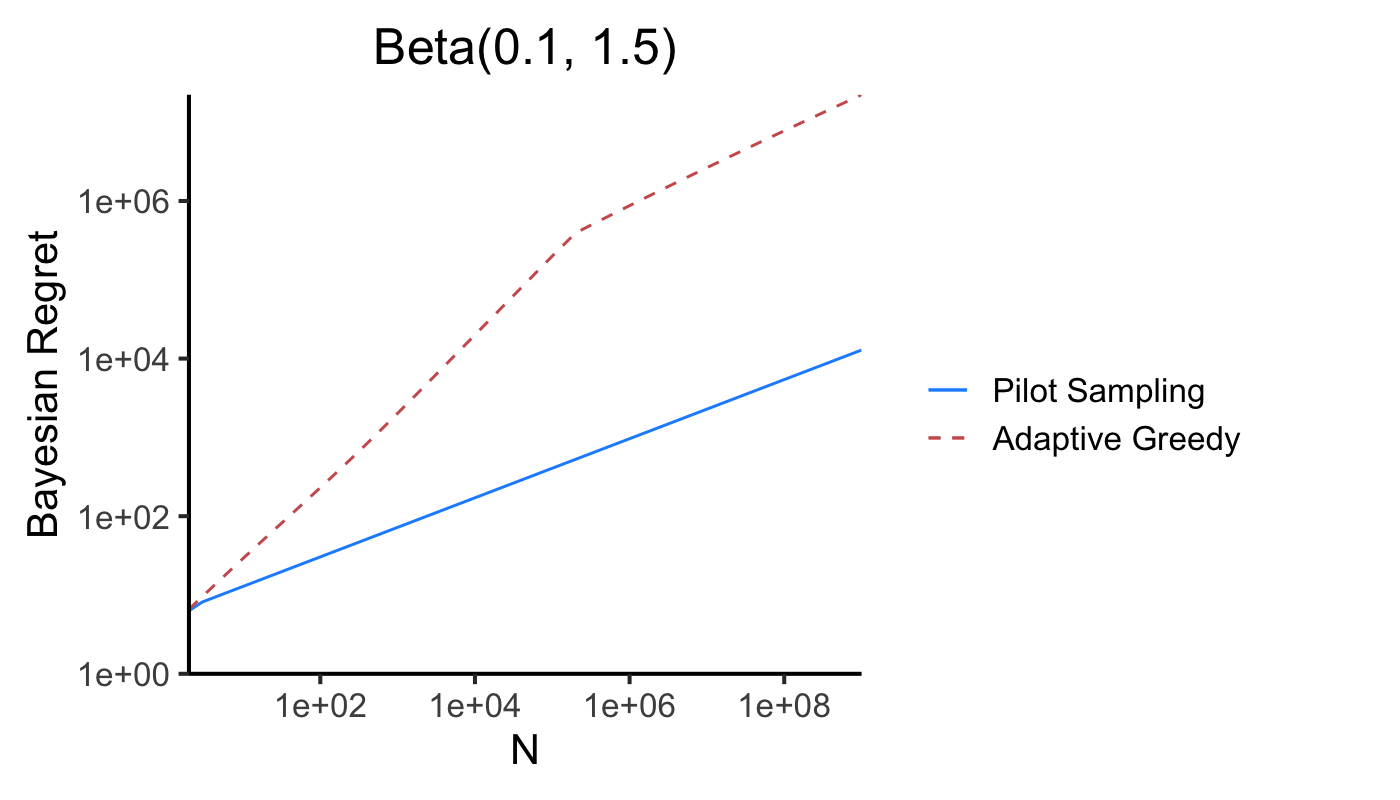}
    \end{subfigure}
    \\
    
    \begin{subfigure}[t]{0.5\textwidth}
        \centering
        \includegraphics[height=2in]{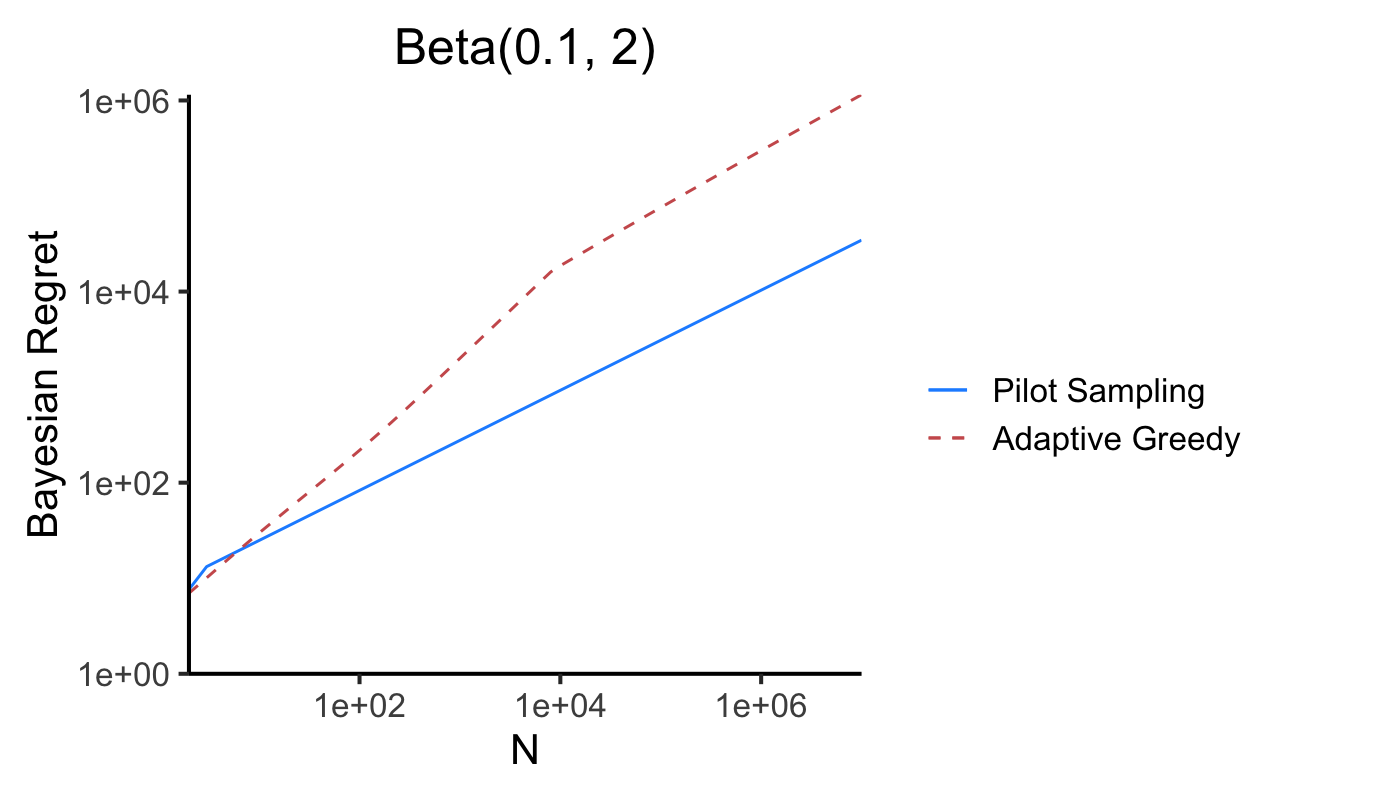}
    \end{subfigure}%
    ~
    \begin{subfigure}[t]{0.5\textwidth}
        \centering
        \includegraphics[height=2in]{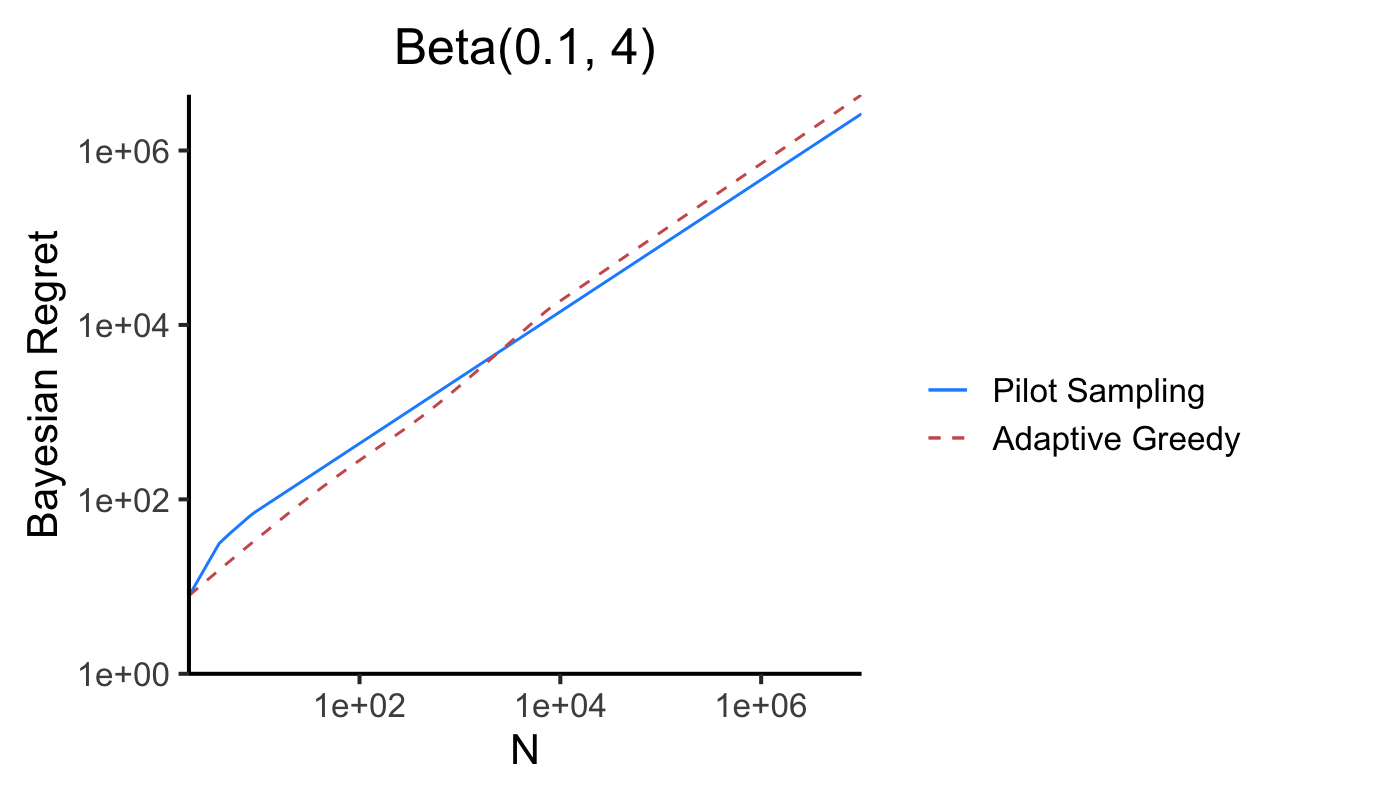}
    \end{subfigure}
    
    \caption{In these figures, we fix $\alpha = 0.1$ and vary $\beta \in \{0.5, 1.5, 2, 4\}$. Except for the case in the top left with $\alpha < \beta < 1$, the asymptotic behavior is reached at a reasonable $N = 10^4$. Based on this, we can conclude that our datasets lie in this asymptotic regime. Here, we fixed $T / N = 0.5$.}
    \label{fig:regret-asymp-small-alpha}
\end{figure*}

\begin{figure*}[!tbh]

    \centering
    \begin{subfigure}[t]{0.5\textwidth}
        \centering
        \includegraphics[height=2in]{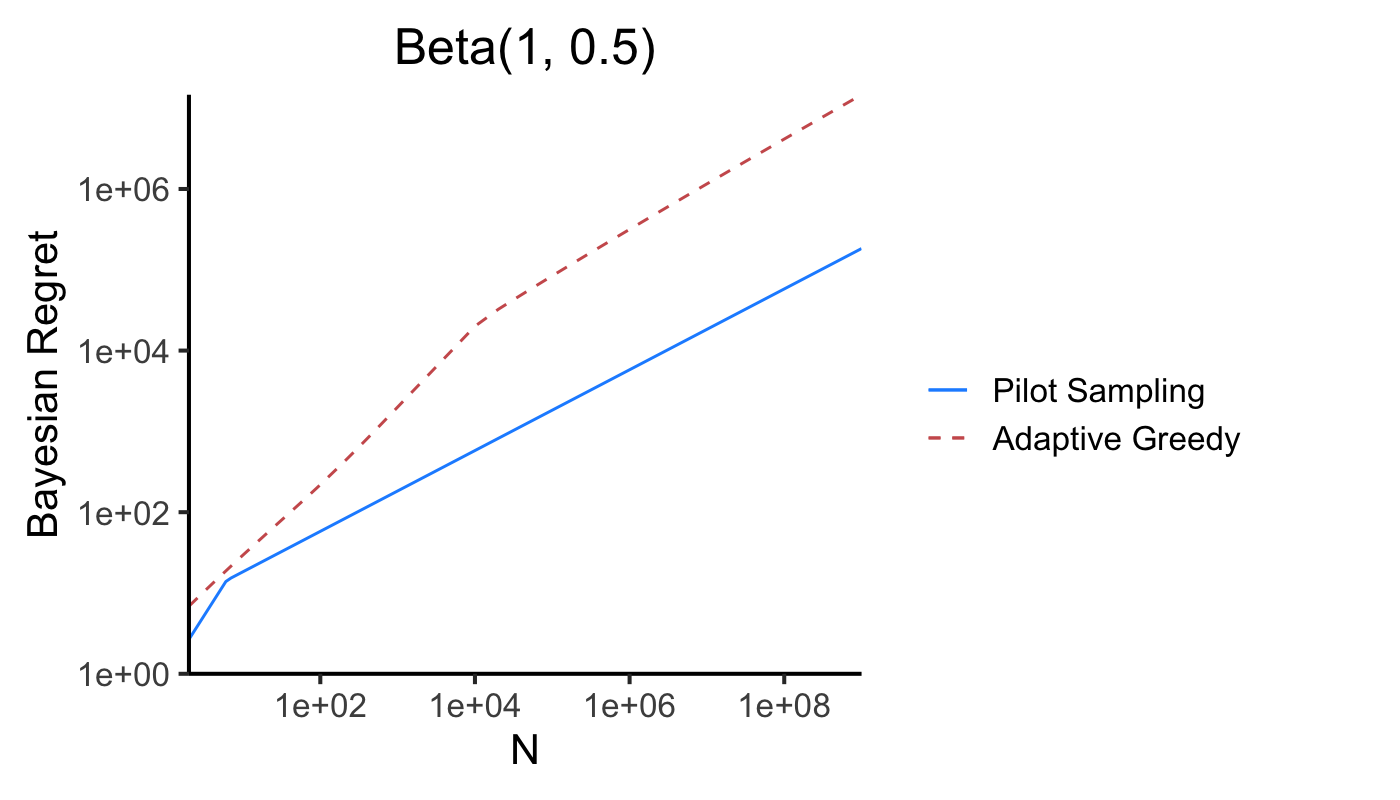}
    \end{subfigure}%
    ~
    \begin{subfigure}[t]{0.5\textwidth}
        \centering
        \includegraphics[height=2in]{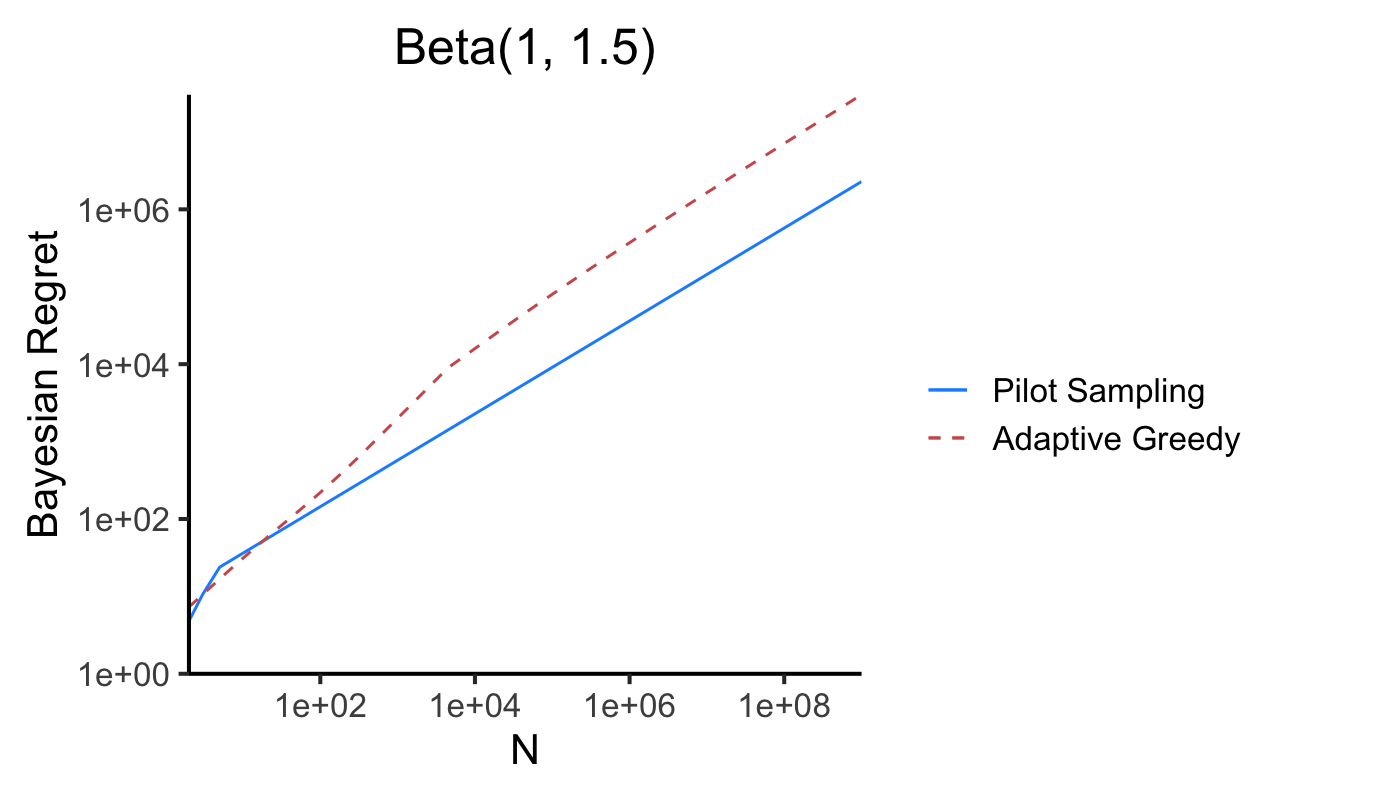}
    \end{subfigure}
    \\
    
    \begin{subfigure}[t]{0.5\textwidth}
        \centering
        \includegraphics[height=2in]{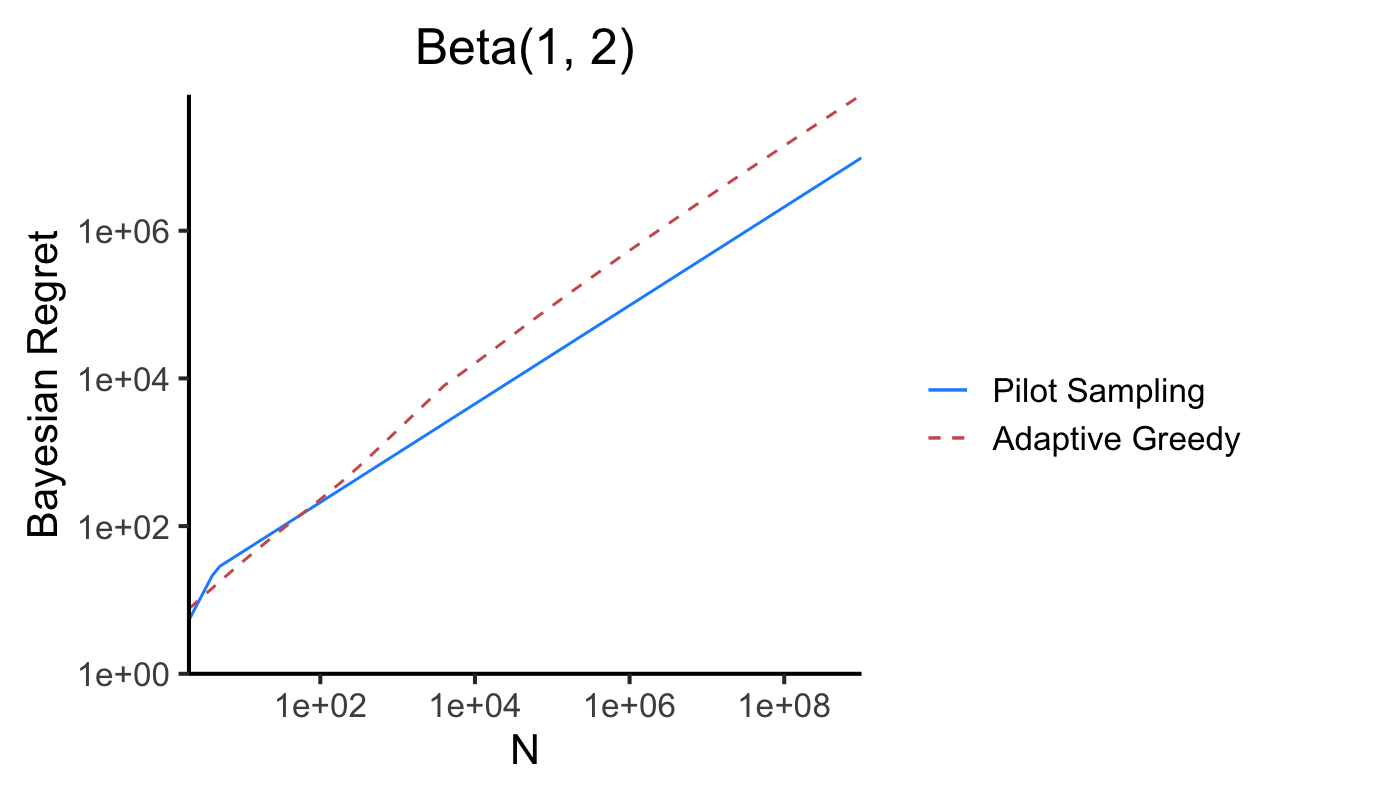}
    \end{subfigure}%
    ~
    \begin{subfigure}[t]{0.5\textwidth}
        \centering
        \includegraphics[height=2in]{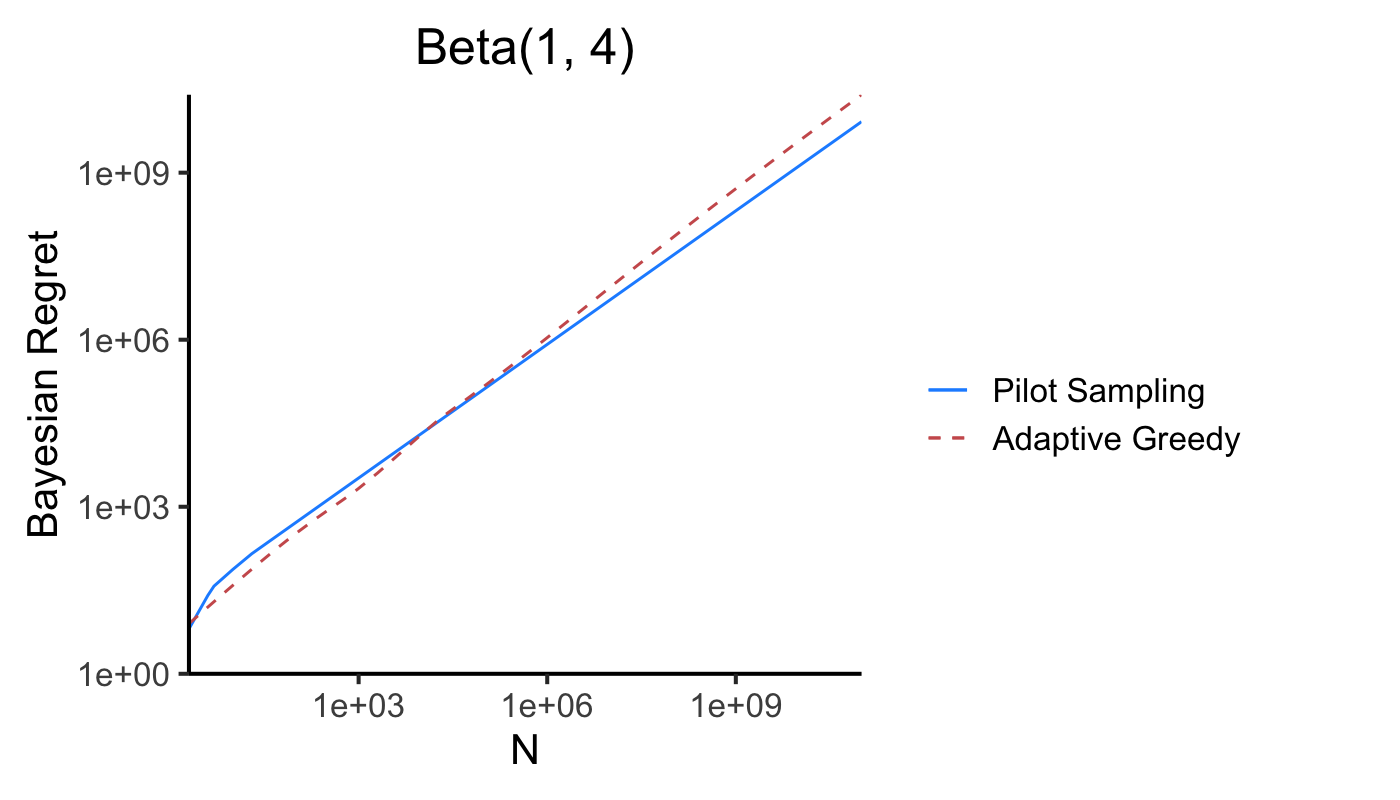}
    \end{subfigure}
    
    \caption{In these figures, we fix $\alpha = 1$ and vary $\beta \in \{0.5, 1.5, 2, 4\}$. Again, we can conclude that the asymptotic behavior is achieved quickly. Here, we fixed $T / N = 0.5$.}
    \label{fig:regret-asymp-alpha-1}
\end{figure*}

\begin{figure*}[!tbh]

    \centering
    \begin{subfigure}[t]{0.5\textwidth}
        \centering
        \includegraphics[height=2in]{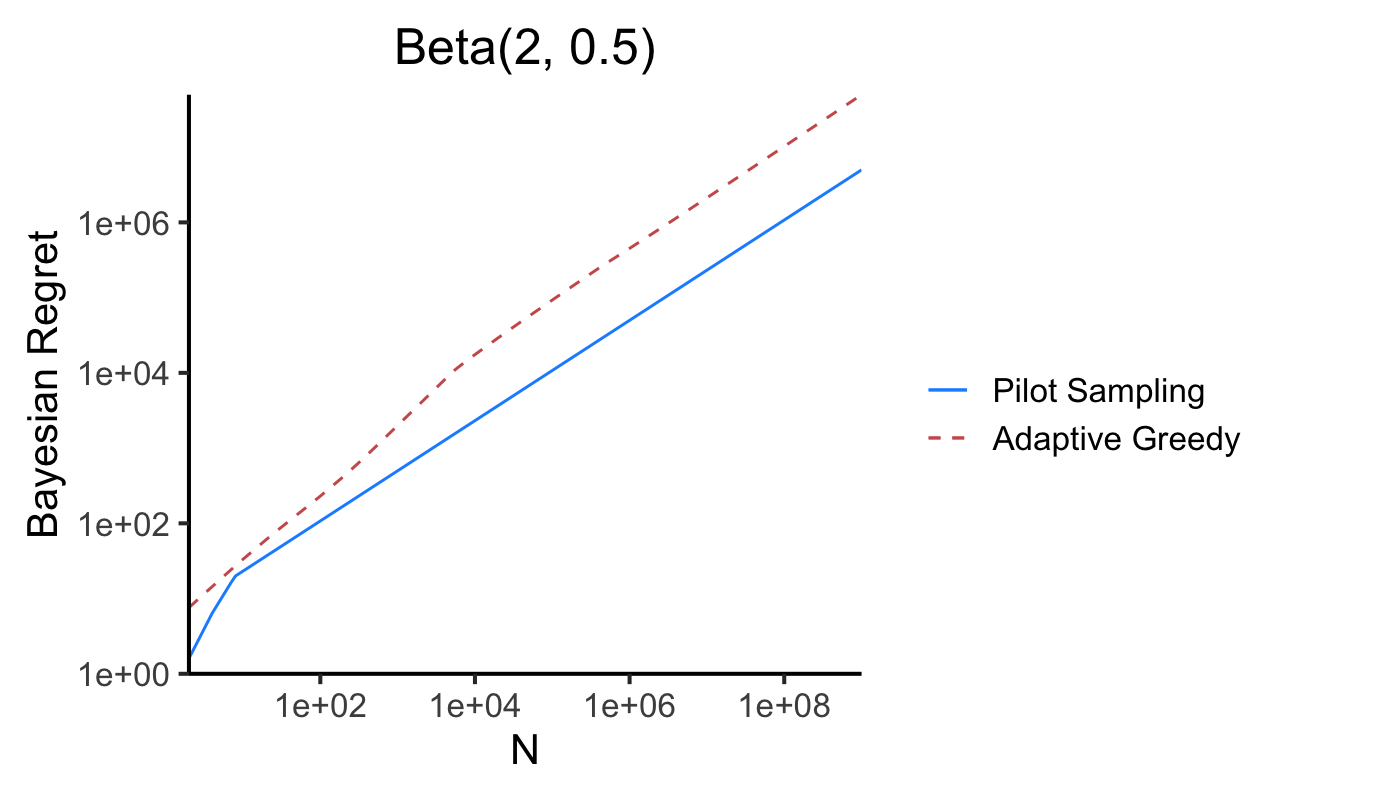}
    \end{subfigure}%
    ~
    \begin{subfigure}[t]{0.5\textwidth}
        \centering
        \includegraphics[height=2in]{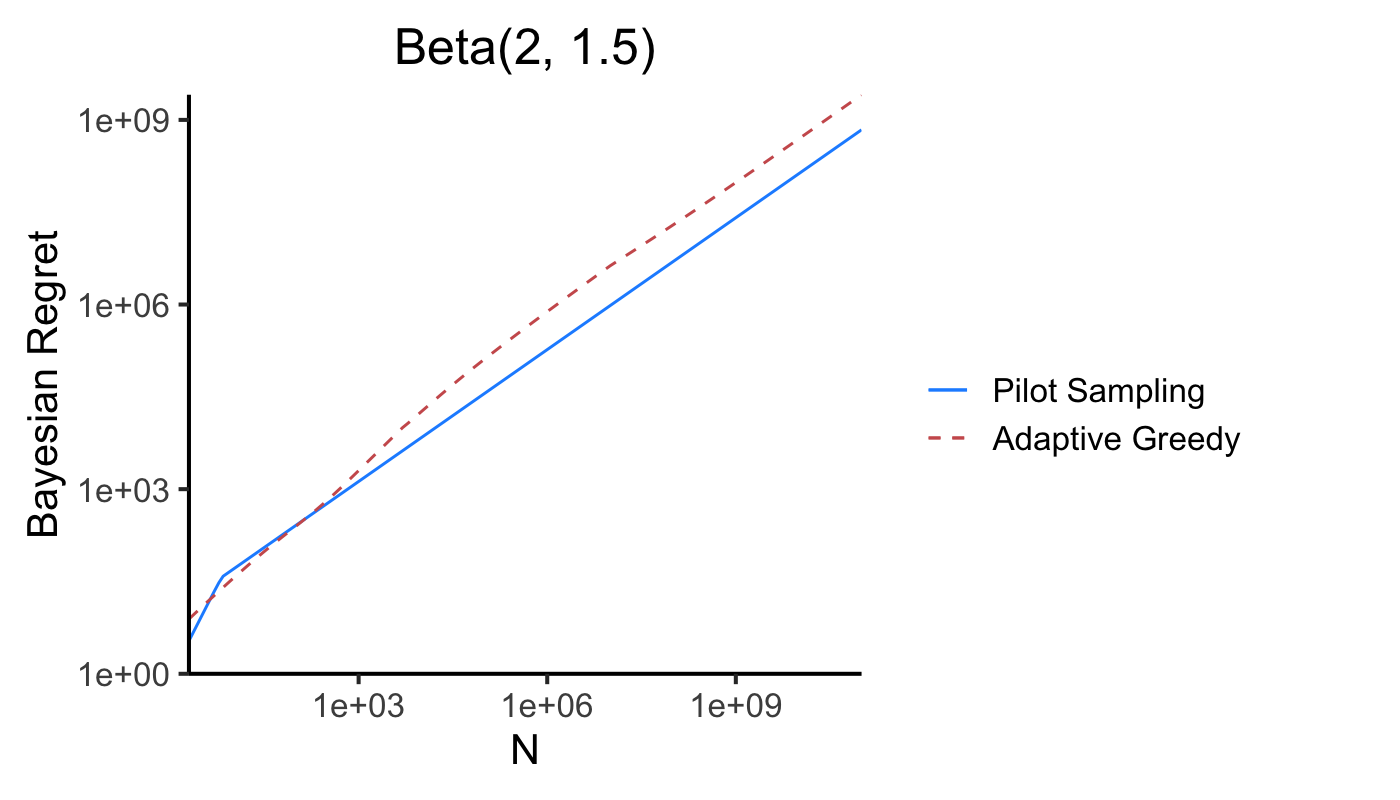}
    \end{subfigure}
    \\
    
    \begin{subfigure}[t]{0.5\textwidth}
        \centering
        \includegraphics[height=2in]{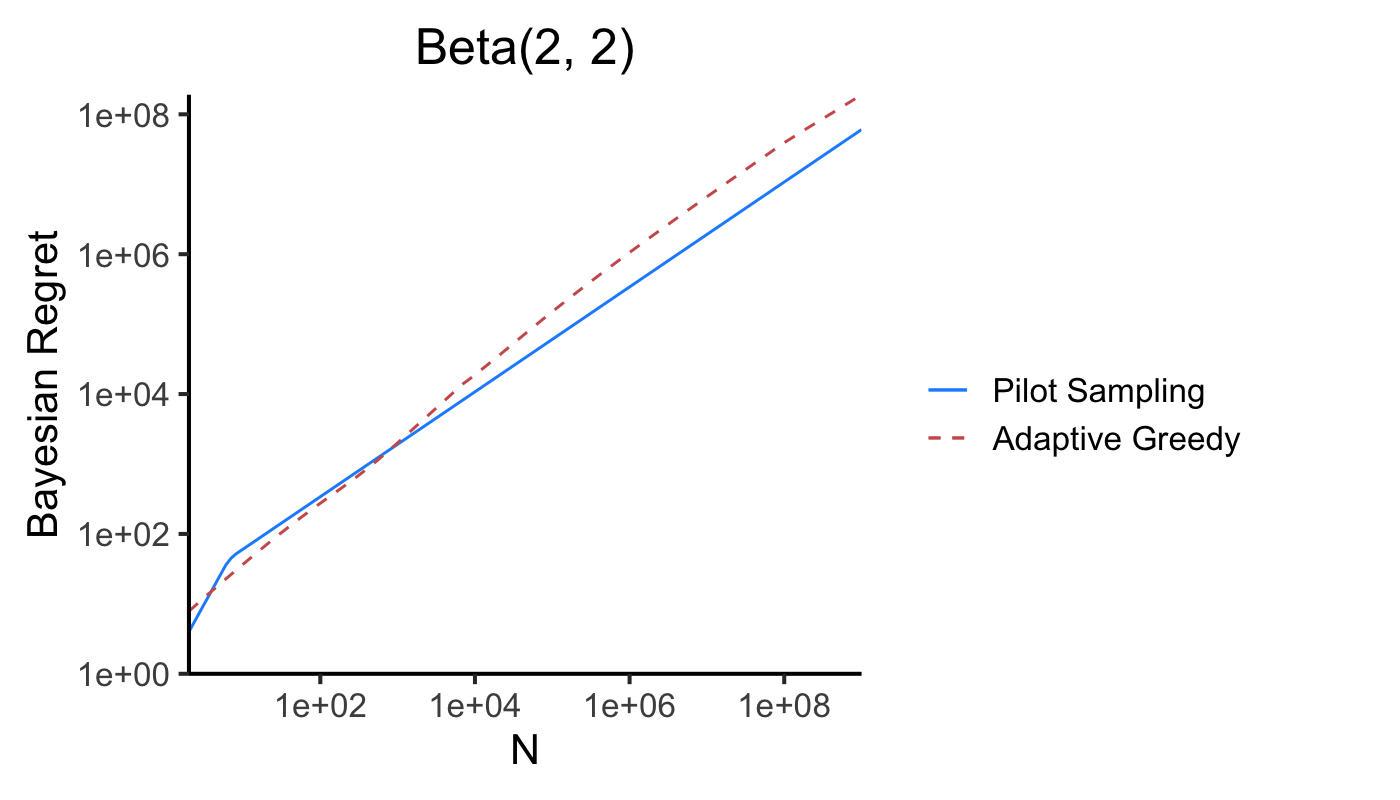}
    \end{subfigure}%
    ~
    \begin{subfigure}[t]{0.5\textwidth}
        \centering
        \includegraphics[height=2in]{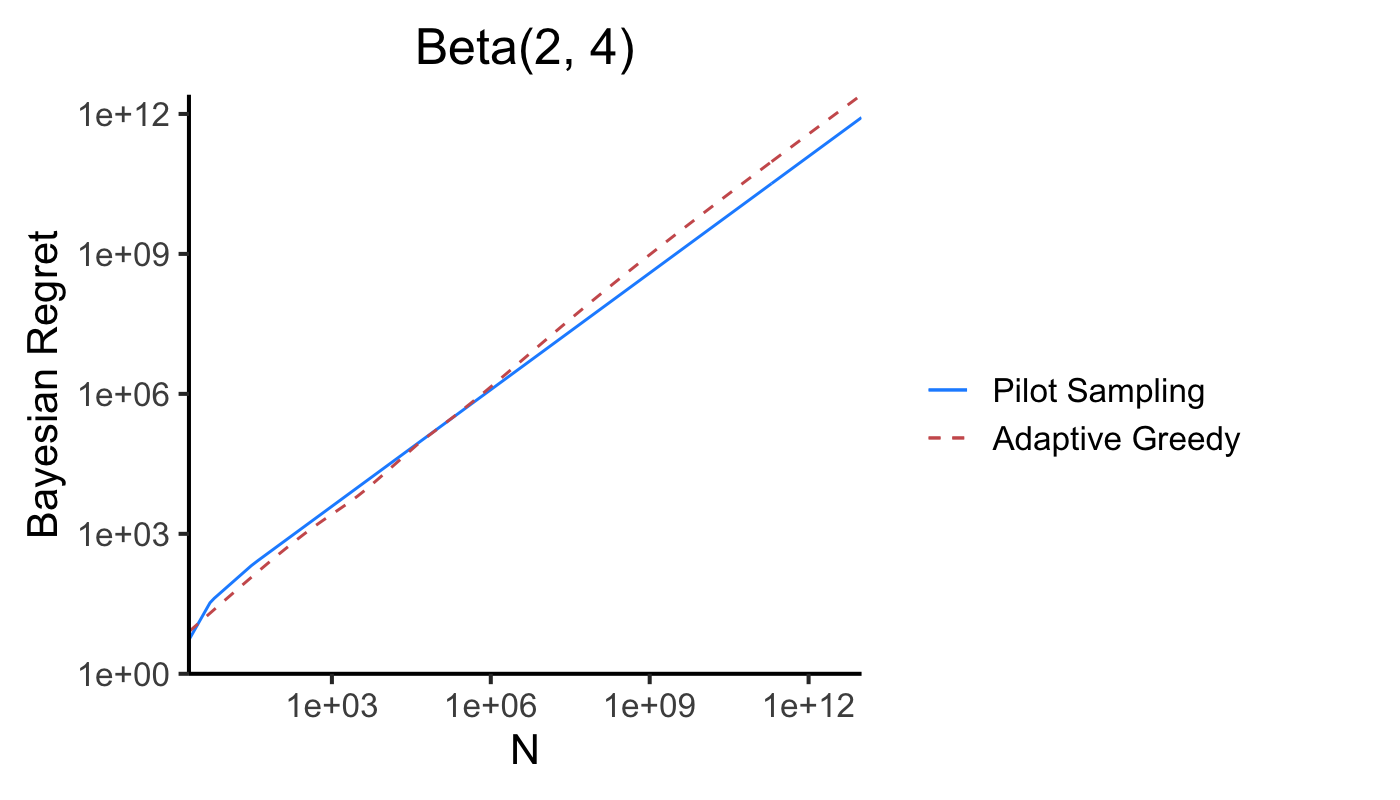}
    \end{subfigure}
    
    \caption{In these figures, we fix $\alpha = 2$ and vary $\beta \in \{0.5, 1.5, 2, 4\}$. Again, the conclusions remain the same. Here, we fixed $T / N = 0.5$.}
    \label{fig:regret-asymp-alpha-2}
\end{figure*}

\begin{figure*}[!tbh]

    \centering
    \begin{subfigure}[t]{0.5\textwidth}
        \centering
        \includegraphics[height=2in]{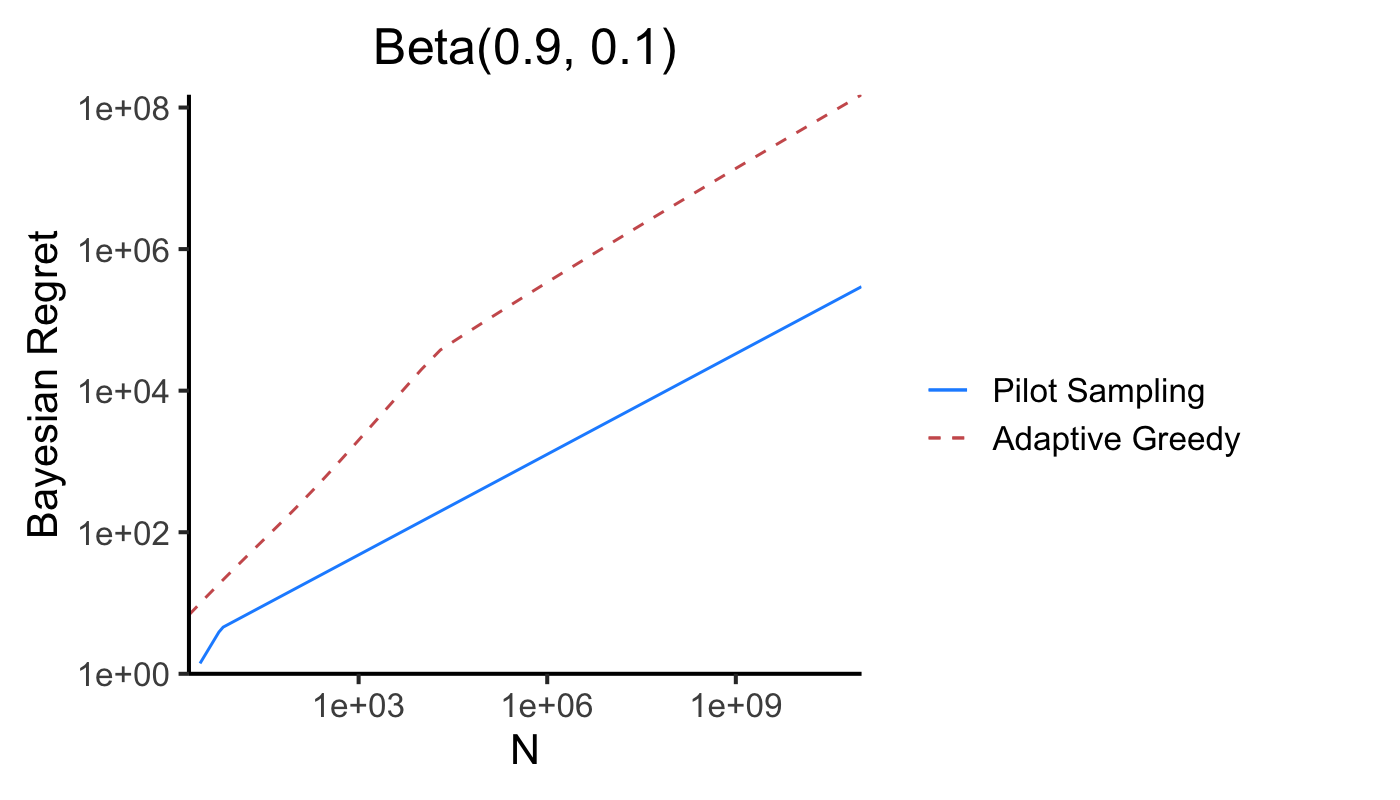}
    \end{subfigure}%
    ~
    \begin{subfigure}[t]{0.5\textwidth}
        \centering
        \includegraphics[height=2in]{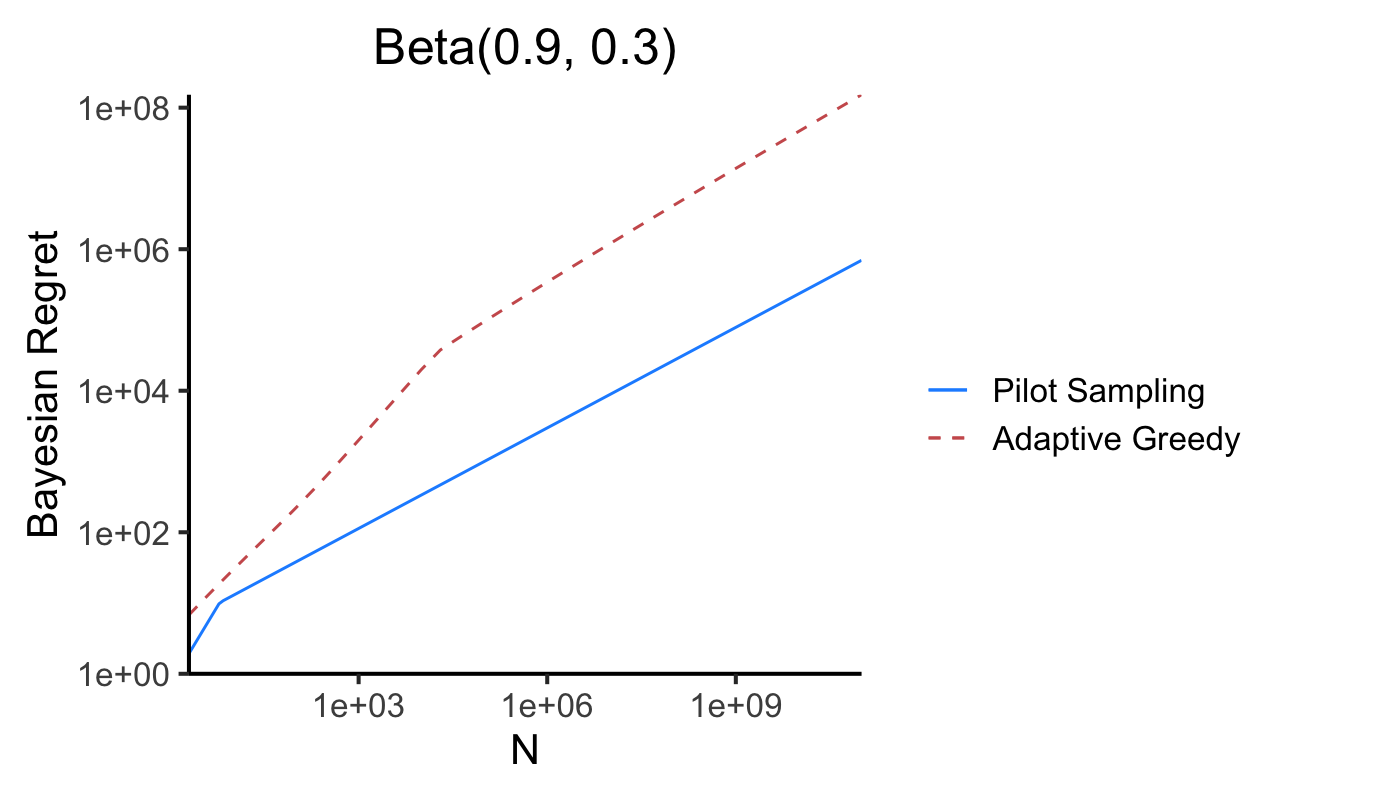}
    \end{subfigure}
    \\
    
    \begin{subfigure}[t]{0.5\textwidth}
        \centering
        \includegraphics[height=2in]{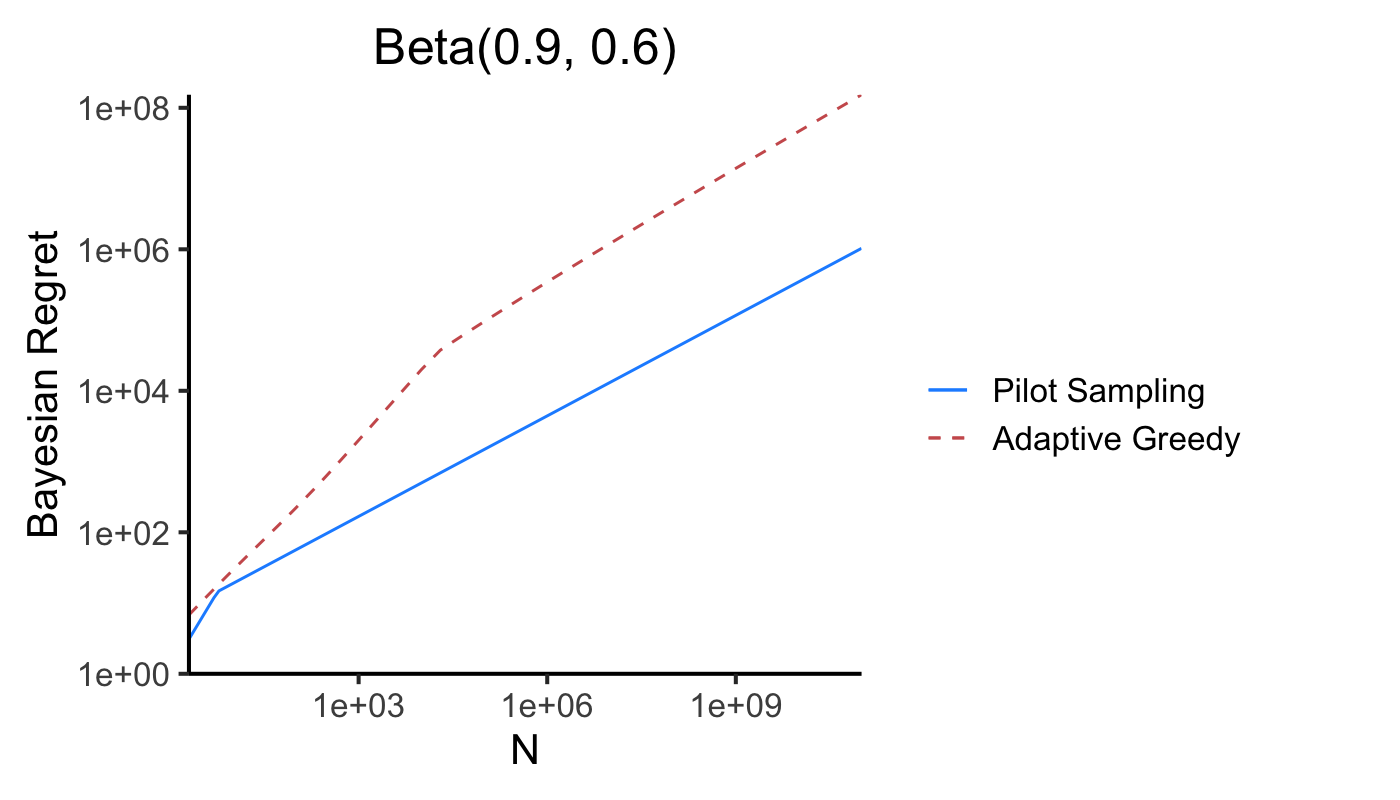}
    \end{subfigure}%
    ~
    \begin{subfigure}[t]{0.5\textwidth}
        \centering
        \includegraphics[height=2in]{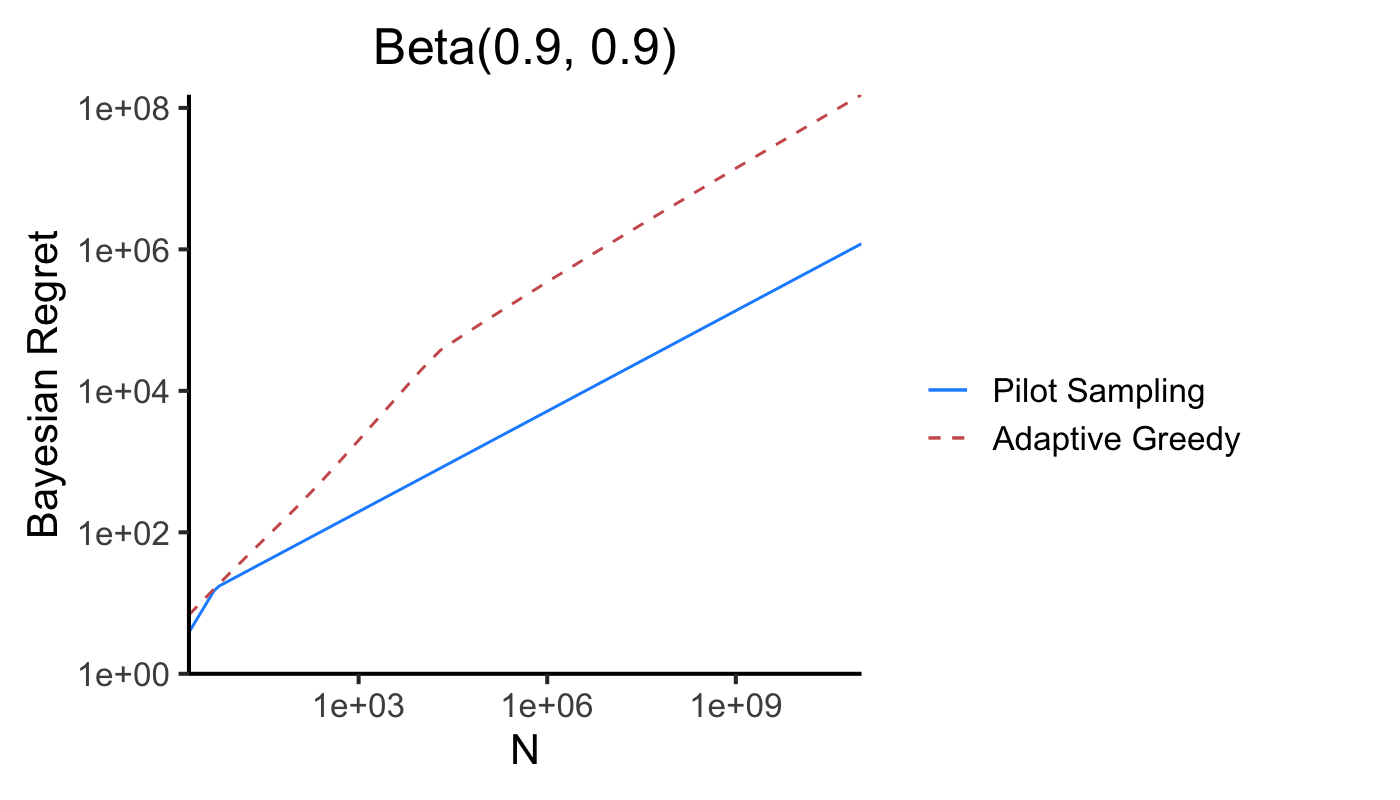}
    \end{subfigure}
    
    \caption{In these figures, we fix $\alpha = 0.9$ and vary $\beta \in \{0.1, 0.3, 0.6, 0.9\}$. Again, the conclusions remain the same. Here, we fixed $T / N = 0.5$.}
    \label{fig:regret-asymp-left-skew}
\end{figure*}

Here, we study the asymptotic behavior of the Bayesian regret bounds for a variety of Beta priors. In particular, we look at four different scenarios: (i) $\alpha < 1$, (ii) $\alpha = 1$, (iii) $\alpha > 1$, and (iv) $\beta < \alpha < 1$. These are shown in Figures \ref{fig:regret-asymp-small-alpha}-\ref{fig:regret-asymp-left-skew} respectively. In these plots, we used the bounds presented in Corollaries \ref{thm:br-adaptive-greedy-beta} and \ref{thm:bayesian-regret-pilot-beta}. In all of these plots, we fixed the ratio $T / N = 0.5$ and varied $N$ to identify when the asymptotic behavior is achieved.

From these simulations, we can immediately learn three things. First, for a fixed $\alpha$, as we increase $\beta$, the upper bounds of Pilot sampling and Adaptive Greedy get closer to each other. Second, for a fixed $\beta$, as we increase $\alpha$, the upper bounds of Pilot sampling Adaptive Greedy get closer to each other. Third, the asymptotic behavior occurs in the range of $N$ between $10^3$ (generally for smaller $\alpha$) and $10^6$ (generally for larger $\alpha$) depending on the prior distribution. The only exception is when $\alpha < \beta < 1$ where sometimes we need $N > 10^{27}$ for the asymptotic behavior to occur.

Now, we attempt to translate some of the patterns revealed by the asymptotic bounds to our empirical simulations. First, the size of the ``asymptotic gap'' between Pilot sampling and Adaptive Greedy tells us how they compare. Note that this gap is primarily composed of the logarithmic factors present in Theorem \ref{thm:ag-br}. Observe that, usually, the size of the gap corresponds to how much better Pilot sampling is. Particularly, consider $\alpha < \beta < 1$, and $\alpha = 1, \beta = 4$. Compared with simulations, we see that, in the former case, both Pilot sampling is better and the gap is larger. Of course, as the gap becomes smaller, it is more likely that Adaptive Greedy performs better than Pilot sampling in practice.

The size of this gap also corresponds with the variance of the distribution of the mean rewards. The larger the variance, the larger the gap. When the variability (heterogeneity) is larger, Pilot sampling performs better. This is seen in the case with $\alpha < \beta < 1$. With a smaller variance ($\alpha = 1, \beta = 4$), the heterogeneity is not large enough for Pilot sampling to beat Adaptive Greedy.

%% file: sections/appendix_algorithms.tex
\section{Additional comparison algorithms}
\label{section:comparison-algorithms}

\paragraph{A naive approach.}

We consider a naive approach that one might take when playing the mortal bandit. In this method, an agent may choose an arm uniformly at random and play it until it dies. Then, the agent repeats the procedure on all remaining arms. While this sampling method is very simple to understand and implement in practice, this method is sub-optimal as the agent does not do any kind of learning of the mean rewards and is agnostic to the lifetimes of arms. Hence, we dub this the ``naive sampler.'' This method is described in Algorithm \ref{alg:naive}.

%=================
% Algorithm 1
%=================
\begin{algorithm}[!tbh]
\caption{Naive sampling}\label{alg:naive}
\begin{algorithmic}[1]
\Require $T$ budget, $N$ arms
\State Set $t = 0$
\While {$t < T$}
    \State randomly choose an available arm $i$
    \State Pull arm $i$ until it dies ($L_i$ times) or we exhaust budget
    % \State Add any new infections as an arm
    \State $t = t + L_i$
\EndWhile
\end{algorithmic}
\end{algorithm}

\paragraph{Thompson sampling.}

Thompson sampling is a sequential decision-making algorithm that uses a Bayesian model \citep{thompson1933likelihood}. While this method is very simple, its empirical performance makes it highly competitive and has been shown to theoretically achieve optimal performance by \citet{chapelle2011empirical} and \citet{agrawal2012analysis} in the standard bandit setting.

Consider the Beta-Bernoulli model, $\mu_i \sim \text{Beta}(\alpha_i, \beta_i)$ and $X_i \mid \mu_i \sim \text{Bernoulli}(\mu_i)$. In our problem, this models the infectiousness of a person $i$ with $\mu$ representing their per-contact infectivity and $X$ representing whether a random contact of person $i$ gets infected.

The algorithm works the following way. First, we assign the mean reward of all arms the same Beta prior. This can, and often will, be different from the prior from which $\mu_i$ originally came from. For simplicity, we may choose the uniform prior i.e., $\text{Beta}(1, 1)$. Then, we draw $\widetilde{\mu}_i$ for every arm. We pick the arm with the largest $\widetilde{\mu}_i$ and pull it once. Using the observed $X_i$, we update arm $i$'s prior. Algorithm \ref{alg:thompson} formally describes the method.

%=================
% Algorithm 2
%=================
\begin{algorithm}[!tbh]
\caption{Thompson sampling}\label{alg:thompson}
\begin{algorithmic}[1]
\Require $T$ budget, $N$ arms, $(\alpha, \beta)$ initial priors
\State For every person $i \in \mathcal{I}$, assign them prior parameters $(\alpha, \beta)$
\State $t = 0$
\While {$t < T$}
    \State $t = t + 1$
    \State Sample $\widetilde{\mu}_i \sim \text{Beta}(\alpha_i, \beta_i)$ for arm $i$ that is alive
    \State Find $k = \argmax_i \widetilde{\mu}_i$
    \State Pull arm $k$ to get reward $X_k$
    \If {$X_k = 1$}
        \State Add infection as a new arm and assign priors $(\alpha, \beta)$
        \State Update $\alpha_k \gets \alpha_k + 1$
    \Else
        \State Update $\beta_k \gets \beta_k + 1$
    \EndIf
    \State Remove arm $k$ if is dead
\EndWhile
\end{algorithmic}
\end{algorithm}

Thompson sampling leverages heterogeneity in mean rewards by assigning it a prior distribution. This allows us to model each arm's mean reward independently. By randomly sampling from and updating the Beta model during each iteration, Thompson sampling enables us to efficiently explore the pool of arms. 

Despite being mathematically simple, Thompson sampling can be cumbersome to implement in practice. Theoretically, Thompson sampling invests a lot of resources in pulling randomly chosen arms, one at a time, before identifying the arm with the largest reward. In a setting such as contact tracing. it can be logistically difficult to test randomly during an outbreak of a disease.

%% file: sections/appendix_additional_sims.tex
\section{Additional Simulations}

\subsection{Visualizing Beta distributions}

\begin{figure*}[!tbh]

    \centering
    \begin{subfigure}[t]{0.5\textwidth}
        \centering
        \includegraphics[height=2in]{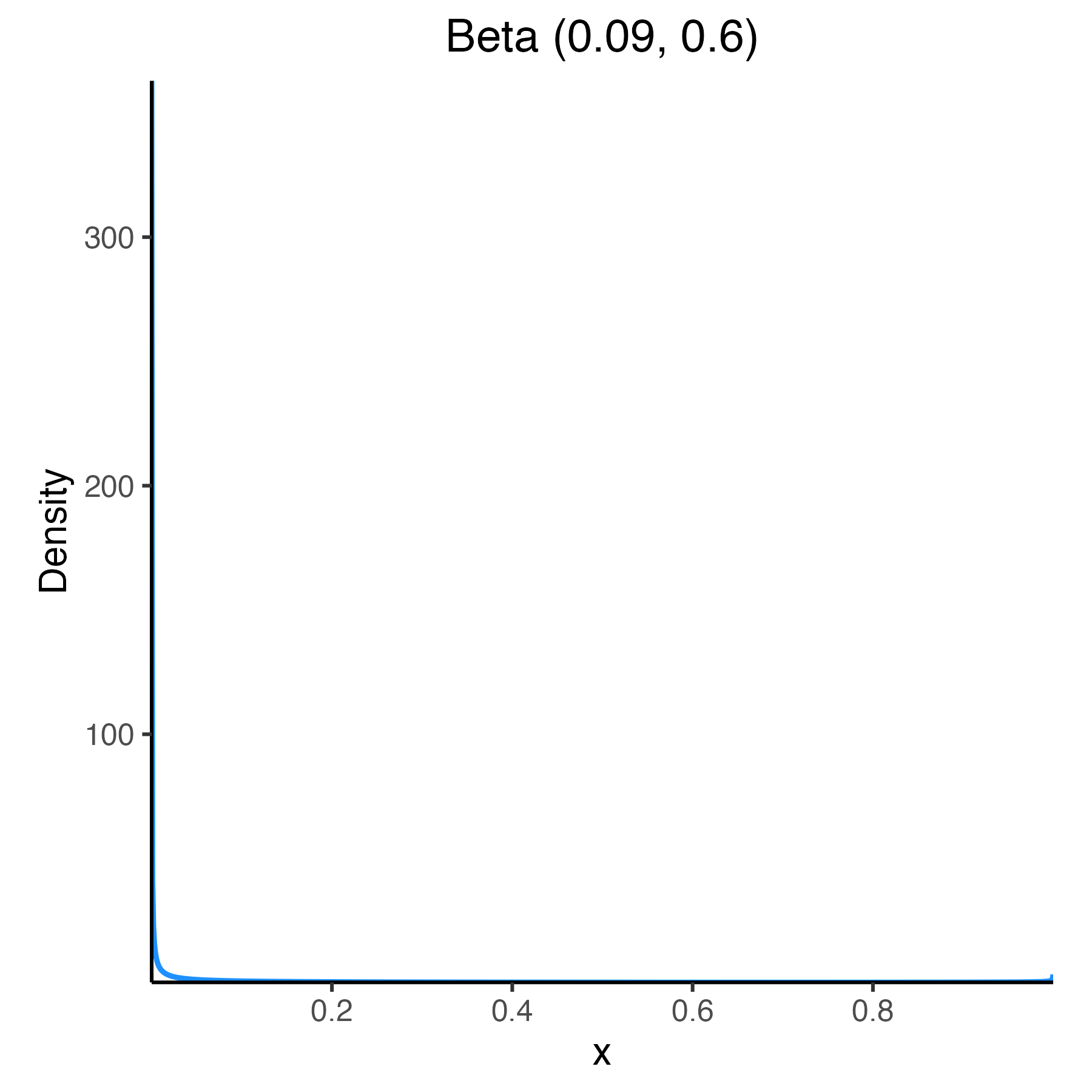}
    \end{subfigure}%
    ~
    \begin{subfigure}[t]{0.5\textwidth}
        \centering
        \includegraphics[height=2in]{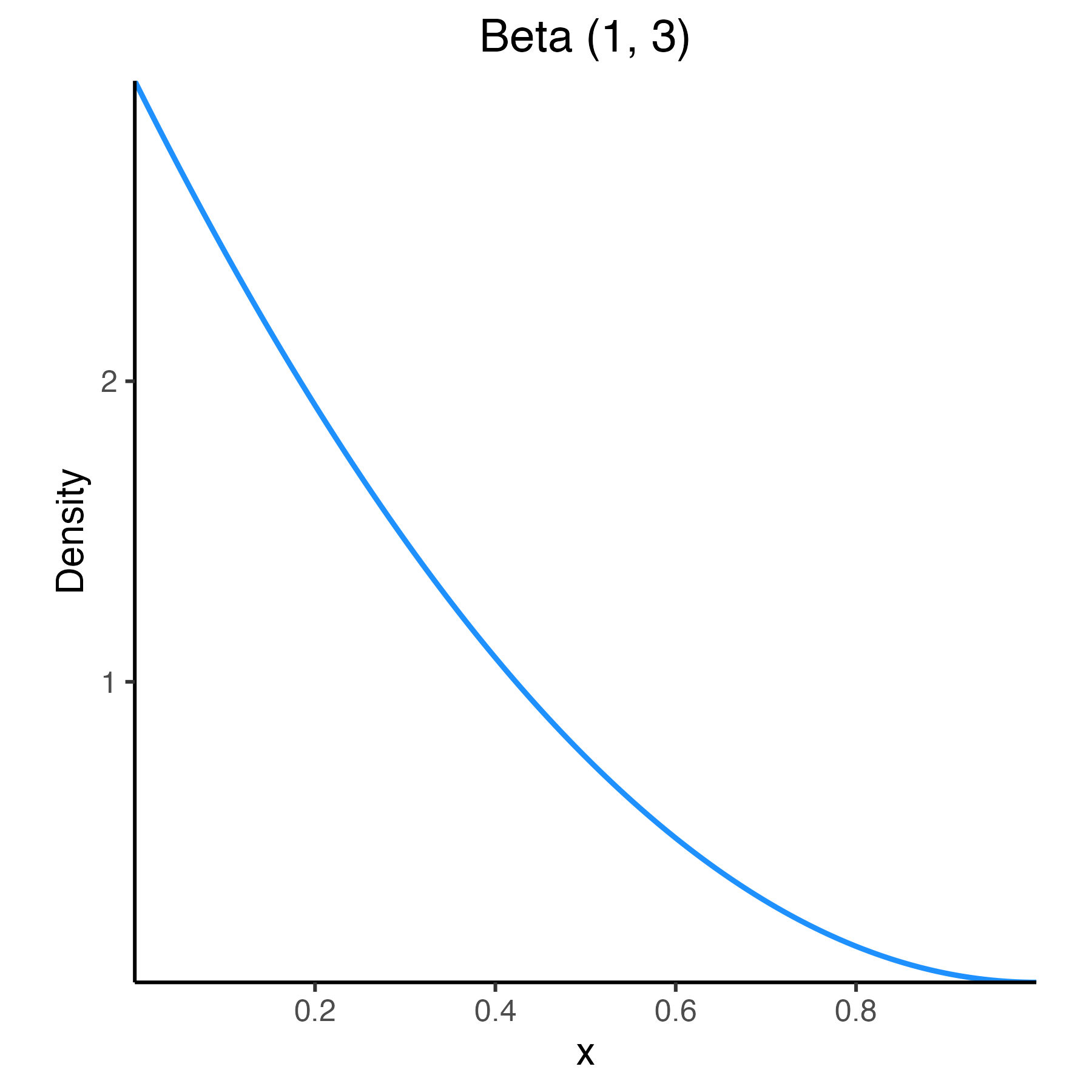}
    \end{subfigure}
    \\
    
    \begin{subfigure}[t]{0.5\textwidth}
        \centering
        \includegraphics[height=2in]{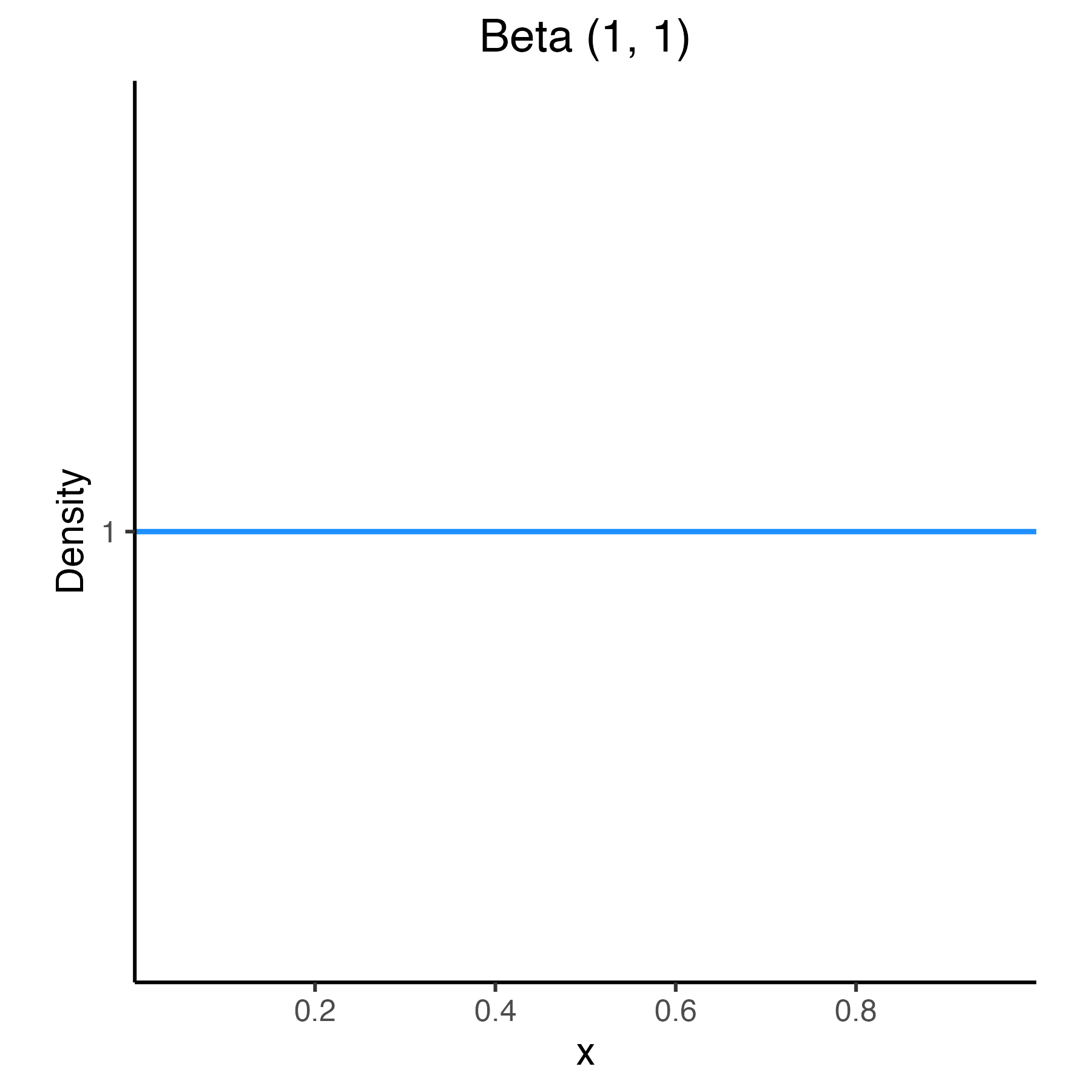}
    \end{subfigure}%
    ~
    \begin{subfigure}[t]{0.5\textwidth}
        \centering
        \includegraphics[height=2in]{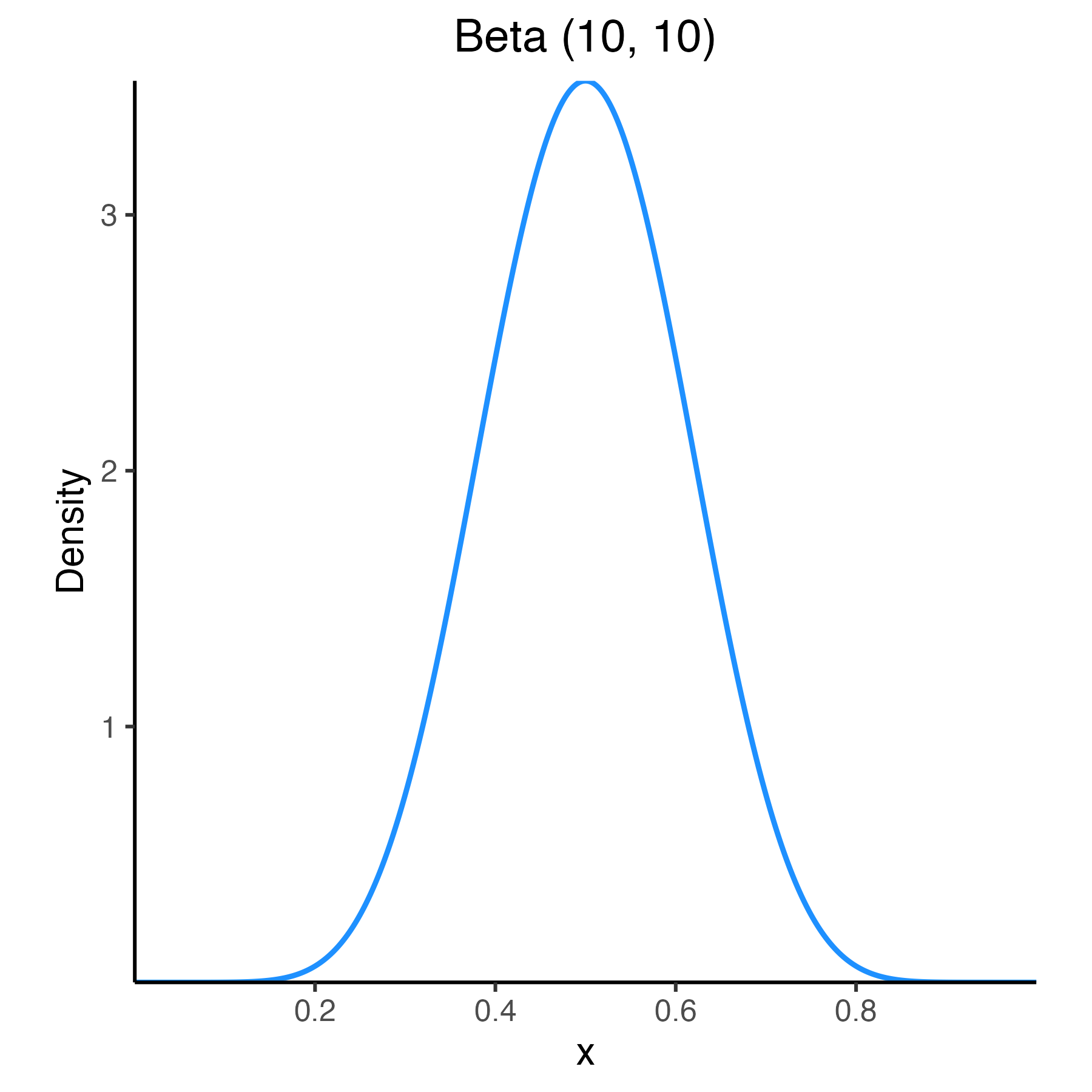}
    \end{subfigure}
    
    \begin{subfigure}[t]{0.5\textwidth}
        \centering
        \includegraphics[height=2in]{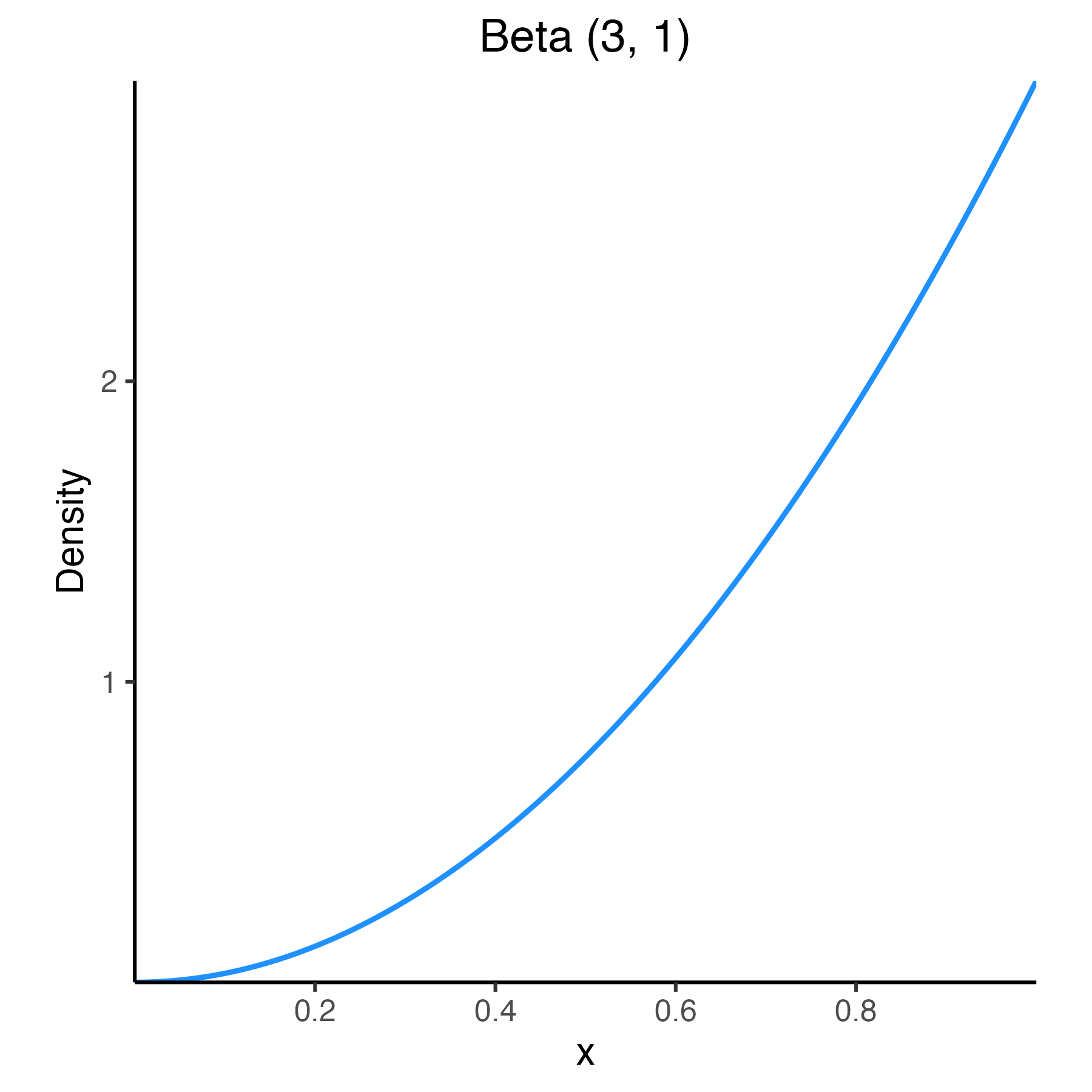}
    \end{subfigure}%
    ~
    \begin{subfigure}[t]{0.5\textwidth}
        \centering
        \includegraphics[height=2in]{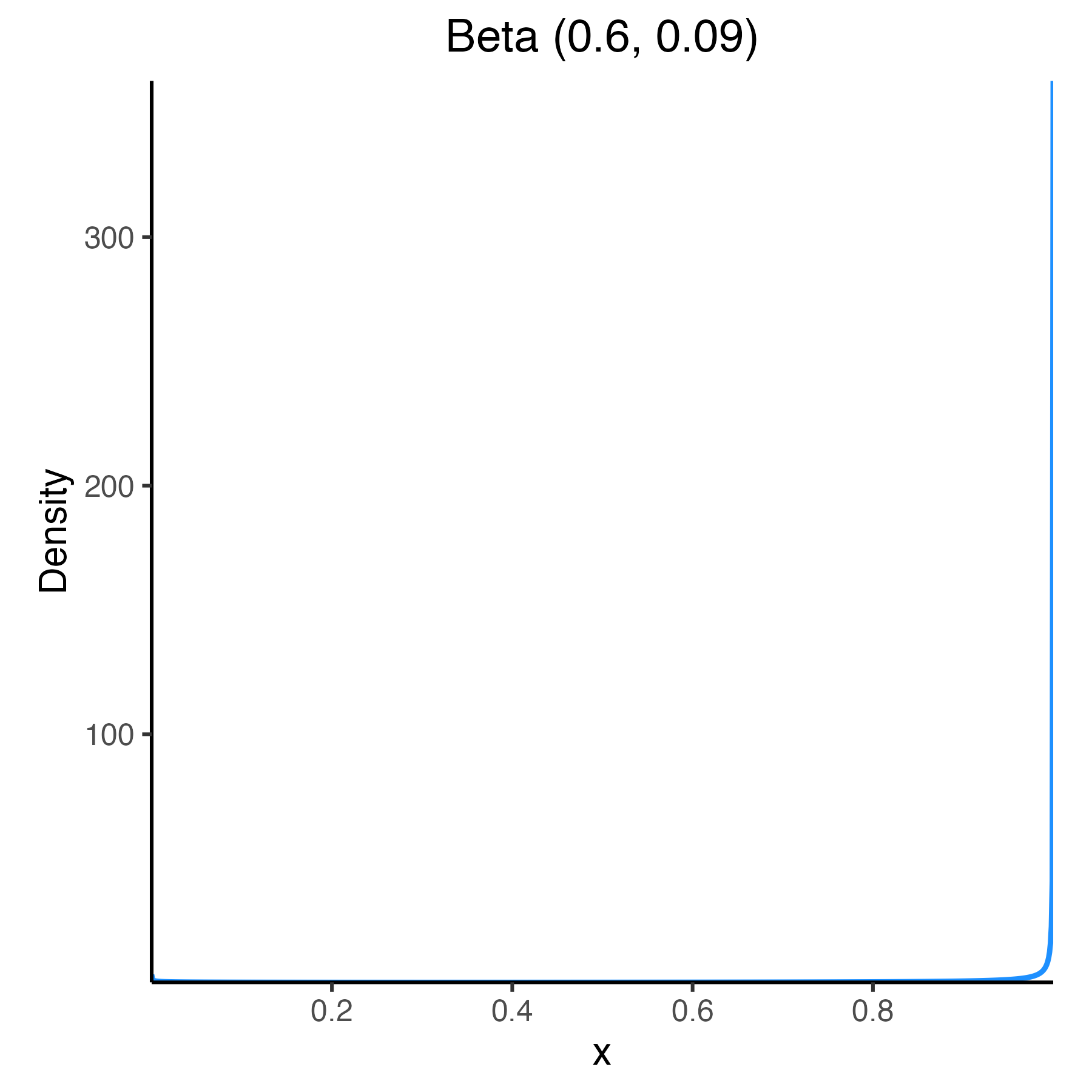}
    \end{subfigure}
    
    \caption{Densities of the various Beta distributions used in the simulations.}
    \label{fig:beta-densities}
\end{figure*}

In Figure \ref{fig:beta-densities}, we visualize the probability densities of different Beta distributions we use in our simulations.

\subsection{Simulations with branching}
\label{section:branching-sims}

Recall that in Section \ref{section:methods}, we worked under the assumption that new arms do not appear for the tractability of the problem. Here, we confront that assumption as well and show that the conclusions we draw are robust. In particular, we perform a second set of simulations where we introduce branching i.e., when we get a reward $X_i = 1$, this creates a new arm. This kind of branching is common in applications such as contact tracing where uncovering an infection presents a new set of contacts to test.

\begin{figure*}[!tbh]

    \centering
    \begin{subfigure}[t]{0.5\textwidth}
        \centering
        \includegraphics[height=2in]{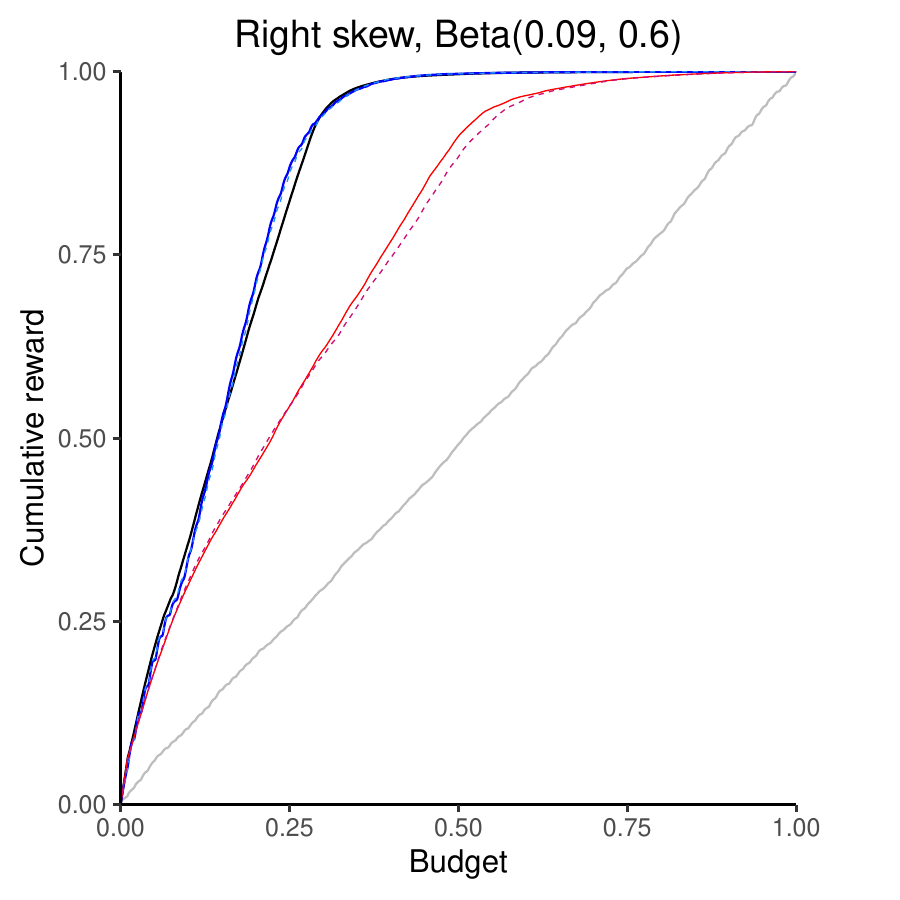}
    \end{subfigure}%
    ~
    \begin{subfigure}[t]{0.5\textwidth}
        \centering
        \includegraphics[height=2in]{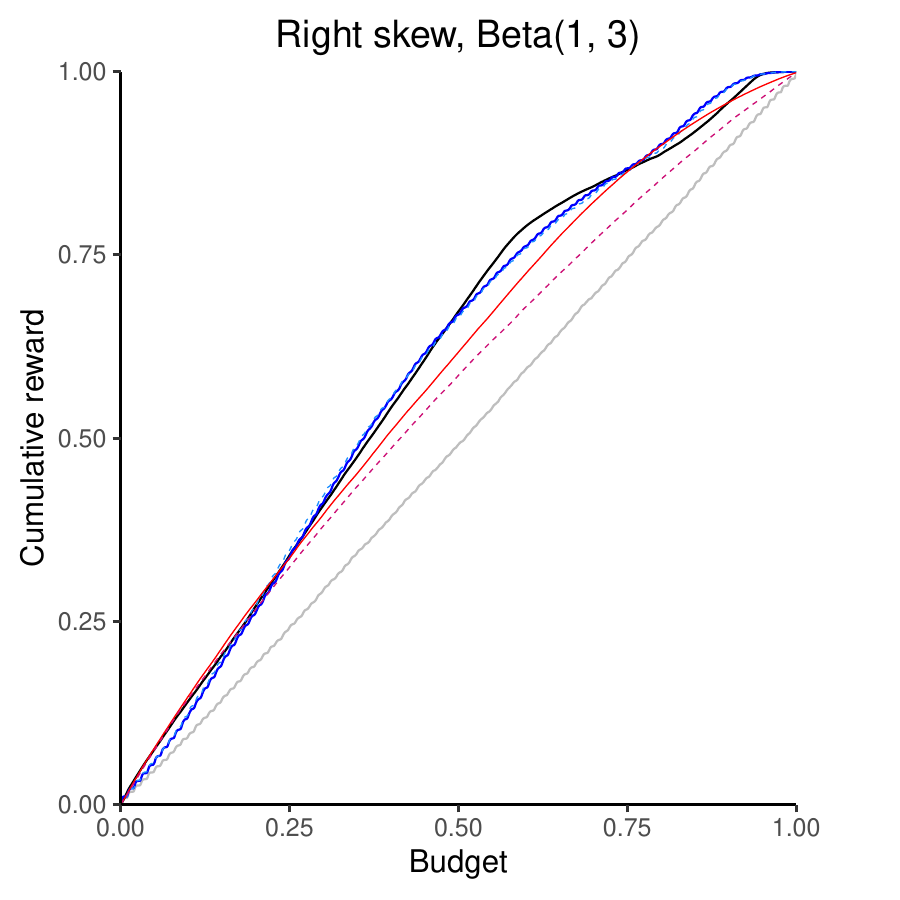}
    \end{subfigure}
    \\
    
    \begin{subfigure}[t]{0.5\textwidth}
        \centering
        \includegraphics[height=2in]{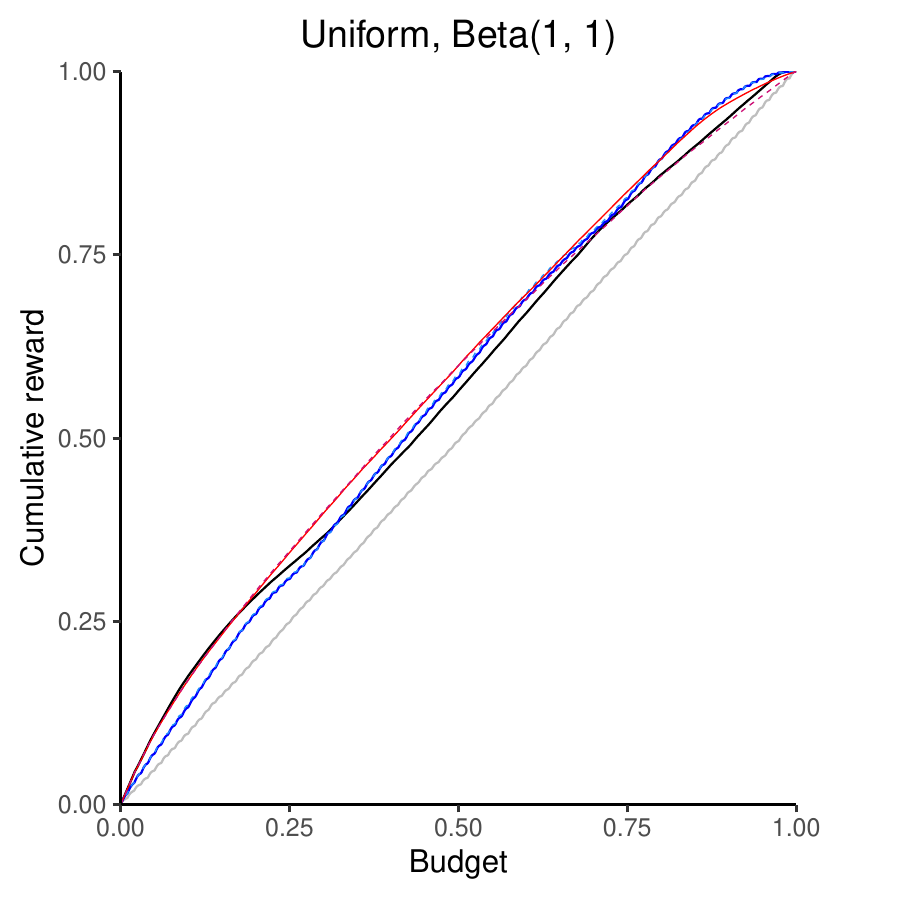}
    \end{subfigure}%
    ~
    \begin{subfigure}[t]{0.5\textwidth}
        \centering
        \includegraphics[height=2in]{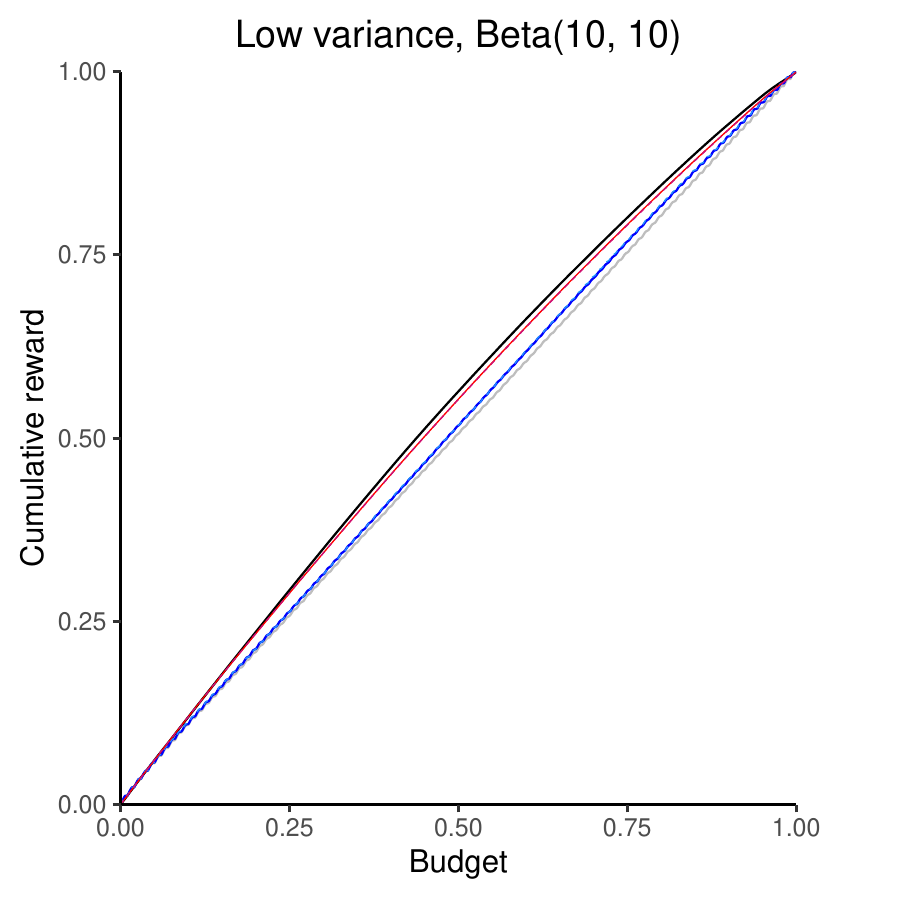}
    \end{subfigure}
    
    \begin{subfigure}[t]{0.5\textwidth}
        \centering
        \includegraphics[height=2in]{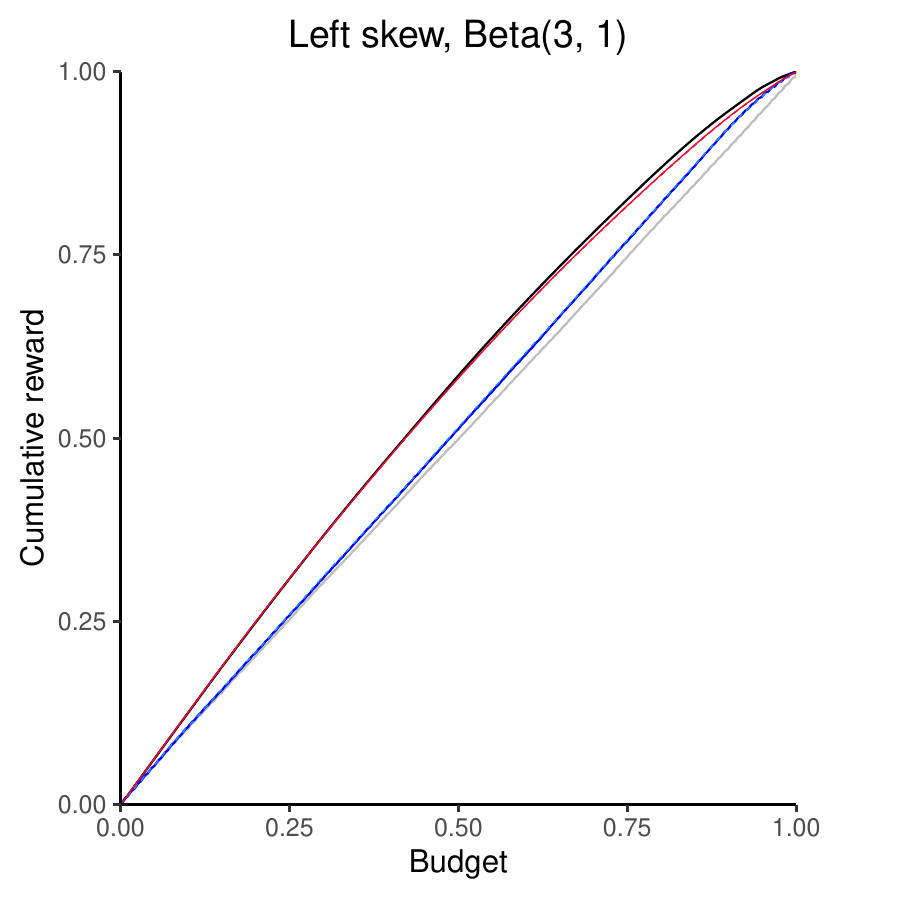}
    \end{subfigure}%
    ~
    \begin{subfigure}[t]{0.5\textwidth}
        \centering
        \includegraphics[height=2in]{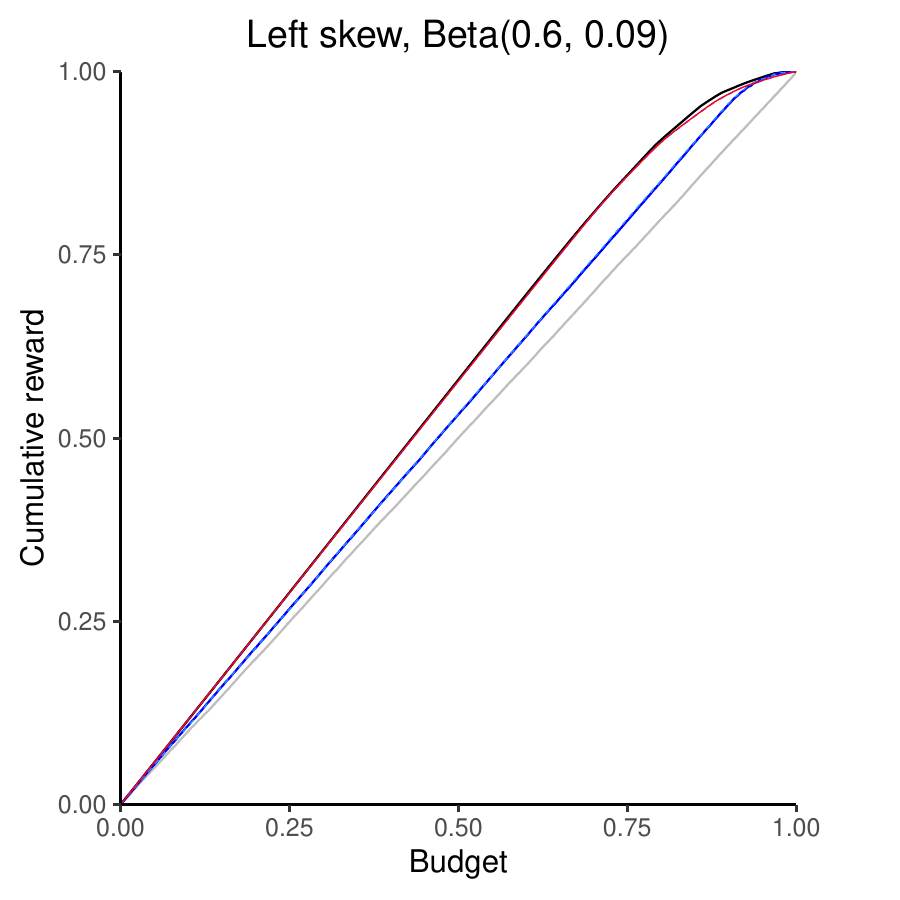}
    \end{subfigure}
    
    \caption{Cumulative reward over the total time horizon for different policies and various reward distributions based on data simulated with branching and Poisson(500) lifetime.  The axes are normalized to facilitate visual comparison. Thompson sampling is in black; pilot sampling with uniform sampling and lifetime sampling are in dark blue and light blue respectively; adaptive greedy with uniform sampling and sampling by lifetime are in red and pink respectively, and; naive sampling is in grey. Pilot sampling performs better when the rewards are right-skewed while adaptive greedy performs better when rewards are left-skewed or rewards have low variance. In other cases, no one policy dominates others. Again, Thompson sampling appears to perform consistently well in all scenarios. In this setup, sampling by degree seems identical to sampling uniformly for pilot sampling while sampling uniformly is better for adaptive greedy. These results are identical to simulations without branching in Figure \ref{fig:comparisons-large-degree}.}
    \label{fig:branching-comparisons-large-degree}
\end{figure*}

As in the simulations without branching, we will consider two regimes for the lifetime of arms -- Poisson and heavy-tailed Pareto. In the first set of simulations, we set the initial number of arms $N = 10$, the average lifetime to $\lambda = 500$, and the budget to $T = 50,000$. The results are shown in Figure \ref{fig:branching-comparisons-large-degree}. The trends largely agree with the simulations without branching in \ref{fig:comparisons-large-degree}. A key distinction we note is that in the simulations without branching, there appeared to be a critical point when Pilot Sampling began outperforming Adaptive Greedy. And as the distribution became more and more right-skewed, that critical point kept moving further toward the end of the time horizon. In the simulations with branching, there does not seem to be such a critical point. This can be attributed to how Adaptive Greedy sampling works. At the beginning of the game, Adaptive Greedy sampling estimates the mean reward for \textit{all} arms that are playable for the entire time horizon (see line \ref{line:estimate-mean} of Algorithm \ref{alg:adaptive-greedy}). This gives it a distinct advantage over the other methods, which do not perform this preparation, allowing it to shine in the non-branching case. This advantage is lost when new arms appear for which Adaptive Greedy does not have a mean reward estimate.

We perform similar simulations with $\lambda = 10$ and $T = 1000$ in Appendix \ref{section:sims-poisson-10}.

\begin{figure*}[!tbh]

    \centering
    \begin{subfigure}[t]{0.5\textwidth}
        \centering
        \includegraphics[height=2in]{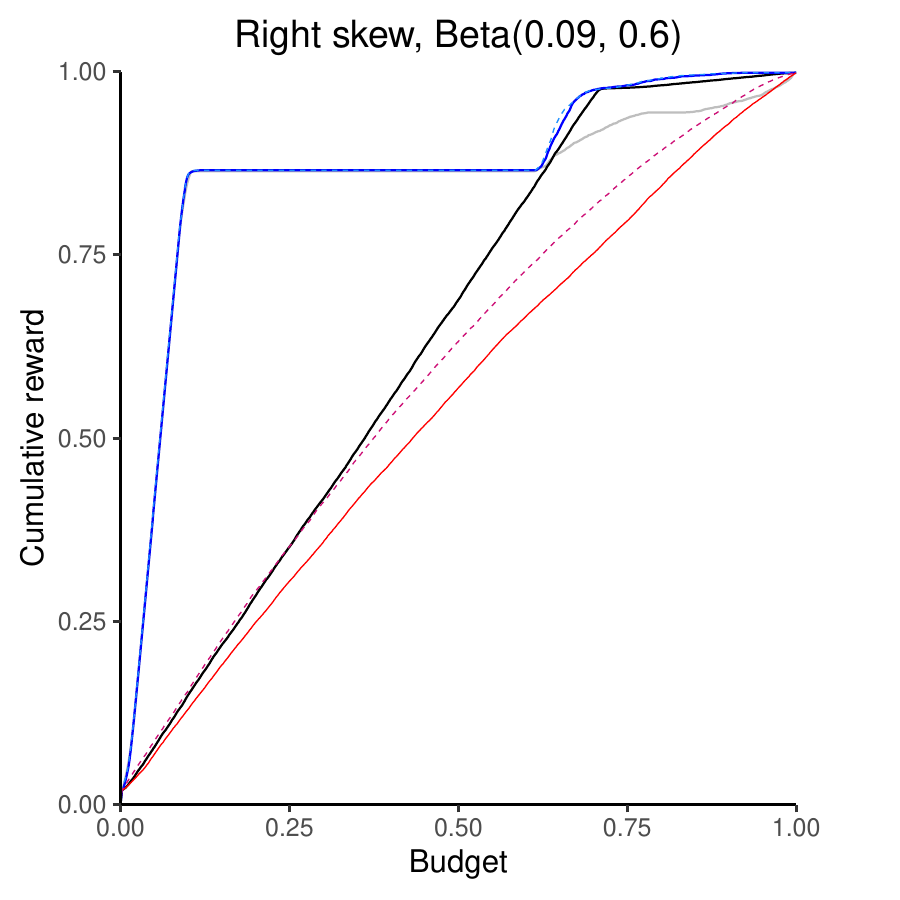}
    \end{subfigure}%
    ~
    \begin{subfigure}[t]{0.5\textwidth}
        \centering
        \includegraphics[height=2in]{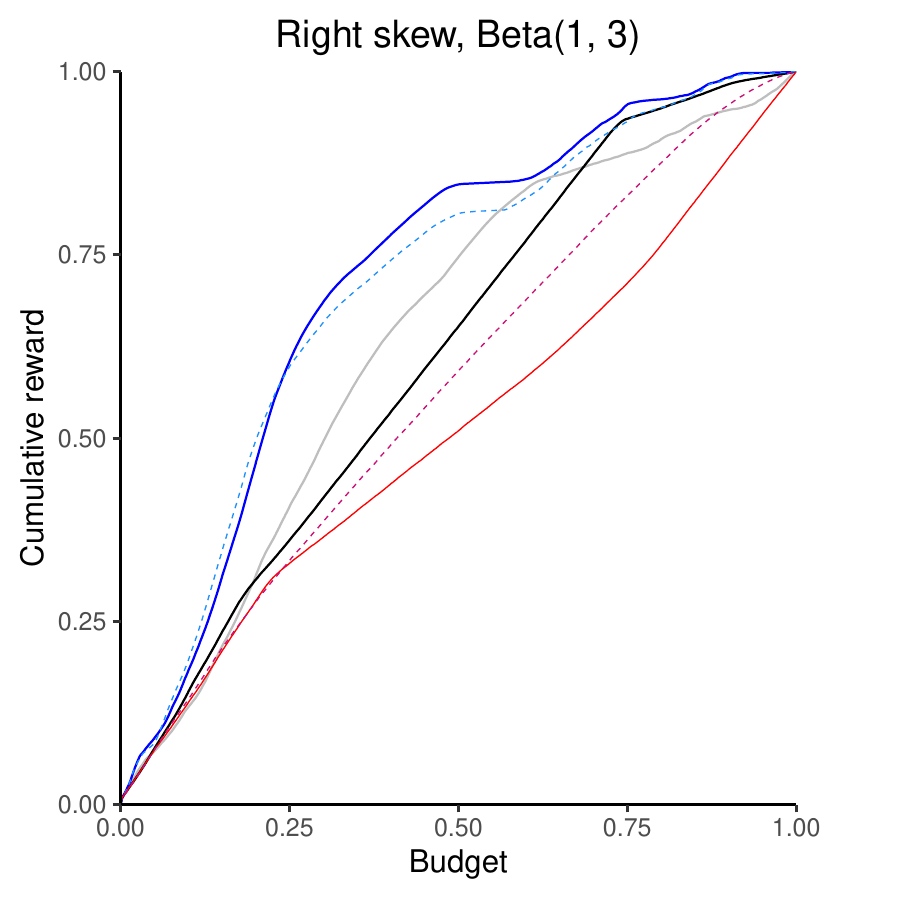}
    \end{subfigure}
    \\
    \begin{subfigure}[t]{0.5\textwidth}
        \centering
        \includegraphics[height=2in]{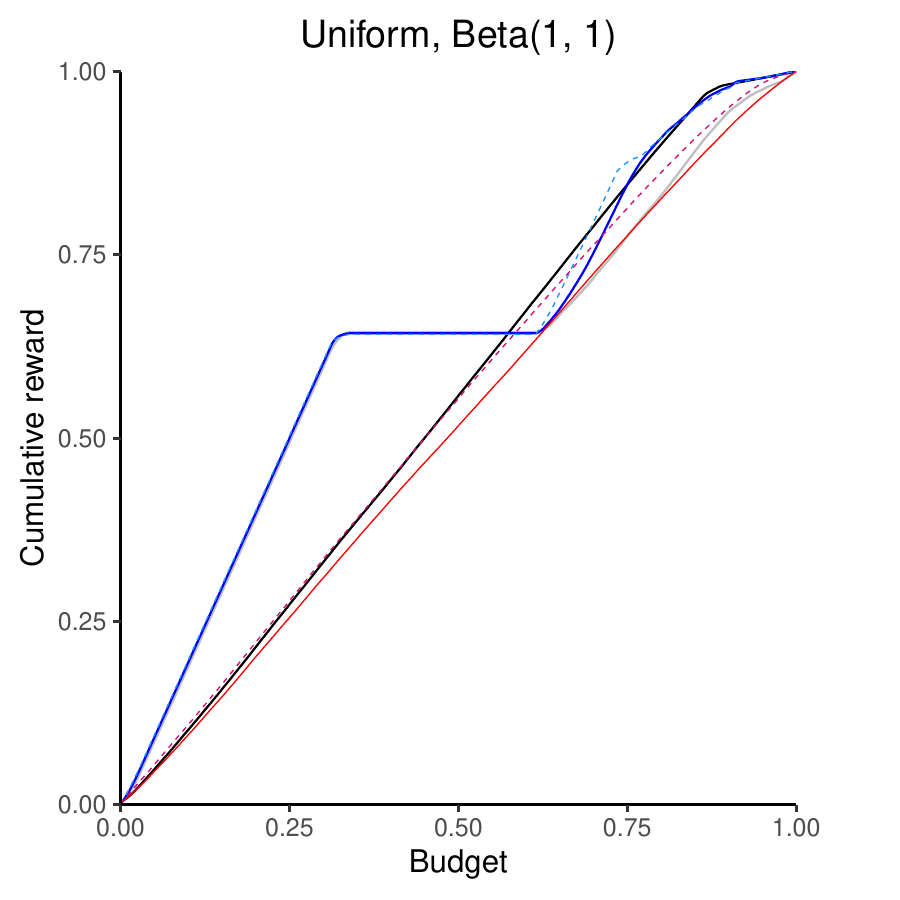}
    \end{subfigure}%
    ~
    \begin{subfigure}[t]{0.5\textwidth}
        \centering
        \includegraphics[height=2in]{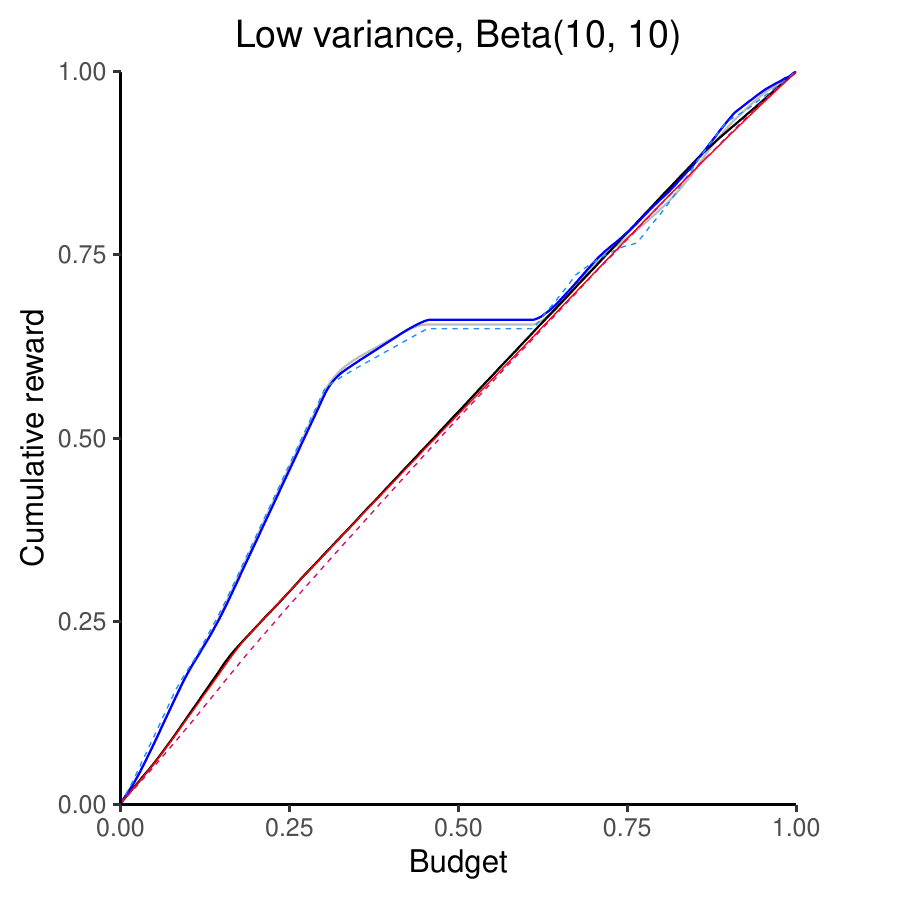}
    \end{subfigure}
    \\
    \begin{subfigure}[t]{0.5\textwidth}
        \centering
        \includegraphics[height=2in]{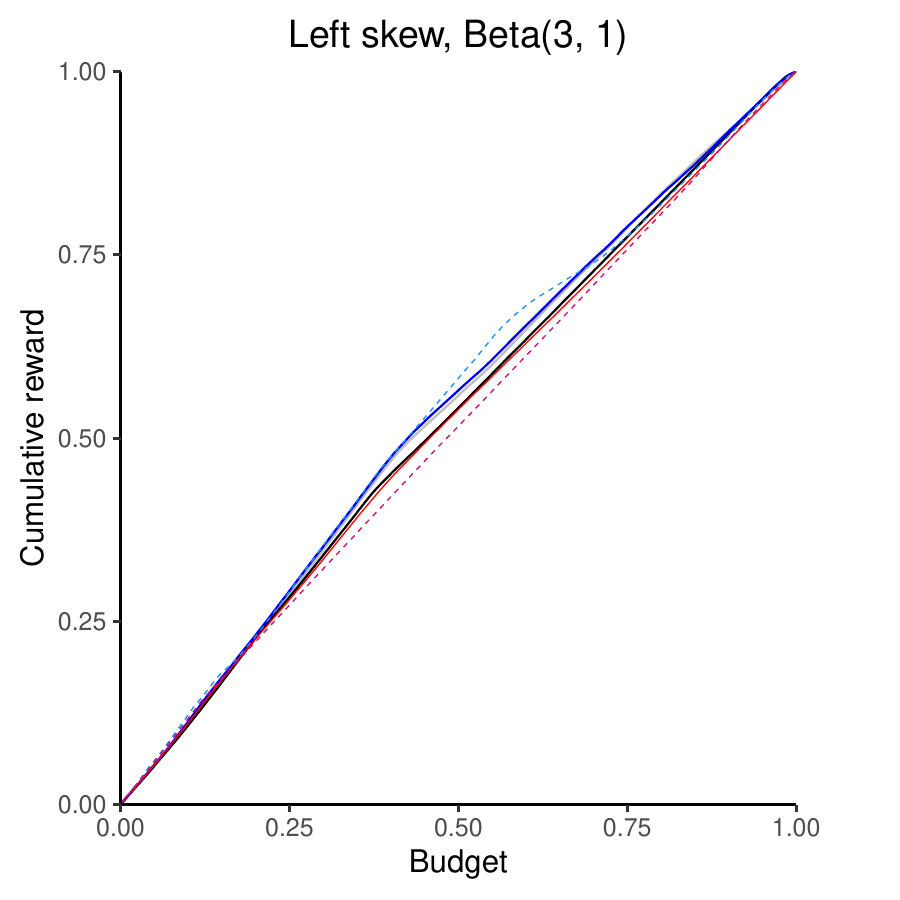}
    \end{subfigure}%
    ~
    \begin{subfigure}[t]{0.5\textwidth}
        \centering
        \includegraphics[height=2in]{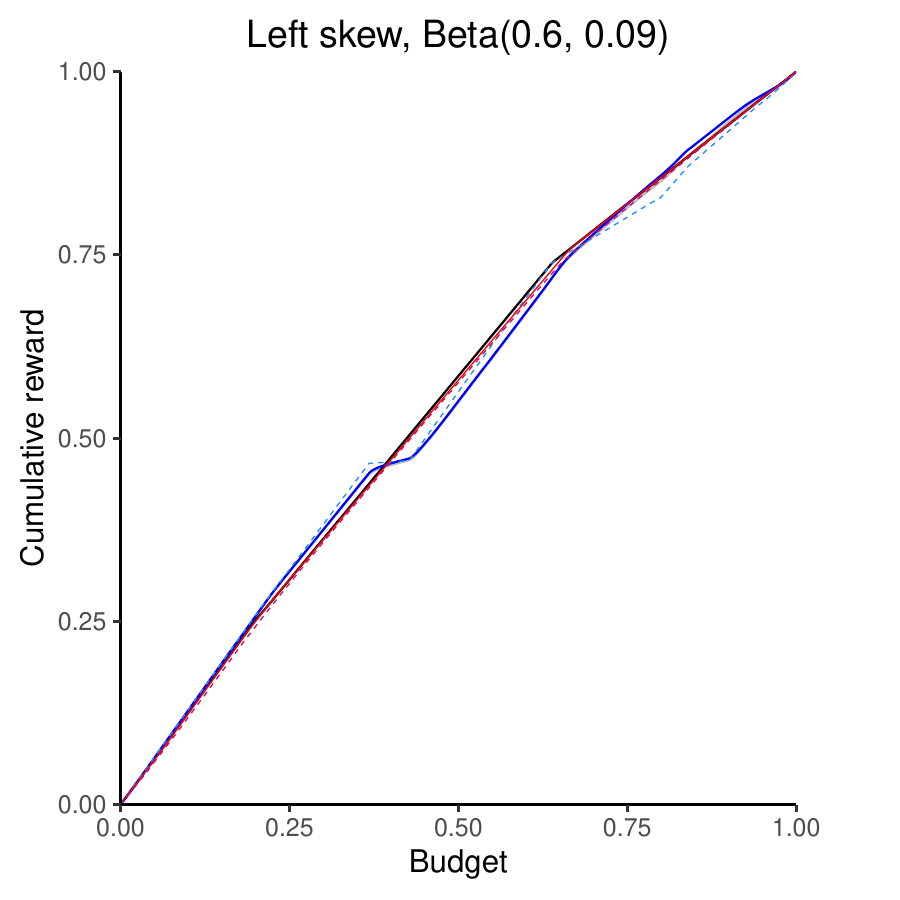}
    \end{subfigure}
    
    \caption{Cumulative reward over the total time horizon for different policies and various reward distributions based on data simulated with branching and Pareto(1, 0.6) lifetime.  The axes are normalized to facilitate visual comparison. Thompson sampling is in black; pilot sampling with uniform sampling and lifetime sampling are in dark blue and light blue respectively; adaptive greedy with uniform sampling and sampling by lifetime are in red and pink respectively, and; naive sampling is in grey. Pilot sampling performs in all scenarios except for heavily left-skewed rewards. In this setup, there is no clear winner between sampling by degree and sampling arms uniformly for pilot sampling while sampling by lifetime is better for adaptive greedy. These results are identical to simulations without branching in Figure \ref{fig:heavy-tail}.}
    \label{fig:branching-heavy-tail}
\end{figure*}

In the final set of simulations, the lifetime was drawn from a Pareto distribution with shape parameter 0.6 and location parameter 1. The results are shown in Figure \ref{fig:branching-heavy-tail} and also agree with the simulations without branching.

\subsection{Choosing a pilot group size with a misspecified prior}
\label{section:pilot-sampling-misspecified-prior}

%=================
% Figure 2
%=================
\begin{figure*}[tbh]
    \centering
    \includegraphics[height=4in]{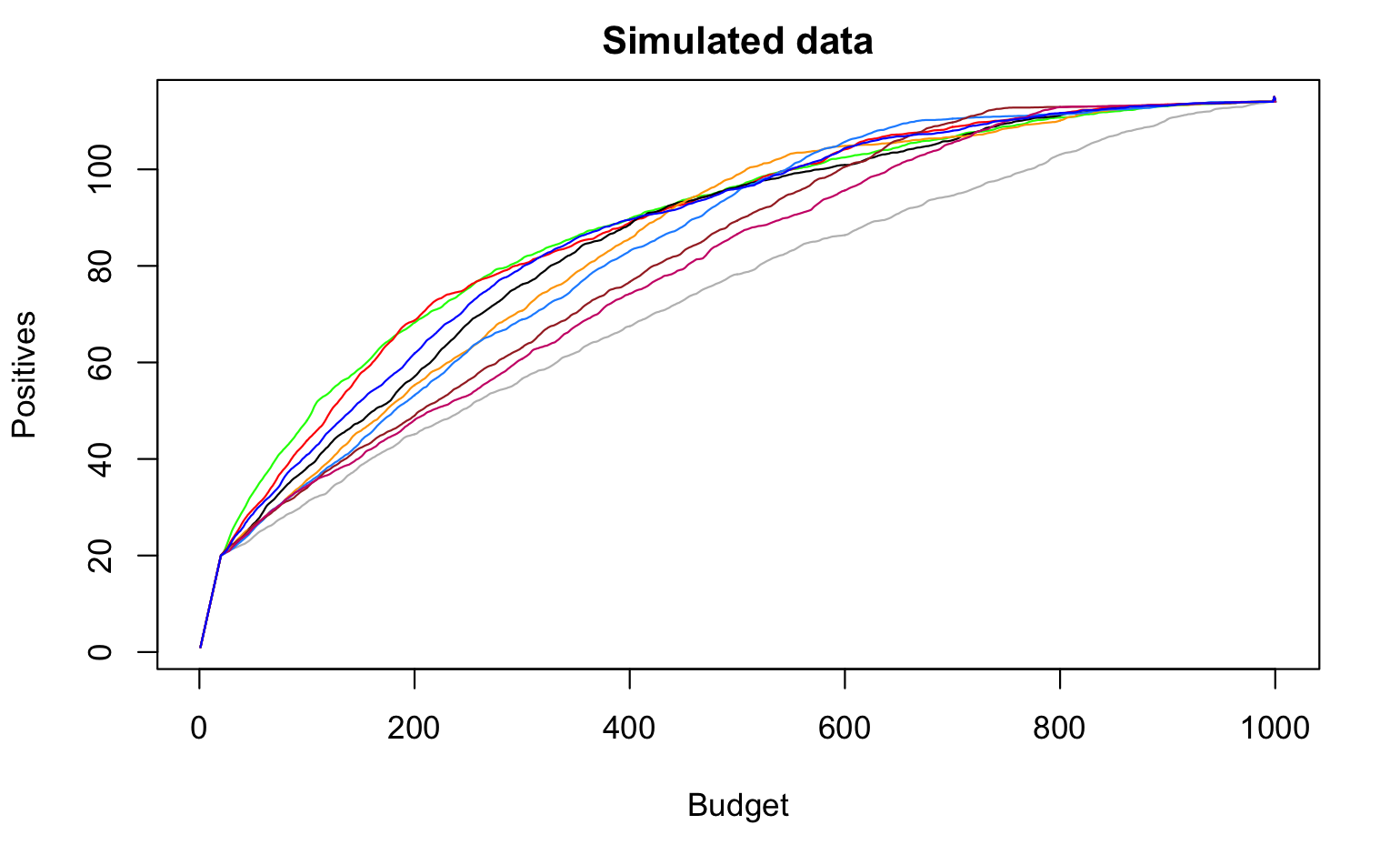}   
    \caption{Performance of Pilot sampling with various group sizes, $K$. Here, $\mu \sim \text{Beta}(0.09, 0.61)$ and average degree was 10. $K = 1$ is green, $K = 2$ is red, $K = 3$ is blue, $K = 4$ is black, $K = 5$ is orange, $K = 6$ is light blue, $K = 7$ is brown, and $K = 8$ is pink. The grey line corresponds to Naive sampling. There is a large variability in our performance as we change the size of the pilot group. So when the priors are incorrectly specified, we might end up choosing a sub-optimal pilot group size.}
    \label{fig:pilot-optimal-group}
\end{figure*}

Now, we turn to a key assumption regarding Pilot sampling. In order to choose an appropriate pilot group size, $K$, we need to know the distribution of the mean rewards. Of course, one could choose an arbitrary $K$ and still successfully run Pilot sampling, but we are not guaranteed to achieve optimal performance. To demonstrate this, we will perform some simulations. Consider the case where Pilot sampling is the best in Figure \ref{fig:comparisons-large-degree}. The mean rewards are drawn from $\text{Beta}(0.09, 0.61)$. We repeat the data generation with $S = 10$, $\lambda = 10$, and $N = 1000$. We compare the performance of Pilot sampling with $K = \{1, \dots, 8\}$ in Figure \ref{fig:pilot-optimal-group}. Although the plot is cluttered, it is easy to see that the difference in performance is large. While they all seem to outperform Naive sampling (increasing $K \to \infty$ will make Pilot sampling identical to Naive sampling), the choice of optimal $K$ is unclear. This is a big drawback of Pilot sampling when we do not know the distribution of mean rewards. In such cases, Thompson sampling or Adaptive Greedy is favorable as they do not need any parameter tuning. (Thompson sampling needs initial conditions for a prior, but in our simulations, we found that the choice of initial conditions does not impact performance, even when the correct priors are provided.)

\subsection{Simulations with smaller average degree}
\label{section:sims-poisson-10}

\begin{figure*}[!tbh]

    \centering
    \begin{subfigure}[t]{0.5\textwidth}
        \centering
        \includegraphics[height=2in]{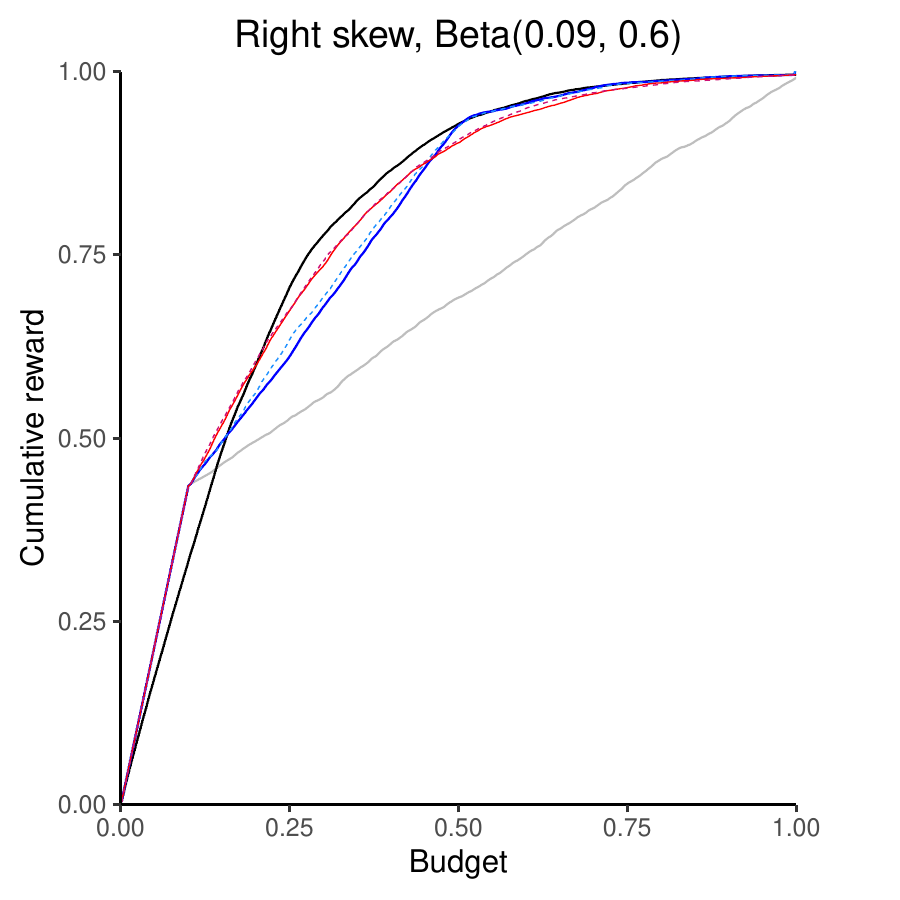}
    \end{subfigure}%
    ~
    \begin{subfigure}[t]{0.5\textwidth}
        \centering
        \includegraphics[height=2in]{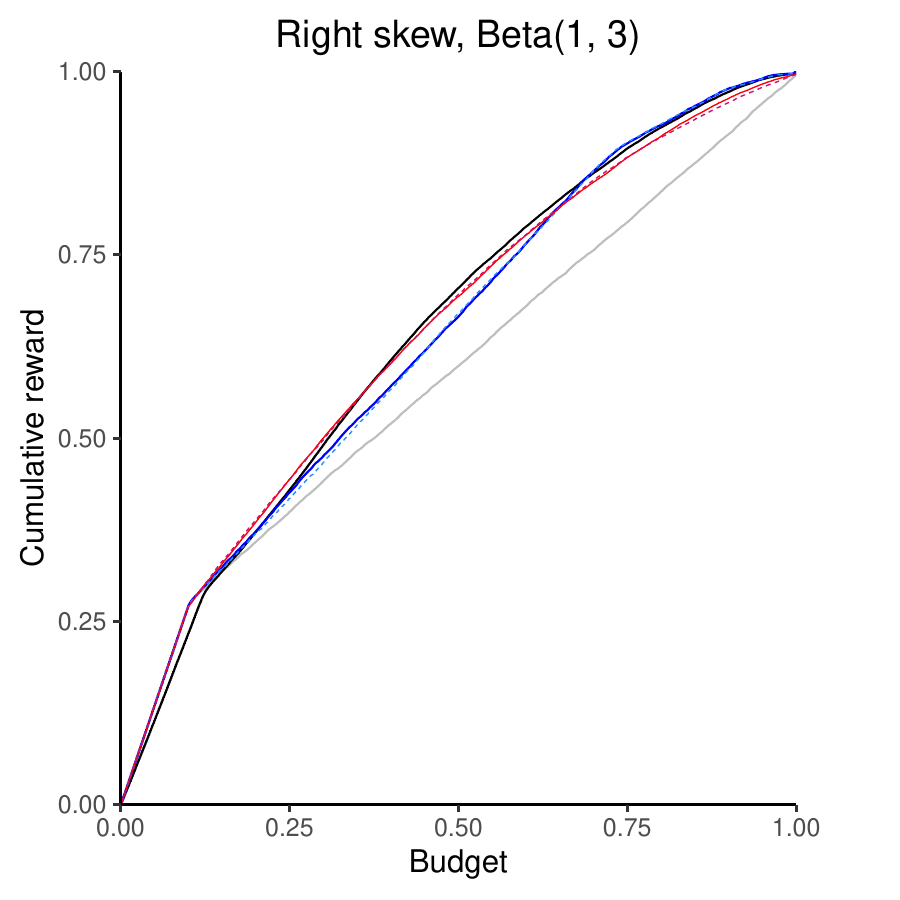}
    \end{subfigure}
    \\
    
    \begin{subfigure}[t]{0.5\textwidth}
        \centering
        \includegraphics[height=2in]{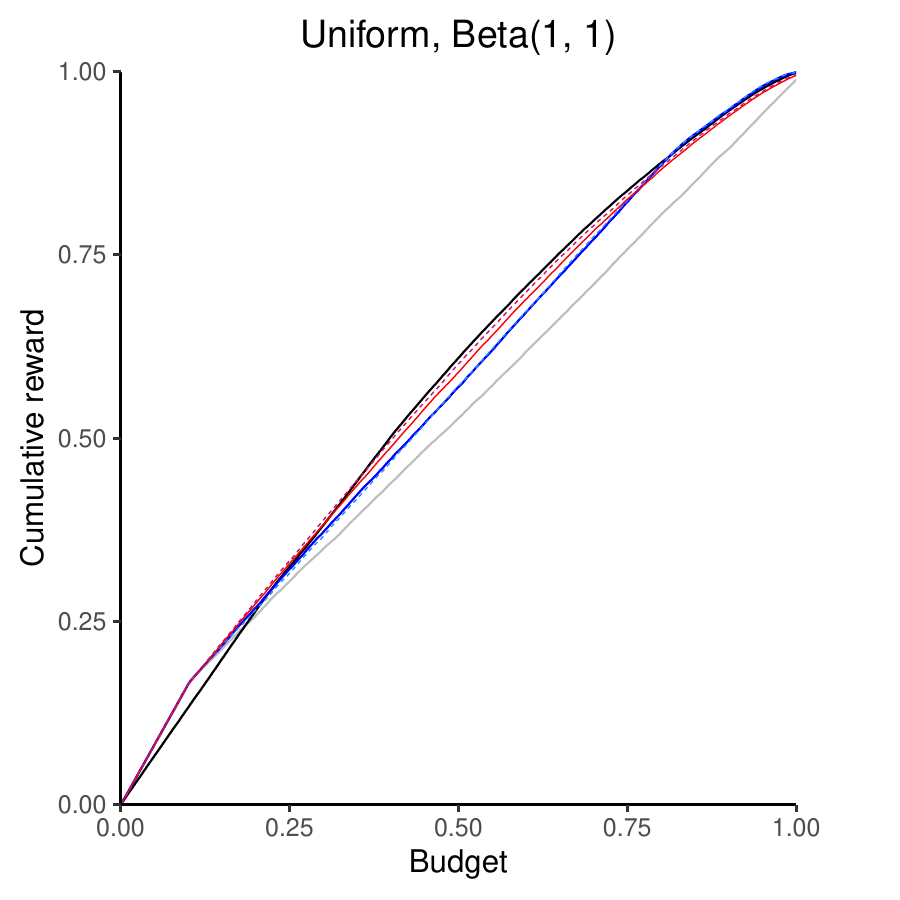}
    \end{subfigure}%
    ~
    \begin{subfigure}[t]{0.5\textwidth}
        \centering
        \includegraphics[height=2in]{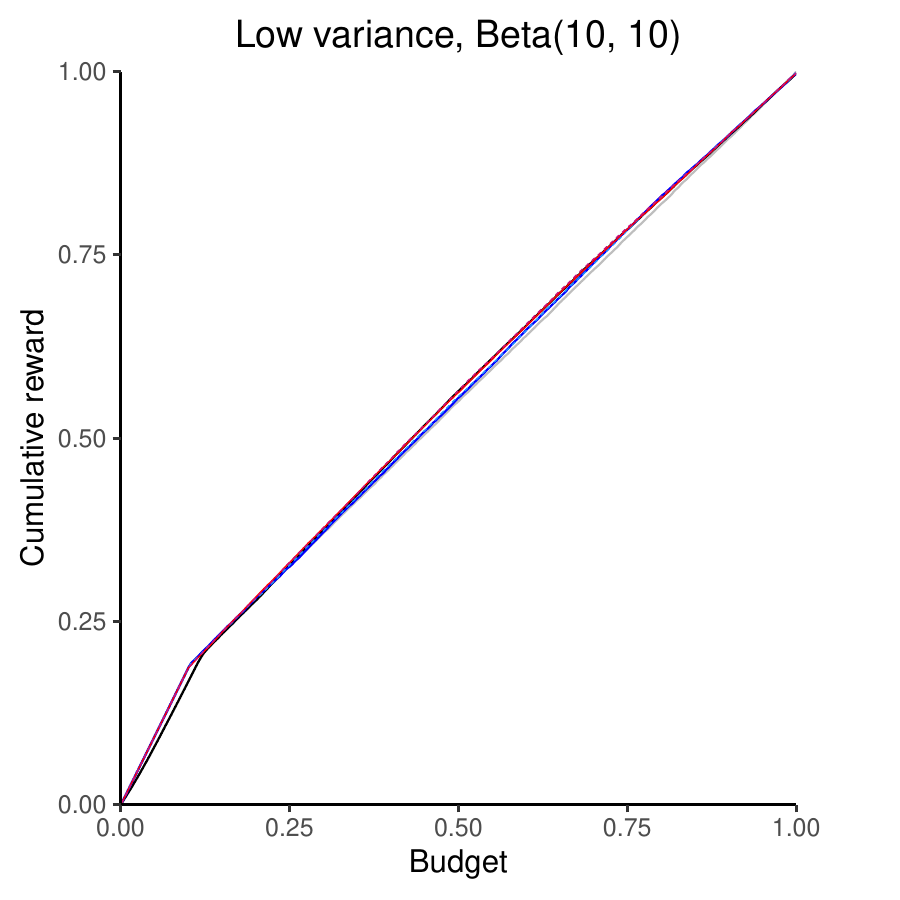}
    \end{subfigure}
    
    \begin{subfigure}[t]{0.5\textwidth}
        \centering
        \includegraphics[height=2in]{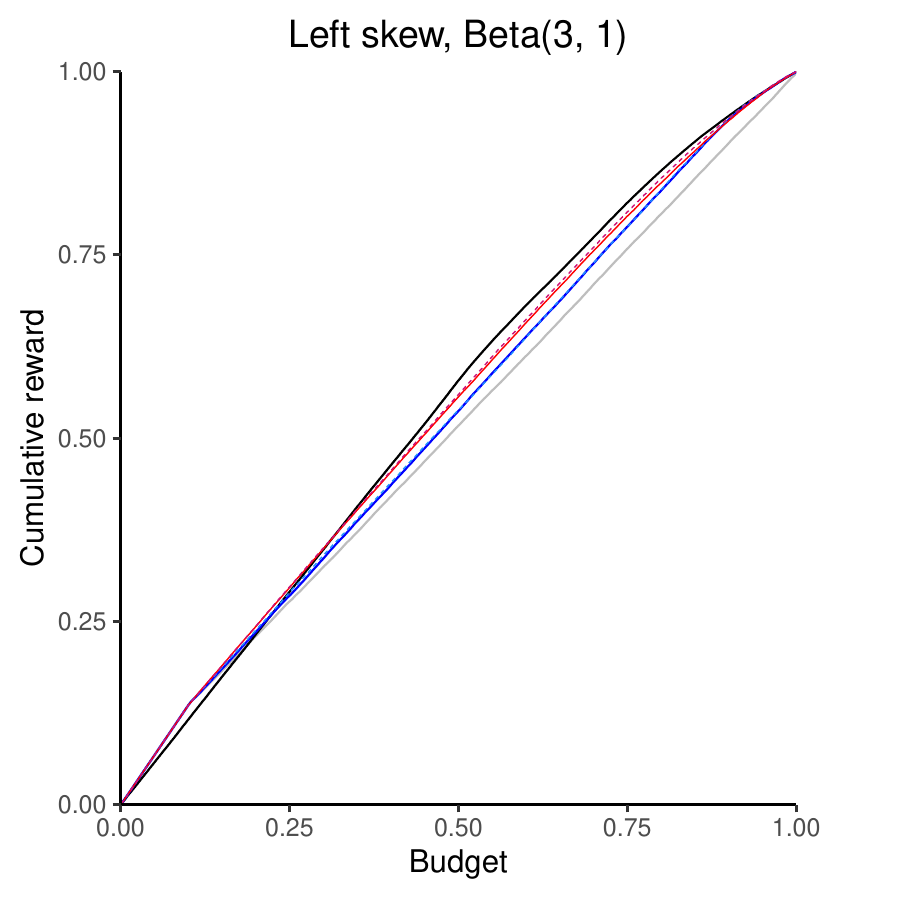}
    \end{subfigure}%
    ~
    \begin{subfigure}[t]{0.5\textwidth}
        \centering
        \includegraphics[height=2in]{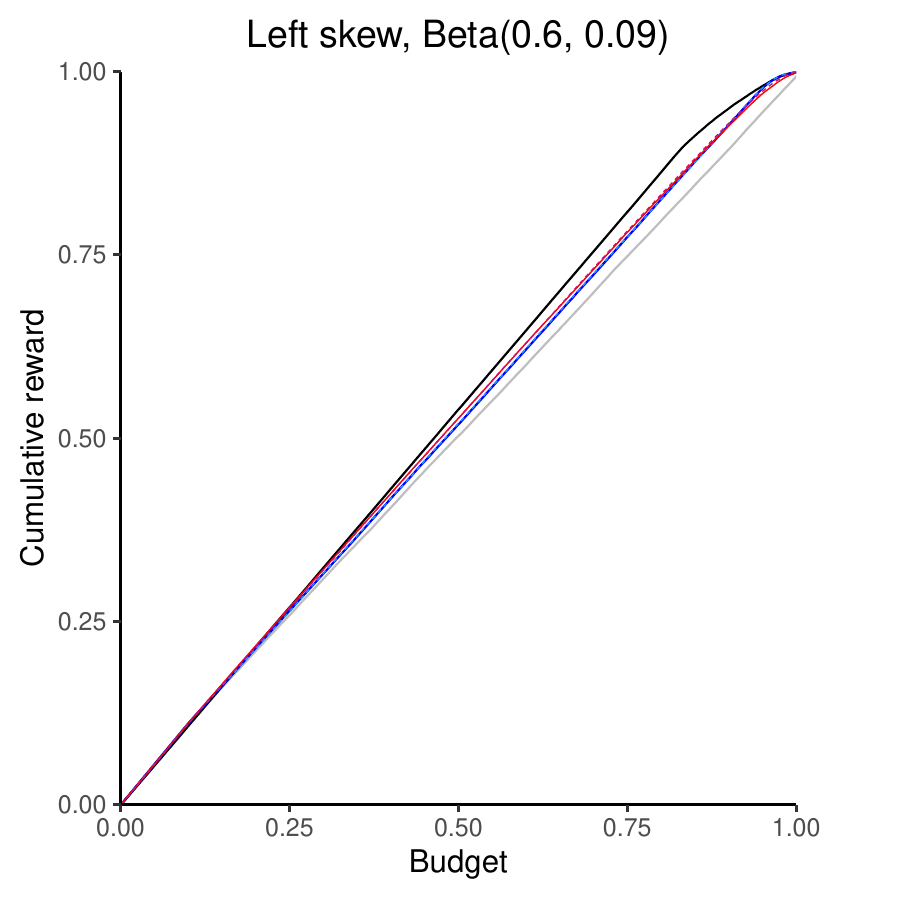}
    \end{subfigure}
    
    \caption{Cumulative reward over the total time horizon for different policies and various reward distributions based on data simulated without branching and Poisson(10) lifetime.  The axes are normalized to facilitate visual comparison. Thompson sampling is in black; pilot sampling with uniform sampling and lifetime sampling are in dark blue and light blue respectively; adaptive greedy with uniform sampling and sampling by lifetime are in red and pink respectively, and; naive sampling is in grey. These results are identical to simulations with a larger mean degree in Figure \ref{fig:comparisons-large-degree}.}
    \label{fig:comparisons-small-degree}
\end{figure*}

\begin{figure*}[!tbh]

    \centering
    \begin{subfigure}[t]{0.5\textwidth}
        \centering
        \includegraphics[height=2in]{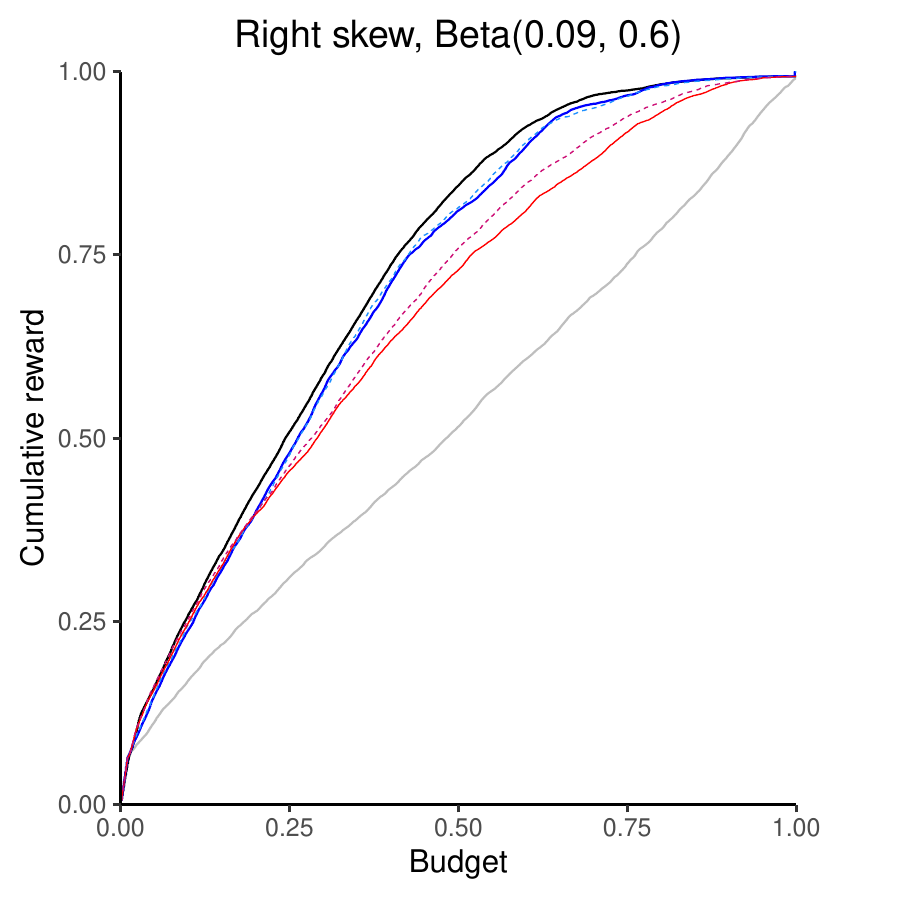}
    \end{subfigure}%
    ~
    \begin{subfigure}[t]{0.5\textwidth}
        \centering
        \includegraphics[height=2in]{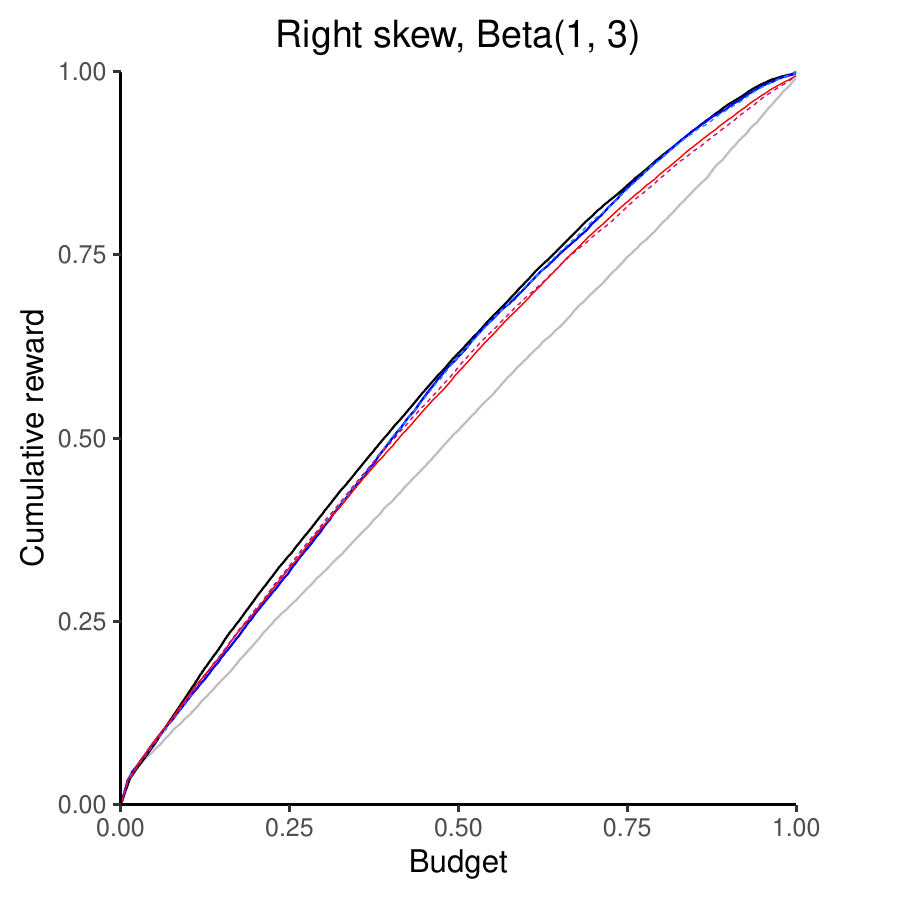}
    \end{subfigure}
    \\
    
    \begin{subfigure}[t]{0.5\textwidth}
        \centering
        \includegraphics[height=2in]{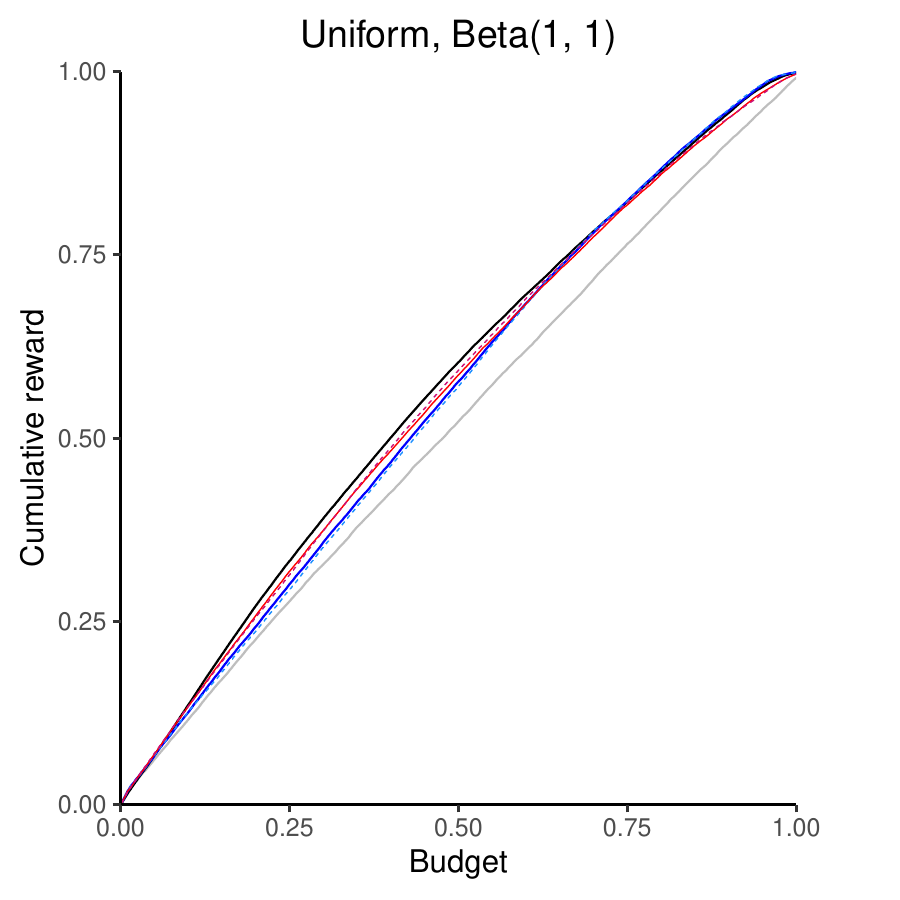}
    \end{subfigure}%
    ~
    \begin{subfigure}[t]{0.5\textwidth}
        \centering
        \includegraphics[height=2in]{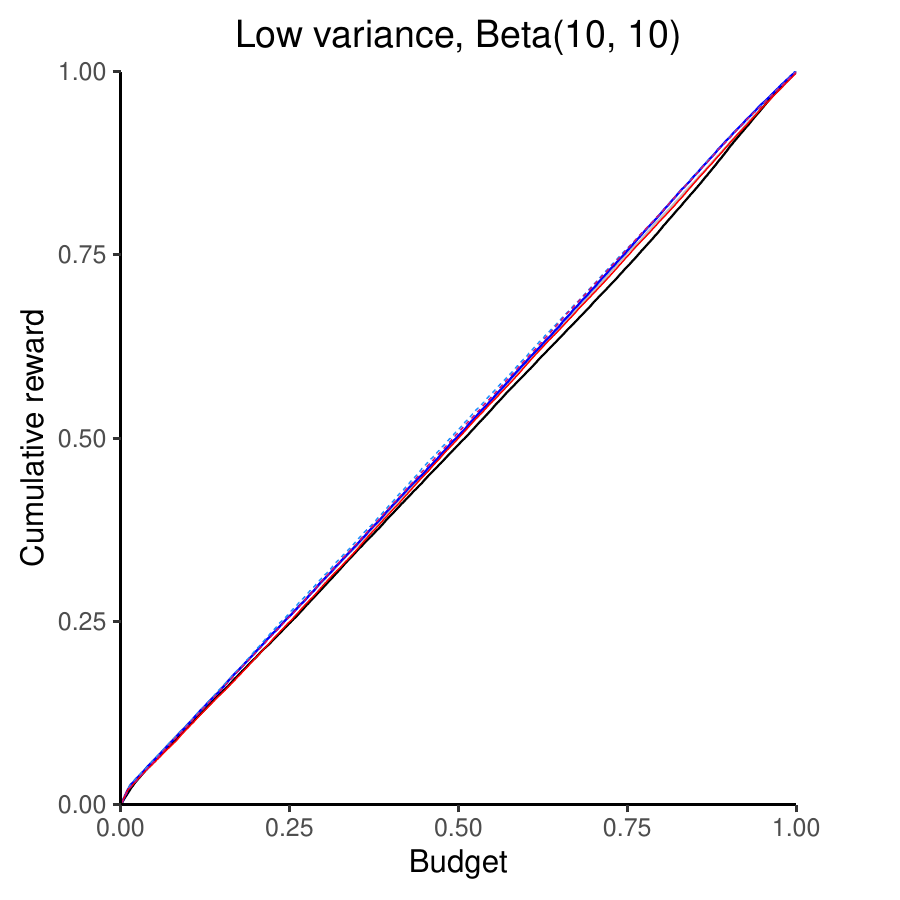}
    \end{subfigure}
    
    \begin{subfigure}[t]{0.5\textwidth}
        \centering
        \includegraphics[height=2in]{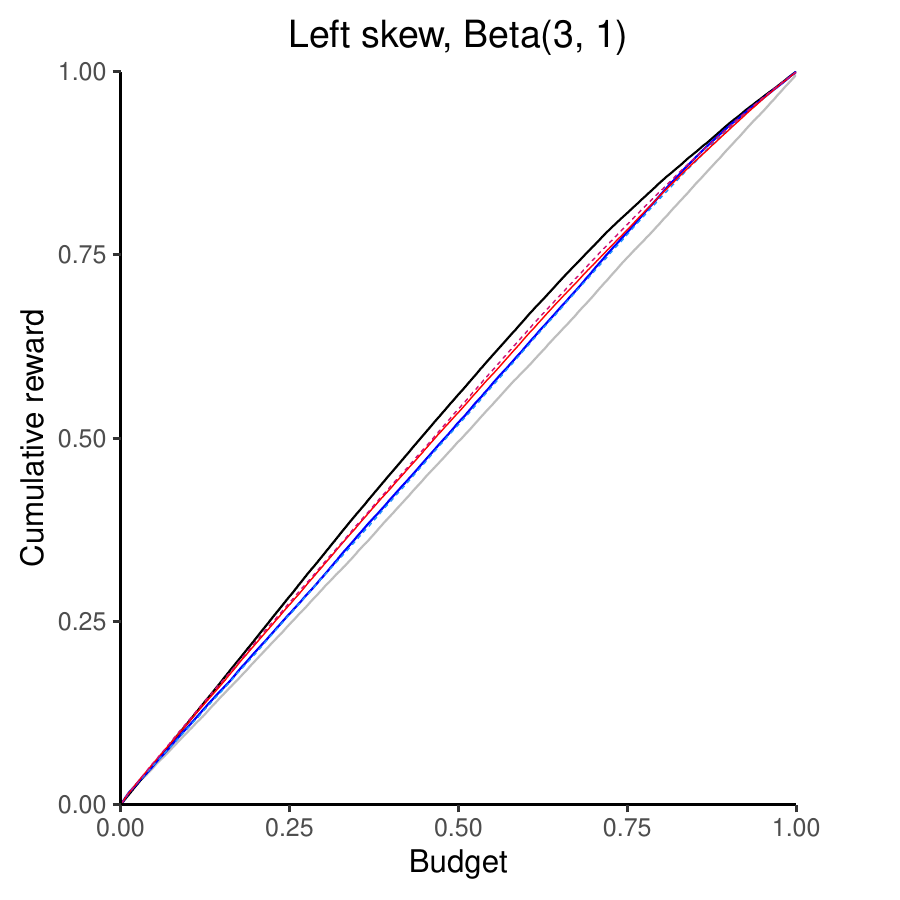}
    \end{subfigure}%
    ~
    \begin{subfigure}[t]{0.5\textwidth}
        \centering
        \includegraphics[height=2in]{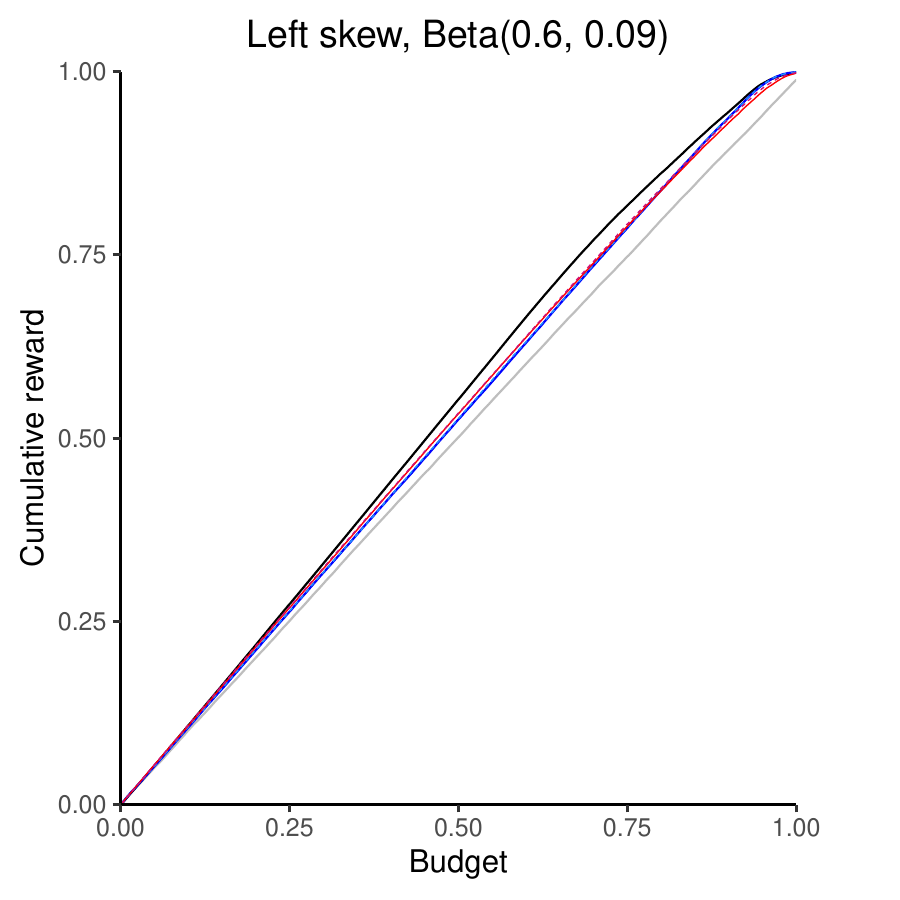}
    \end{subfigure}
    
    \caption{Cumulative reward over the total time horizon for different policies and various reward distributions based on data simulated with branching and Poisson(10) lifetime. The axes are normalized to facilitate visual comparison. Thompson sampling is in black; pilot sampling with uniform sampling and lifetime sampling are in dark blue and light blue respectively; adaptive greedy with uniform sampling and sampling by lifetime are in red and pink respectively, and; naive sampling is in grey. These results are identical to simulations with a larger mean degree in Figure \ref{fig:branching-comparisons-large-degree}.}
    \label{fig:branching-comparisons-small-degree}
\end{figure*}

In these simulations, we fix $N = 10$, $\lambda = 10$, and $T = 1000$. We varied the parameters of the Beta distribution. For Thompson sampling, we initialized the prior distribution at $\text{Beta}(1, 1)$, a uniform prior. The results without branching are shown in Figure \ref{fig:comparisons-small-degree}. And the results with branching are shown in Figure \ref{fig:branching-comparisons-small-degree}.

%% file: sections/appendix_real_data.tex
\section{Real data}
\label{section:real-data-appendix}

Properties of these datasets are summarized in Table \ref{tab:data}. Estimated PCI parameters are described in Table \ref{tab:estimated-params}. We estimated the PCI using the Bayes shrinkage estimator and used the method of moments to estimate the parameters of the Beta model. The optimal pilot group size was chosen as described in Appendix \ref{section:pilot-group-size}.

\begin{table}[!tbh]
    \centering
    \renewcommand{\arraystretch}{1.2}
    \begin{tabular}{c | c c c}
        \hline
        Region & People traced & Tests administered & Positive infections \\
        \hline
        Punjab, Pakistan & 165,072 & 1,911,669 & 36,868 \\
        Punjab, India & 2,077 & 18,284 & 1,620 \\
        Southern India & 88,616 & 649,990 & 27,196 \\
        \hline
    \end{tabular}
    \caption{Properties of data collected as a part of contact tracing efforts of COVID-19 in 2020}
    \label{tab:data}
\end{table}

\begin{table}[!p]
    \centering
    \renewcommand{\arraystretch}{1.2}
    \begin{tabular}{c | c c c}
        \hline
        Region & Beta parameters $(\alpha, \beta)$ & Optimal $x^* \ (\Gamma(x^*))$ & Pilot group size  \\
        \hline
         Punjab, Pakistan & (0.0877, 2.0681) & 0.1902 (0.1902) & 6 \\
         Punjab, India & (0.1382, 0.8830) & 0.5902 (0.5902) & 2  \\
         Southern India & (0.1228 0.9709) & 0.3415 (0.3415) & 3  \\
         \hline
    \end{tabular}
    \caption{Estimated parameters for each dataset. All PCI distributions are right-skewed.}
    \label{tab:estimated-params}
\end{table}

\begin{figure*}[!p]
    \centering
    \begin{subfigure}[t]{0.5\textwidth}
        \centering
        \includegraphics[width=3in]{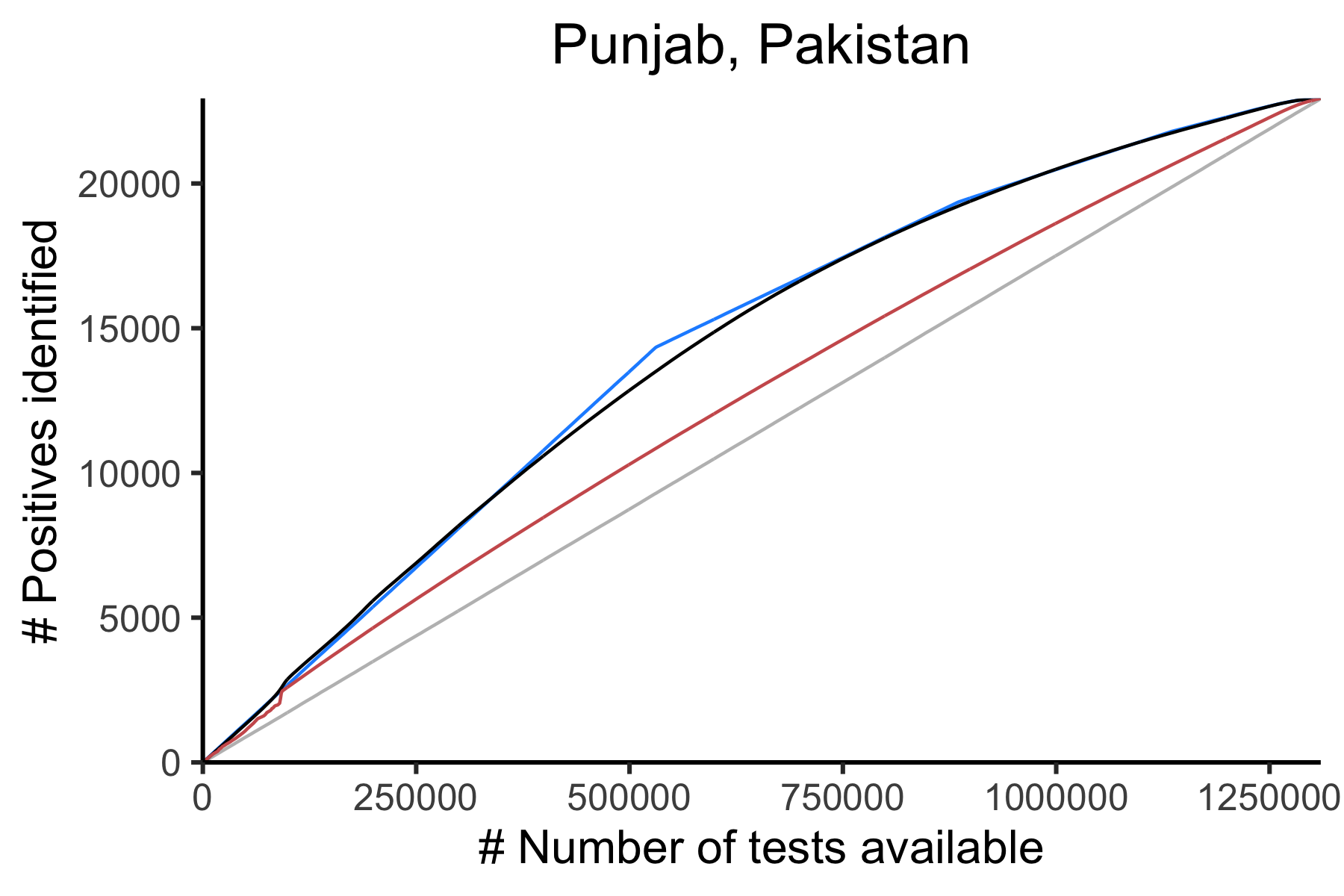}
    \end{subfigure}%
    ~ 
    \begin{subfigure}[t]{0.5\textwidth}
        \centering
        \includegraphics[width=3in]{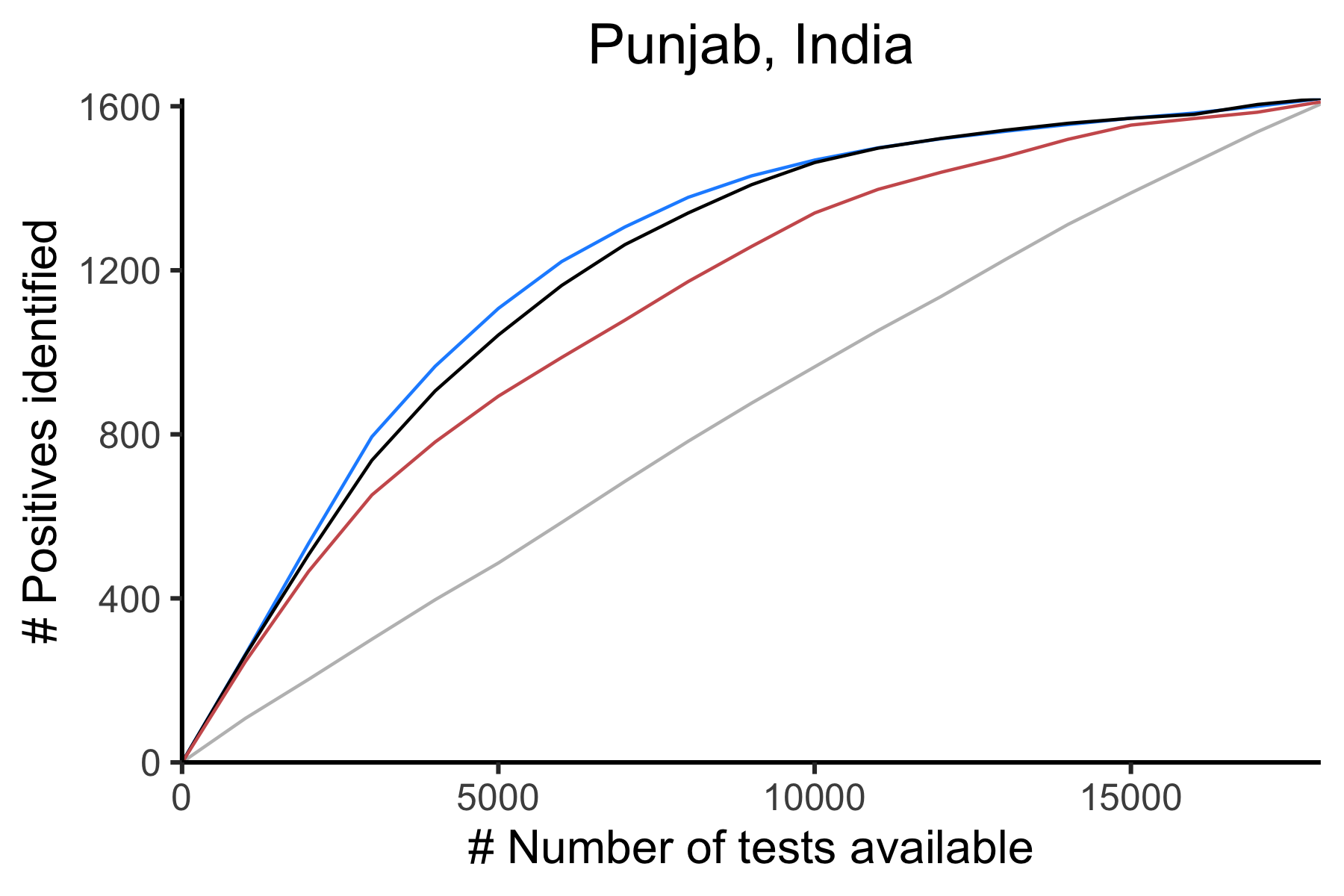}
    \end{subfigure}%
    \
    \begin{subfigure}[t]{0.4\textwidth}
        \centering
        \includegraphics[width=4in]{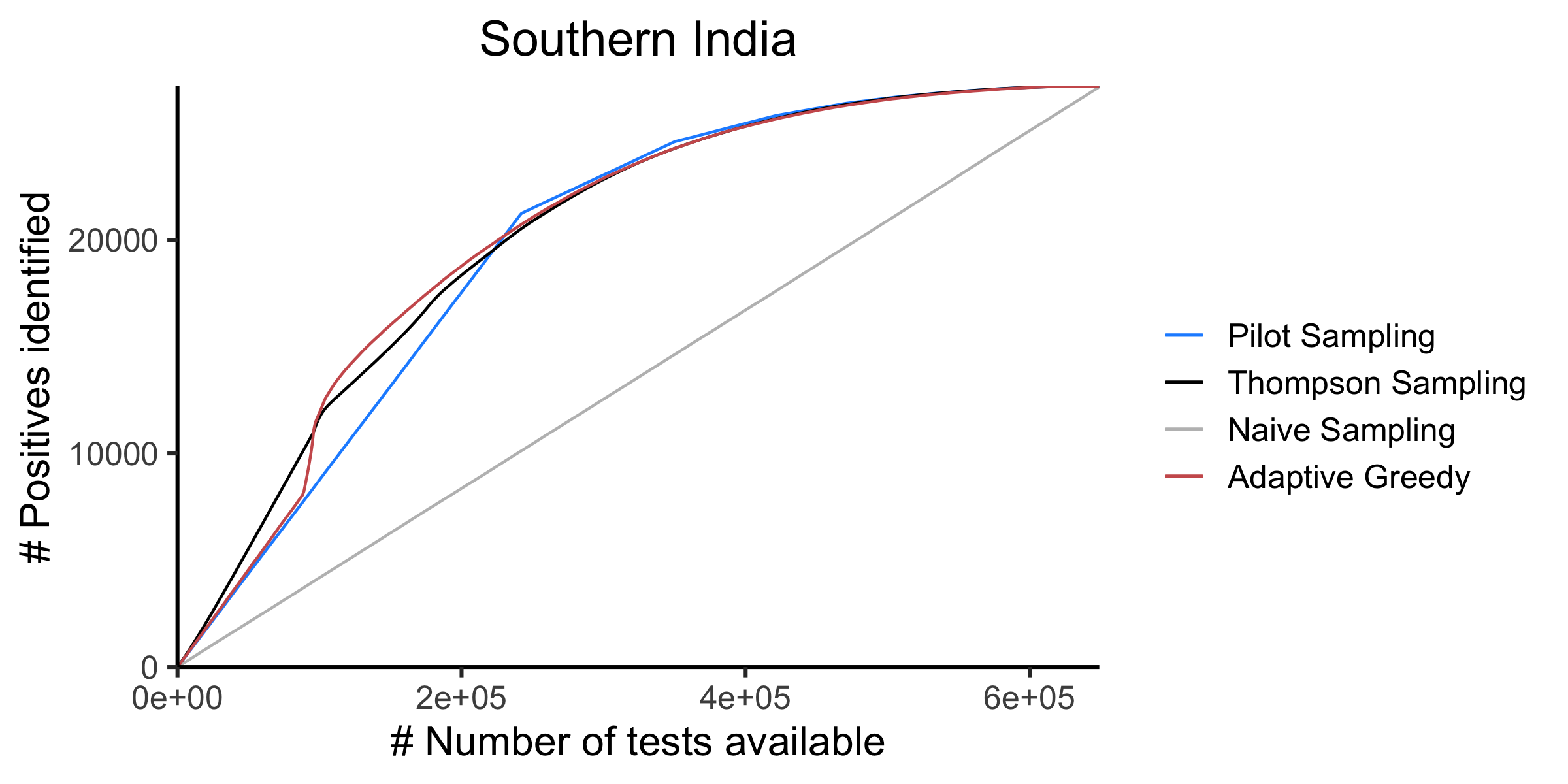}
    \end{subfigure}
    \caption{The top left figure shows results on Punjab, Pakistan dataset. The top right figure shows results on the Punjab, India dataset. The bottom figure shows results on the southern India dataset. Pilot sampling and Thompson sampling are clearly doing better when we have branching data (both Punjab datasets). However, Thompson sampling may be logistically difficult making pilot sampling favorable when implementing contact tracing. This is identical to Figure \ref{fig:real-data} except for the axes' scales.}
    \label{fig:real-data-absolute}
\end{figure*}

Figure \ref{fig:real-data-absolute} is identical to Figure \ref{fig:real-data} but the axes reflect absolute numbers i.e., they are not normalized.

\begin{table}[tb]
    \centering
    \renewcommand{\arraystretch}{1.2}
    \begin{tabular}{c | c c }
        \hline
        Region & Estimated Poisson mean & Estimated Pareto location and shape \\
        \hline
         Punjab, Pakistan &  10.58 & (1, 0.51) \\
         Punjab, India &  26.62 & (1, 0.39) \\
         Southern India &  6.36 & (1, 0.71) \\
         \hline
    \end{tabular}
    \caption{Estimated lifetime parameters for each dataset.}
    \label{tab:estimated-lifetime-params}
\end{table}

\begin{figure*}[!tbh]
    \centering
    \begin{subfigure}[t]{0.5\textwidth}
        \centering
        \includegraphics[width=3.5in]{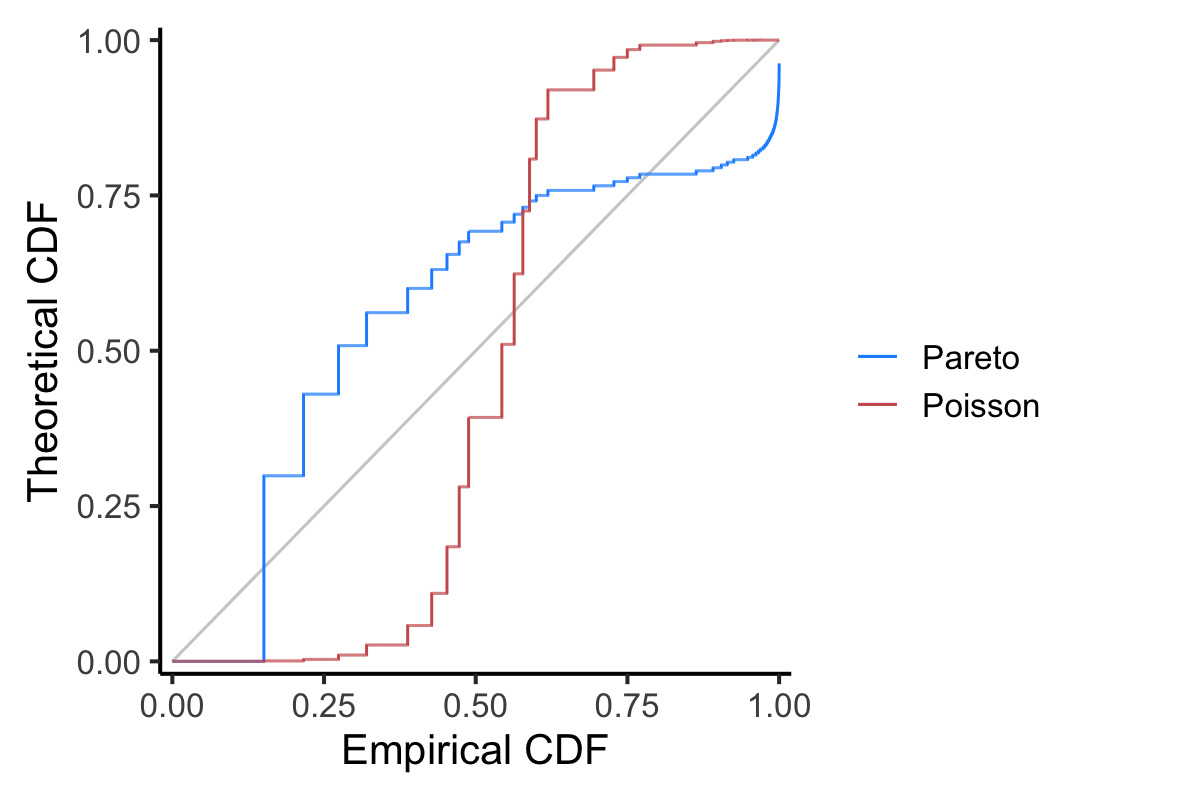}
    \end{subfigure}%
    ~ 
    \begin{subfigure}[t]{0.5\textwidth}
        \centering
        \includegraphics[width=3.5in]{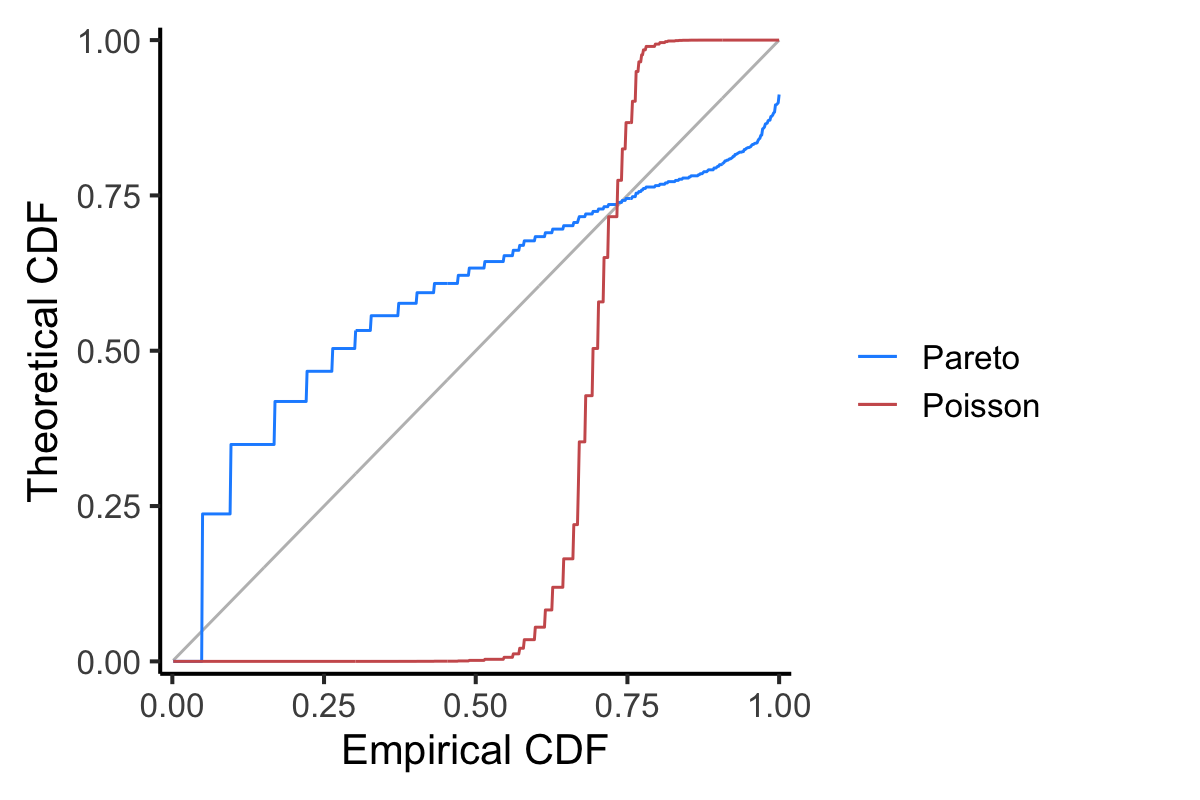}
    \end{subfigure}%
    \
    \begin{subfigure}[t]{0.4\textwidth}
        \centering
        \includegraphics[width=3.5in]{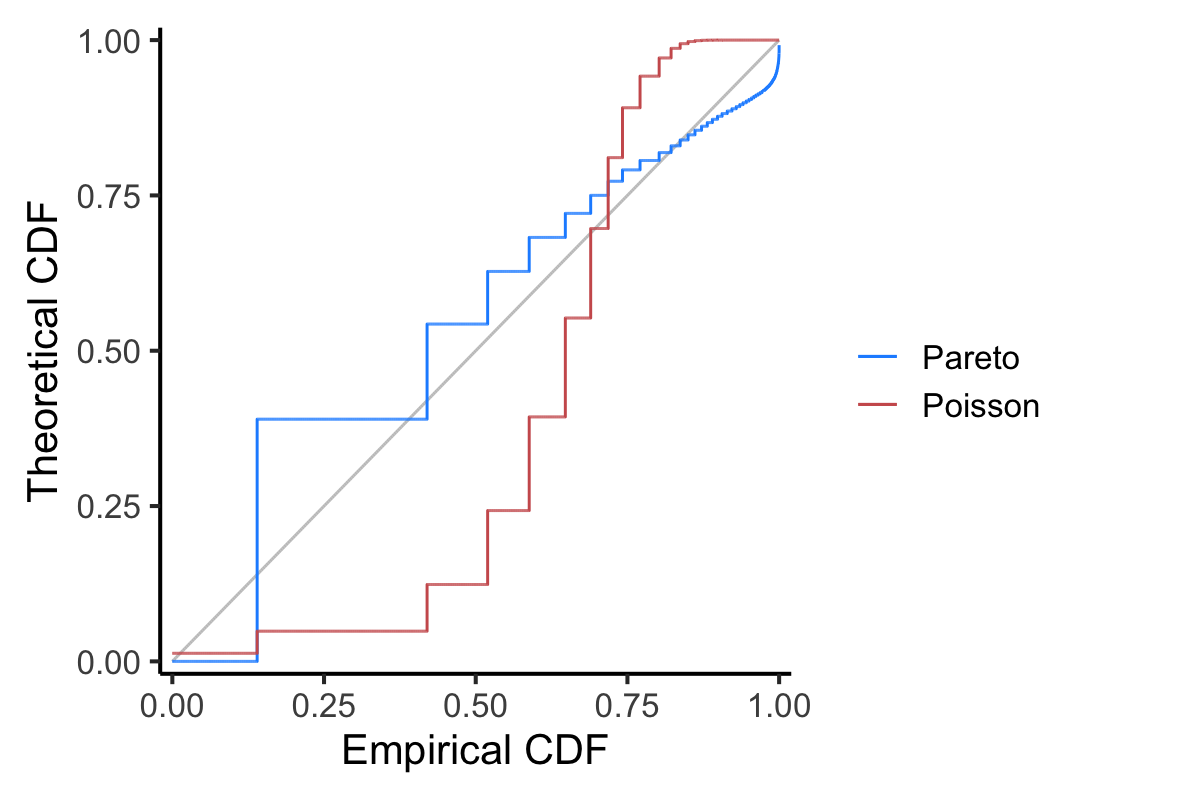}
    \end{subfigure}
    \caption{The top left figure shows the P-P plot for the Punjab, Pakistan dataset. The top right figure shows the P-P plot for the Punjab, India dataset. The bottom figure shows the P-P plot for the southern India dataset. Pareto distribution seems to be a better fit than the Poisson distribution in all datasets reinforcing our belief that real-world networks tend to have heavy tails.}
    \label{fig:pp-plots}
\end{figure*}

To compare with the Poisson and Pareto lifetime distributions we used in the simulations in Section \ref{section:sims}, we found the maximum likelihood estimates of the parameters in Poisson and Pareto distributions. These are shown in Table \ref{tab:estimated-lifetime-params}. To visualize how well they compare to the observed lifetimes, we use a P-P plot to compare the theoretical and empirical CDFs in Figure \ref{fig:pp-plots}. Just from these plots, it is unclear if either distribution is a good fit for the observed data. For instance, the Pareto distribution captures the third quantile in all cases and Poisson captures the median in the Punjab, Pakistan data but only the third quantile for the other two. The Poisson distribution tries to capture the median of the observation but is unable to account for the overdispersion. Pareto distribution does a better job of accounting for the overdispersion but loses out on the behavior around the median. It does appear to approximate the dataset from southern India well. Overall, these plots indicate that the distribution of the observed lifetimes does not have a heavy tail, which is indicated by the blue curve (corresponding to the Pareto distribution) lying below the grey curve as we approach 1 in all panels of Figure \ref{fig:pp-plots}.

%% file: sections/appendix_useful_results.tex
\section{Some useful results}

\begin{lemma}[Lemma 4.2 of \citet{bayati2020unreasonable}]
\label{lemma:bound-bernoulli-bad-event}
Let $\{X_i\}_{i=1}^{\infty}$ be an i.i.d. sequence of $\text{Bern}(\mu)$ random variables. Let $\sum_{i=1}^n X_i / n$ as the sample mean of the first $n$ random variables. For $\epsilon < 1/6$ and $\mu \geq 1 - \epsilon$,
\begin{align*}
    \prob(\exists n : M_n < 1 - 2 \epsilon) &\leq 1 - \frac{\exp(-0.3)}{2}
\end{align*}
\end{lemma}

\begin{lemma}[Lemma 4.3 of \citet{bayati2020unreasonable}]
\label{lemma:bound-bad-event}
Suppose that $P_{\mu}$ is a distribution with mean $\mu$ and support $[0, 1]$. Let $\{X_i\}_{i=1}^{\infty}$ be a sequence of i.i.d. random variables with distribution $P_{\mu}$. Define $M_n = \sum_{i=1}^n X_i / n$ as the sample mean of the first $n$ random variables. If $P_{\mu}$ is 1-subgaussian, then for any $\epsilon > 0$,
\begin{align*}
    \prob( \exists n : M_n < \mu - \epsilon) &\leq \exp(-\epsilon^2 / 2)
\end{align*}
\end{lemma}

\begin{lemma}[Lemma D.3 of \citet{bayati2020unreasonable}]
\label{lemma:bound-good-event}
Suppose that the distribution of $\mu \sim Q$ is $\gamma$-regular i.e., $\prob(\mu > 1 - \epsilon) = \Theta(\epsilon^{\gamma})$. Then,
\begin{align*}
    \E_{Q} \left[ \mathbb{I}(1 - \mu > 3 \epsilon) \min \left\{ \left( 1  + \frac{1}{1 - \mu - 2 \epsilon} \right), T(1 - \mu) \right\} \right] &\leq C_0 \begin{cases}
        5 + \log(1 / \epsilon), & \gamma = 1 \\
        C(\gamma), & \gamma > 1 \\
        C(\gamma) \min \left(\sqrt{T}, 1 / \epsilon \right)^{1 - \gamma}, & \gamma < 1
    \end{cases}
\end{align*}
where $C(\gamma)$ is a constant that depends only on $\gamma$.
\end{lemma}

%% file: feasible_contact_tracing.bbl
\begin{thebibliography}{49}
\providecommand{\natexlab}[1]{#1}
\providecommand{\url}[1]{\texttt{#1}}
\expandafter\ifx\csname urlstyle\endcsname\relax
  \providecommand{\doi}[1]{doi: #1}\else
  \providecommand{\doi}{doi: \begingroup \urlstyle{rm}\Url}\fi

\bibitem[Agrawal(1995)]{agrawal1995continuum}
R.~Agrawal.
\newblock The continuum-armed bandit problem.
\newblock \emph{SIAM journal on control and optimization}, 33\penalty0
  (6):\penalty0 1926--1951, 1995.

\bibitem[Agrawal and Goyal(2012)]{agrawal2012analysis}
S.~Agrawal and N.~Goyal.
\newblock Analysis of {T}hompson sampling for the multi-armed bandit problem.
\newblock In \emph{Conference on learning theory}, pages 39--1. JMLR {W}orkshop
  and {C}onference {P}roceedings, 2012.

\bibitem[Arinaminpathy et~al.(2020)Arinaminpathy, Das, McCormick, Mukhopadhyay,
  and Sircar]{arinaminpathy2020quantifying}
N.~Arinaminpathy, J.~Das, T.~McCormick, P.~Mukhopadhyay, and N.~Sircar.
\newblock Quantifying heterogeneity in {SARS-CoV-2} transmission during the
  lockdown in {I}ndia.
\newblock \emph{Medrxiv}, 2020.

\bibitem[Athey and Imbens(2019)]{athey2019machine}
S.~Athey and G.~W. Imbens.
\newblock Machine learning methods that economists should know about.
\newblock \emph{Annual Review of Economics}, 11:\penalty0 685--725, 2019.

\bibitem[Auer and Cesa-Bianchi(1998)]{auer1998line}
P.~Auer and N.~Cesa-Bianchi.
\newblock On-line learning with malicious noise and the closure algorithm.
\newblock \emph{Annals of mathematics and artificial intelligence},
  23:\penalty0 83--99, 1998.

\bibitem[Bastani et~al.(2021)Bastani, Drakopoulos, Gupta, Vlachogiannis,
  Hadjicristodoulou, Lagiou, Magiorkinis, Paraskevis, and
  Tsiodras]{bastani2021efficient}
H.~Bastani, K.~Drakopoulos, V.~Gupta, I.~Vlachogiannis, C.~Hadjicristodoulou,
  P.~Lagiou, G.~Magiorkinis, D.~Paraskevis, and S.~Tsiodras.
\newblock Efficient and targeted {COVID}-19 border testing via reinforcement
  learning.
\newblock \emph{Nature}, 599\penalty0 (7883):\penalty0 108--113, 2021.

\bibitem[Bayati et~al.(2020)Bayati, Hamidi, Johari, and
  Khosravi]{bayati2020unreasonable}
M.~Bayati, N.~Hamidi, R.~Johari, and K.~Khosravi.
\newblock Unreasonable effectiveness of greedy algorithms in multi-armed bandit
  with many arms.
\newblock \emph{Advances in Neural Information Processing Systems},
  33:\penalty0 1713--1723, 2020.

\bibitem[Besbes et~al.(2014)Besbes, Gur, and Zeevi]{besbes2014stochastic}
O.~Besbes, Y.~Gur, and A.~Zeevi.
\newblock Stochastic multi-armed-bandit problem with non-stationary rewards.
\newblock \emph{Advances in neural information processing systems}, 27, 2014.

\bibitem[Bolzoni et~al.(2007)Bolzoni, Real, and
  De~Leo]{bolzoni2007transmission}
L.~Bolzoni, L.~Real, and G.~De~Leo.
\newblock Transmission heterogeneity and control strategies for infectious
  disease emergence.
\newblock \emph{PLoS One}, 2\penalty0 (8):\penalty0 e747, 2007.

\bibitem[Bubeck et~al.(2012)Bubeck, Cesa-Bianchi, et~al.]{bubeck2012regret}
S.~Bubeck, N.~Cesa-Bianchi, et~al.
\newblock Regret analysis of stochastic and nonstochastic multi-armed bandit
  problems.
\newblock \emph{Foundations and Trends{\textregistered} in Machine Learning},
  5\penalty0 (1):\penalty0 1--122, 2012.

\bibitem[Carpentier and Kim(2015)]{carpentier2015adaptive}
A.~Carpentier and A.~K. Kim.
\newblock Adaptive and minimax optimal estimation of the tail coefficient.
\newblock \emph{Statistica Sinica}, pages 1133--1144, 2015.

\bibitem[Carpentier and Valko(2015)]{carpentier2015simple}
A.~Carpentier and M.~Valko.
\newblock Simple regret for infinitely many armed bandits.
\newblock In \emph{International Conference on Machine Learning}, pages
  1133--1141. PMLR, 2015.

\bibitem[Chakrabarti et~al.(2008)Chakrabarti, Kumar, Radlinski, and
  Upfal]{chakrabarti2008mortal}
D.~Chakrabarti, R.~Kumar, F.~Radlinski, and E.~Upfal.
\newblock Mortal multi-armed bandits.
\newblock \emph{Advances in {N}eural {I}nformation {P}rocessing {S}ystems},
  21:\penalty0 273--280, 2008.

\bibitem[Chapelle and Li(2011)]{chapelle2011empirical}
O.~Chapelle and L.~Li.
\newblock An empirical evaluation of {T}hompson sampling.
\newblock \emph{Advances in {N}eural {I}nformation {P}rocessing {S}ystems}, 24,
  2011.

\bibitem[Chugg and Ho(2021)]{chugg2021reconciling}
B.~Chugg and D.~E. Ho.
\newblock Reconciling risk allocation and prevalence estimation in public
  health using batched bandits.
\newblock \emph{arXiv preprint arXiv:2110.13306}, 2021.

\bibitem[Cohen et~al.(2007)Cohen, McClure, and Yu]{cohen2007should}
J.~D. Cohen, S.~M. McClure, and A.~J. Yu.
\newblock Should i stay or should i go? how the human brain manages the
  trade-off between exploitation and exploration.
\newblock \emph{Philosophical Transactions of the Royal Society B: Biological
  Sciences}, 362\penalty0 (1481):\penalty0 933--942, 2007.

\bibitem[Currie and MacLeod(2020)]{currie2020understanding}
J.~M. Currie and W.~B. MacLeod.
\newblock Understanding doctor decision making: The case of depression
  treatment.
\newblock \emph{Econometrica}, 88\penalty0 (3):\penalty0 847--878, 2020.

\bibitem[Danquah et~al.(2019)Danquah, Hasham, MacFarlane, Conteh, Momoh,
  Tedesco, Jambai, Ross, and Weiss]{danquah2019use}
L.~O. Danquah, N.~Hasham, M.~MacFarlane, F.~E. Conteh, F.~Momoh, A.~A. Tedesco,
  A.~Jambai, D.~A. Ross, and H.~A. Weiss.
\newblock Use of a mobile application for {E}bola contact tracing and
  monitoring in northern {S}ierra {L}eone: {A} proof-of-concept study.
\newblock \emph{BMC {I}nfectious {D}iseases}, 19\penalty0 (1):\penalty0 1--12,
  2019.

\bibitem[DiPrete et~al.(2011)DiPrete, Gelman, McCormick, Teitler, and
  Zheng]{diprete2011segregation}
T.~A. DiPrete, A.~Gelman, T.~McCormick, J.~Teitler, and T.~Zheng.
\newblock Segregation in social networks based on acquaintanceship and trust.
\newblock \emph{American journal of sociology}, 116\penalty0 (4):\penalty0
  1234--1283, 2011.

\bibitem[Frank and Zeckhauser(2007)]{frank2007custom}
R.~G. Frank and R.~J. Zeckhauser.
\newblock Custom-made versus ready-to-wear treatments: Behavioral propensities
  in physicians’ choices.
\newblock \emph{Journal of health economics}, 26\penalty0 (6):\penalty0
  1101--1127, 2007.

\bibitem[Gittins(1979)]{gittins1979bandit}
J.~C. Gittins.
\newblock Bandit processes and dynamic allocation indices.
\newblock \emph{Journal of the Royal Statistical Society Series B: Statistical
  Methodology}, 41\penalty0 (2):\penalty0 148--164, 1979.

\bibitem[Grushka-Cohen et~al.(2020)Grushka-Cohen, Cohen, Shapira, Moran-Gilad,
  and Rokach]{grushka2020framework}
H.~Grushka-Cohen, R.~Cohen, B.~Shapira, J.~Moran-Gilad, and L.~Rokach.
\newblock A framework for optimizing {COVID}-19 testing policy using a {M}ulti
  {A}rmed {B}andit approach.
\newblock \emph{arXiv preprint arXiv:2007.14805}, 2020.

\bibitem[Haan and Ferreira(2006)]{haan2006extreme}
L.~Haan and A.~Ferreira.
\newblock \emph{Extreme value theory: an introduction}, volume~3.
\newblock Springer, 2006.

\bibitem[Hagenaars et~al.(2004)Hagenaars, Donnelly, and
  Ferguson]{hagenaars2004spatial}
T.~Hagenaars, C.~Donnelly, and N.~Ferguson.
\newblock Spatial heterogeneity and the persistence of infectious diseases.
\newblock \emph{Journal of theoretical biology}, 229\penalty0 (3):\penalty0
  349--359, 2004.

\bibitem[Hill(1975)]{hill1975simple}
B.~M. Hill.
\newblock A simple general approach to inference about the tail of a
  distribution.
\newblock \emph{The annals of statistics}, pages 1163--1174, 1975.

\bibitem[Huerta and Tsimring(2002)]{huerta2002contact}
R.~Huerta and L.~S. Tsimring.
\newblock Contact tracing and epidemics control in social networks.
\newblock \emph{Physical Review E}, 66\penalty0 (5):\penalty0 056115, 2002.

\bibitem[Hyman et~al.(2003)Hyman, Li, and Stanley]{hyman2003modeling}
J.~M. Hyman, J.~Li, and E.~A. Stanley.
\newblock Modeling the impact of random screening and contact tracing in
  reducing the spread of {HIV}.
\newblock \emph{Mathematical {B}iosciences}, 181\penalty0 (1):\penalty0 17--54,
  2003.

\bibitem[Kim et~al.(2022)Kim, Vojnovic, and Yun]{kim2022rotting}
J.-h. Kim, M.~Vojnovic, and S.-Y. Yun.
\newblock Rotting infinitely many-armed bandits.
\newblock In \emph{International Conference on Machine Learning}, pages
  11229--11254. PMLR, 2022.

\bibitem[Lai and Robbins(1985)]{lai1985asymptotically}
T.~L. Lai and H.~Robbins.
\newblock Asymptotically efficient adaptive allocation rules.
\newblock \emph{Advances in applied mathematics}, 6\penalty0 (1):\penalty0
  4--22, 1985.

\bibitem[Langford and Zhang(2007)]{langford2007epoch}
J.~Langford and T.~Zhang.
\newblock The epoch-greedy algorithm for contextual multi-armed bandits.
\newblock \emph{Advances in neural information processing systems}, 20\penalty0
  (1):\penalty0 96--1, 2007.

\bibitem[Levine et~al.(2017)Levine, Crammer, and Mannor]{levine2017rotting}
N.~Levine, K.~Crammer, and S.~Mannor.
\newblock Rotting bandits.
\newblock \emph{Advances in neural information processing systems}, 30, 2017.

\bibitem[Lloyd-Smith et~al.(2005)Lloyd-Smith, Schreiber, Kopp, and
  Getz]{lloyd2005superspreading}
J.~O. Lloyd-Smith, S.~J. Schreiber, P.~E. Kopp, and W.~M. Getz.
\newblock Superspreading and the effect of individual variation on disease
  emergence.
\newblock \emph{Nature}, 438\penalty0 (7066):\penalty0 355--359, 2005.

\bibitem[McCormick et~al.(2010)McCormick, Salganik, and
  Zheng]{mccormick2010many}
T.~H. McCormick, M.~J. Salganik, and T.~Zheng.
\newblock How many people do you know?: {E}fficiently estimating personal
  network size.
\newblock \emph{Journal of the {A}merican {S}tatistical {A}ssociation},
  105\penalty0 (489):\penalty0 59--70, 2010.

\bibitem[Meister and Kleinberg(2021)]{meister2021optimizing}
M.~Meister and J.~Kleinberg.
\newblock Optimizing the order of actions in contact tracing.
\newblock \emph{arXiv preprint arXiv:2107.09803}, 2021.

\bibitem[Miller(2007)]{miller2007epidemic}
J.~C. Miller.
\newblock Epidemic size and probability in populations with heterogeneous
  infectivity and susceptibility.
\newblock \emph{Physical Review E}, 76\penalty0 (1):\penalty0 010101, 2007.

\bibitem[Newman and Park(2003)]{newman2003social}
M.~E. Newman and J.~Park.
\newblock Why social networks are different from other types of networks.
\newblock \emph{Physical review E}, 68\penalty0 (3):\penalty0 036122, 2003.

\bibitem[Pickands~III(1975)]{pickands1975statistical}
J.~Pickands~III.
\newblock Statistical inference using extreme order statistics.
\newblock \emph{the Annals of Statistics}, pages 119--131, 1975.

\bibitem[Reverdy et~al.(2014)Reverdy, Srivastava, and
  Leonard]{reverdy2014modeling}
P.~B. Reverdy, V.~Srivastava, and N.~E. Leonard.
\newblock Modeling human decision making in generalized gaussian multiarmed
  bandits.
\newblock \emph{Proceedings of the IEEE}, 102\penalty0 (4):\penalty0 544--571,
  2014.

\bibitem[Robbins(1952)]{robbins1952some}
H.~Robbins.
\newblock Some aspects of the sequential design of experiments.
\newblock \emph{Bulletin of the American Mathematical Society}, 58\penalty0
  (5):\penalty0 527--535, 1952.

\bibitem[Russo and Van~Roy(2014)]{russo2014learning}
D.~Russo and B.~Van~Roy.
\newblock Learning to optimize via posterior sampling.
\newblock \emph{Mathematics of Operations Research}, 39\penalty0 (4):\penalty0
  1221--1243, 2014.

\bibitem[Saurabh and Prateek(2017)]{saurabh2017role}
S.~Saurabh and S.~Prateek.
\newblock Role of contact tracing in containing the 2014 {E}bola outbreak: {A}
  {R}eview.
\newblock \emph{African {H}ealth {S}ciences}, 17\penalty0 (1):\penalty0
  225--236, 2017.

\bibitem[Seznec et~al.(2019)Seznec, Locatelli, Carpentier, Lazaric, and
  Valko]{seznec2019rotting}
J.~Seznec, A.~Locatelli, A.~Carpentier, A.~Lazaric, and M.~Valko.
\newblock Rotting bandits are no harder than stochastic ones.
\newblock In \emph{The 22nd International Conference on Artificial Intelligence
  and Statistics}, pages 2564--2572. PMLR, 2019.

\bibitem[Sutton and Barto(2018)]{sutton2018reinforcement}
R.~S. Sutton and A.~G. Barto.
\newblock \emph{Reinforcement learning: An introduction}.
\newblock MIT press, 2018.

\bibitem[Thompson(1933)]{thompson1933likelihood}
W.~R. Thompson.
\newblock On the likelihood that one unknown probability exceeds another in
  view of the evidence of two samples.
\newblock \emph{Biometrika}, 25\penalty0 (3-4):\penalty0 285--294, 1933.

\bibitem[Trac{\`a} et~al.(2020)Trac{\`a}, Rudin, and Yan]{traca2020reducing}
S.~Trac{\`a}, C.~Rudin, and W.~Yan.
\newblock Reducing exploration of dying arms in mortal bandits.
\newblock In \emph{Uncertainty in Artificial Intelligence}, pages 156--163.
  PMLR, 2020.

\bibitem[Wang et~al.(2008)Wang, Audibert, and Munos]{wang2008algorithms}
Y.~Wang, J.-Y. Audibert, and R.~Munos.
\newblock Algorithms for infinitely many-armed bandits.
\newblock \emph{Advances in Neural Information Processing Systems}, 21, 2008.

\bibitem[Wang et~al.(2020)Wang, Yahav, and Padmanabhan]{wang2020whom}
Y.~Wang, I.~Yahav, and B.~Padmanabhan.
\newblock {W}hom to {T}est? {A}ctive {S}ampling {S}trategies for {M}anaging
  {COVID}-19.
\newblock \emph{arXiv preprint arXiv:2012.13483}, 2020.

\bibitem[Wu et~al.(2018)Wu, Schulz, Speekenbrink, Nelson, and
  Meder]{wu2018generalization}
C.~M. Wu, E.~Schulz, M.~Speekenbrink, J.~D. Nelson, and B.~Meder.
\newblock Generalization guides human exploration in vast decision spaces.
\newblock \emph{Nature human behaviour}, 2\penalty0 (12):\penalty0 915--924,
  2018.

\bibitem[Zhao et~al.(2020)Zhao, Wen, Lin, Xuan, and Shroff]{zhao2020accuracy}
Q.~Zhao, H.~Wen, Z.~Lin, D.~Xuan, and N.~Shroff.
\newblock On the accuracy of measured proximity of bluetooth-based contact
  tracing apps.
\newblock In \emph{International Conference on Security and Privacy in
  Communication Systems}, pages 49--60. Springer, 2020.

\end{thebibliography}
